\documentclass[prd,nofootinbib,preprintnumbers]{revtex4-2}
\usepackage{epsfig,float}
\usepackage{amsfonts}
\usepackage{amsmath}
\usepackage{longtable}
\usepackage[colorlinks=true,linkcolor=red,citecolor=blue]{hyperref}

\input{epsf}
\newcommand{\be}{\begin{eqnarray}}
\newcommand{\ee}{\end{eqnarray}}
\newcommand{\ba}{\begin{array}}
\newcommand{\ea}{\end{array}}

\newcommand{\bea}{\begin{eqnarray}}
\newcommand{\eea
}{\end{eqnarray}}

\newcommand{\bi}{\begin{itemize}}
\newcommand{\ei}{\end{itemize}}
\newcommand{\nn}{{\nonumber}}

\restylefloat{figure}
\restylefloat{table}

\begin{document}

\title{Pion and photon beam initiated backward charmonium or lepton pair production}

\author{Bernard~Pire$^1$,  Kirill~M.~Semenov-Tian-Shansky$^{2,3,4}$,
Alisa~A. Shaikhutdinova$^{4}$,  Lech~Szymanowski$^{5}$ }
\affiliation{
$^1$ CPHT, CNRS, \'{E}cole Polytechnique, I.P. Paris,  91128 Palaiseau, France  \\
$^2$ Department of Physics, Kyungpook National University, Daegu 41566, Korea \\
$^3$ National Research Centre Kurchatov Institute: Petersburg Nuclear Physics Institute, 188300 Gatchina, Russia \\
$^4$ Higher School of Economics,
National Research University, 194100 St. Petersburg, Russia \\
$^5$ National Centre for Nuclear Research, NCBJ, 02-093 Warsaw, Poland
}

\preprint{CPHT-RR063.122022}

\begin{abstract}
Hard exclusive reactions initiated by pion or photon beams within the near-backward kinematical regime specified by the small Mandelstam variable $-u$
can be studied to access pion-to-nucleon and photon-to-nucleon Transition Distribution Amplitudes (TDAs). Checking the validity of
collinear factorized description of pion and photon induced reactions in terms of TDAs allows to test the universality of TDAs  
between the space-like and time-like regimes that is the indispensable feature of the QCD collinear factorization approach.

In this short review we consider the exclusive pion- and photo-production off nucleon of a highly virtual lepton pair (or heavy quarkonium) in the near-backward region.
We first employ a simplistic cross channel nucleon exchange model  of  pion-to-nucleon
TDAs  to estimate the magnitude of the corresponding
cross sections for the kinematical conditions of J-PARC. We then illustrate the flexibility of our approach by building a two parameter model for the photon-to-nucleon
TDAs based on recent results for near threshold
$J/\psi$ photoproduction at JLab and provide our estimates 
for near-backward $J/\psi$ photoproduction and Timelike Compton Scattering
cross sections for the kinematical conditions of JLab and of
future EIC and EIcC.

\end{abstract}

\maketitle
\thispagestyle{empty}
\renewcommand{\thesection}{\arabic{section}}
\renewcommand{\thesubsection}{\arabic{subsection}}



\section{Introduction}

Complementary to deep electroproduction reactions such as the much studied near-forward Deeply Virtual Compton Scattering (DVCS) and
Deeply Virtual Meson Production (DVMP), the  validity of the collinear factorization approach for hard exclusive reactions
and the relevance of the leading twist approximation analysis
can be challenged in studies of time-like reactions such as
near-forward photoproduction of lepton pairs\cite{Muller:1994ses, Berger:2001xd} and the exclusive limit of the Drell-Yan process in $\pi$ beam experiments \cite{Berger:2001zn,Sawada:2016mao}.
Analyticity (in $Q^2$) properties of the leading twist scattering amplitudes \cite{Mueller:2012sma}
exressed within the collinear factorization framework
in terms of Generalized Parton Distribution (GPDs) and meson distribution amplitudes (DAs) relate these time-like and space-like reactions.

The first experimental study of Timelike Compton Scattering (TCS) performed  at JLab was recently reported by the CLAS collaboration
\cite{CLAS:2021lky}.
The new results for $J/\psi$ photoproduction reactions near threshold 
\cite{GlueX:2019mkq} attracted much attention in the context of the studies of GPDs (see \cite{Lee:2022ymp} for a review)  and the energy momentum tensor of the nucleon
\cite{Sun:2021pyw}.

A natural extension of this research program is the study of the crossed ($t \leftrightarrow u$) channel counterpart reactions, \emph{i.e.} the same hard exclusive reactions in the
near-backward kinematics,   admitting a description in terms of
meson-to-nucleon~\cite{Pire:2016gut} or photon-to-nucleon~\cite{Pire:2022fbi} Transition Distribution Amplitudes (TDAs).
Nucleon-to-meson (and nucleon-to-photon) TDAs, as well as their crossed version, are defined as  transition matrix elements between a nucleon and a meson (or a photon)
states of the same non-local three-quark operator on the light-cone occurring in the definition of baryon DAs
(see Ref.~\cite{Pire:2021hbl} for a review). They
 arise within the collinear  factorization framework for hard exclusive electroproduction reactions
in near-backward kinematics~\cite{Gayoso:2021rzj}.

The physical contents of nucleon-to-meson and nucleon-to-photon TDAs is conceptually similar to that of GPDs enriched by the three-quark structure of the QCD operator which defines them. Since this operator  carries the quantum numbers of baryons, it provides access to the momentum  distribution of baryonic number inside hadrons. It intrinsically gives access to the non-minimal Fock components of hadronic light-front wave functions. Similarly to GPDs, by switching to the impact parameter space, one can address the distribution of the baryonic charge inside hadrons in the transverse plane. This also enables to study  the mesonic and electromagnetic clouds surrounding hadrons and provides new tools for \emph{microsurgery} and \emph{femtophotography} of hadrons.

In this paper we focus on the photon or pion beam induced hard exclusive reactions, being motivated by the
intensive experimental studies at JLab
\cite{Li:2020nsk,Gayoso:2021rzj}
and prospects to access backward hard exclusive reactions at J-PARC 
\cite{Aoki:2021cqa}
and future EIC
\cite{AbdulKhalek:2021gbh,Burkert:2022hjz} and EIcC \cite{Anderle:2021wcy}.
Also recent measurements of   the $J/\psi$ photoproduction cross section over the full near-threshold
kinematic region \cite{Adhikari:2023fcr} provide some more hints in favor of the manifestation of
a backward peak in exclusive cross sections.

We review the existing results for near-backward TCS \cite{Pire:2022fbi} and pion-beam-induced near-backward $J/\psi$ production
\cite{Pire:2016gut}
and present new cross section estimates  for near-backward $J/\psi$ photoproduction process and  pion beam induced
near-backward production of lepton pairs.

\section{Kinematics of pion and photon beam initiated backward reactions}
\label{Sec_Kinematics}


In this section we consider the near-backward kinematics regime for hard processes
\be
\left.
\begin{cases}{\pi(p_\pi)}
\\ {\gamma(q, \lambda_\gamma)}\end{cases}  \!\!\!\!\!\!\! \right\}
+ N(p_N,s_N) \to N'(p'_N,s'_N)+
\left.
\begin{cases}{\gamma^*(q',\lambda'_\gamma)
} \\ {J/\psi(p_\psi, \lambda_\psi)}\end{cases}  \!\!\!\!\!\!\!
\right\},
\label{Reaction_cases}
\ee
where $s_N$, $s'_N$ ($\lambda_\gamma$, $\lambda'_\gamma$,  $\lambda_\psi$)
stand for nucleon (photon or charmonium) polarization variables.

For definiteness, we choose to present the necessary formulas for
\be
\pi(p_\pi)+ N(p_N,s_N) \to N'(p'_N,s'_N)+\gamma^*(q', \lambda'_\gamma).
\label{Def_React_kin}
\ee
\bi
\item The expressions for the case of the photoproduction reactions can be obtained
with the obvious change
\be
p_\pi \to q; \ \ \ p_\pi^2= m_\pi^2 \to q^2=0.
\ee
\item The set of kinematical formulas when the final state virtual photon $\gamma^*(q', \lambda'_\gamma)$ is replaced  by the heavy quarkonium $J/\psi(p_\psi, \lambda_\psi)$
can be obtained by changing
\be
q' \to p_\psi; \ \ \ {q'}^2=Q'^2 \to p_\psi^2=  {M_\psi^2}.
\ee
\ei

To describe the $2 \to 2$ hard subprocess
(\ref{Def_React_kin}) we employ the standard Mandelstam variables
\be
s= (p_\pi +p_N)^2 \equiv W^2; \ \ t=(p'_N-p_N)^2; \ \  u=(p'_N-p_\pi)^2;
\ee
with $s+t+u=2m_N^2 +m_\pi^2+Q'^2$.
The
$z$-axis is chosen along the direction of the pion beam in the
meson-nucleon center-of-momentum (CMS) frame.
We introduce the
light-cone vectors
$p, n$ ($p^2=n^2=0$)
satisfying
$ 2p \cdot n =1$.
The Sudakov decomposition of the relevant momenta  reads
\bea
&&
p_\pi = (1+\xi) p +\frac{m_\pi^2}{1+\xi}n\, ;
\nonumber \\ &&
p_N =
\frac{2(1+\xi)m_N^2}{W^2+\Lambda(W^2,m_N^2,m_\pi^2)-m_N^2-m_\pi^2}\,
p +
\frac{W^2+\Lambda(W^2,m_N^2,m_\pi^2)-m_N^2-m_\pi^2}{2(1+\xi)} \,
n\,;   \nonumber \\ &&
\Delta \equiv (p'_N-p_\pi)   =
-2\xi p + \left( \frac{ m_N^2-\Delta_T^2}{1-\xi } - \frac{m_\pi^2}{1+\xi}\right)n+\Delta_T;
\nonumber \\ &&
q' =  p_N - \Delta \, ; \ \ \ \
p'_N= p_\pi +\Delta,
\label{Sudakov_decomposition}
\eea
where
\be
\Lambda(x,y,z)= \sqrt{x^2+y^2+z^2-2xy-2xz-2yz}
\label{Def_Mandelstam_f}
\ee
is the Mandelstam function and
$m_N$
and
$m_\pi$
stand respectively for the nucleon and pion masses.
The transverse direction in
(\ref{Sudakov_decomposition})
is defined with respect to the
$z$
direction and
$\xi$
is the  skewness variable that specifies the longitudinal momentum transfer in the $u$-channel:
\be
\xi \equiv  -\frac{(p'_N-p_\pi) \cdot n}{(p'_N+p_\pi) \cdot n}.
\label{Def_xi}
\ee
The transverse invariant momentum transfer $\Delta_T^2 \le 0$ is expressed as
\be
\Delta_T^2= \frac{1-\xi}{1+\xi} \left( u-2\xi \left[ \frac{m_\pi^2}{1+\xi}-
\frac{m_N^2}{1-\xi} \right] \right).
\label{Def_DT2}
\ee

Within the collinear factorization framework we neglect both the pion
and nucleon masses with respect to the hard scale introduced by  $Q'$ (or $M_\psi$)
and
$W$
and set
$\Delta_T=0$
within the coefficient functions.
This results in the approximate expression for the  skewness variable
(\ref{Def_xi}):
\be
\xi \simeq \frac{Q'^2}{2 W^2-Q'^2} \simeq   \frac{\tau}{2 -\tau},
\label{Xi_collinear}
\ee
where $\tau$ is the time-like analogue of the Bjorken variable
\be
\tau \equiv \frac{Q'^2}{2 p_N \cdot p_\pi}= \frac{Q'^2}{W^2-m_N^2-m_\pi^2}.
\label{Def_tau}
\ee

It  is also instructive to consider the exact kinematics of the reaction
(\ref{Def_React_kin})
in the
$\pi N$
CMS frame. In this frame the relevant momenta read:
\be
&&
p_\pi= \left(     \frac{W^2+m_\pi^2-m_N^2}{2W}    , \, \vec{p}_\pi\right);
\ \ \ \ \
q'= \left(  \frac{W^2+Q'^2-m_N^2}{2W}, \, -\vec{p}'_N \right); \nonumber
\\ &&
p_N= \left( \frac{W^2+m_N^2-m_\pi^2}{2W} , \, -\vec{p}_\pi \right);
\ \ \ \ \
p'_N= \left(  \frac{W^2+m_N^2-Q'^2}{2W}, \, \vec{p}'_N \right),
\ee
where
\be
|\vec{p}_\pi|= \frac{\Lambda(W^2,m_N^2,m_\pi^2)}{2W}; \ \ \
|\vec{p}'_N|=\frac{\Lambda(W^2,m_N^2,Q'^2)}{2W}.
\ee
The CMS scattering angle
$\theta_u^*$
is defined as the angle between
$\vec{p}_\pi$
and
$\vec{p}'_N$:
\be
\cos \theta_u^*= \frac{2W^2(u-m_N^2-m_\pi^2)+(W^2+m_\pi^2-m_N^2)(W^2+m_N^2-Q'^2)}
{\Lambda(W^2,m_N^2,m_\pi^2)\Lambda(W^2,m_N^2,Q'^2)}.
\label{CosTheta_exact}
\ee
The transverse momentum transfer squared
(\ref{Def_DT2})
is then given by
\be
\Delta_T^2=- \frac{\Lambda^2(W^2,Q'^2,m_N^2)}{4W^2}(1-\cos^2 \theta_u^*);
\label{DeltaT2_exact}
\ee
and the physical domain for the reaction
(\ref{Def_React_kin})
is defined from the requirement that
$\Delta_T^2 \le 0 $.

\bi
\item
In particular, the backward kinematics regime
$\theta_u^*=0$
corresponds to
$\vec{p}'_N$
along
$\vec{p}_\pi$,
which means that
$\gamma^*$
is produced along
$-\vec{p}_\pi$ {\it i.e.}
in the backward direction. In this case
$u$
reaches its maximal value
\be
&&
u_{0} \equiv \frac{2 \xi(m_\pi^2(\xi-1)+m_N^2(\xi+1))}{\xi^2-1}
\nonumber \\ &&
=m_N^2+m_\pi^2- \frac{(W^2+m_\pi^2-m_N^2)(W^2+m_N^2-Q'^2)}{2W^2}
+2|\vec{p}_\pi||\vec{q}'|.
\label{Def_umax}
\ee
At the same time
$t=(p'_N-p_N)^2$
reaches its minimal value
$t_{1}$
($W^2+u_{0}+t_{1}=2m_N^2+m_\pi^2+Q'^2$).

Note that $u$ is negative, and therefore
$|u_{0}|$
is the minimal possible absolute value of the momentum transfer squared.
It is for
$u \sim u_{0}$
that one may expect to satisfy the requirement
$|u| \ll W^2, \, Q'^2$
which is crucial for the validity of the factorized description of
(\ref{Def_React_kin})
in terms of
$\pi \to N$
TDAs and nucleon DAs.

\item Another limiting value
 $\theta_u^*=\pi$
corresponds to
$\vec{p}'_N$ along $-\vec{p}_\pi$
{\it i.e}
$\gamma^*(q')$
produced in the forward direction.
In this case $u$ reaches its minimal value
\be
u_{1}=m_N^2+m_\pi^2- \frac{(W^2+m_\pi^2-m_N^2)(W^2+m_N^2-Q'^2)}{2W^2}
-2|\vec{p}_\pi||\vec{q}'|.
\ee
At the same time
$t$
reaches its maximal value
$t_{0}$.
The factorized description in terms of
$\pi \to N$
TDAs does not apply in this case as
$|u|$
turns out to be of order of
$W^2$.
\ei

\section{Pion-to-nucleon and photon-to-nucleon   TDAs}
\label{Sec_def_TDAs}

\begin{figure}[H]
 \begin{center}
\includegraphics[width=0.45\textwidth]{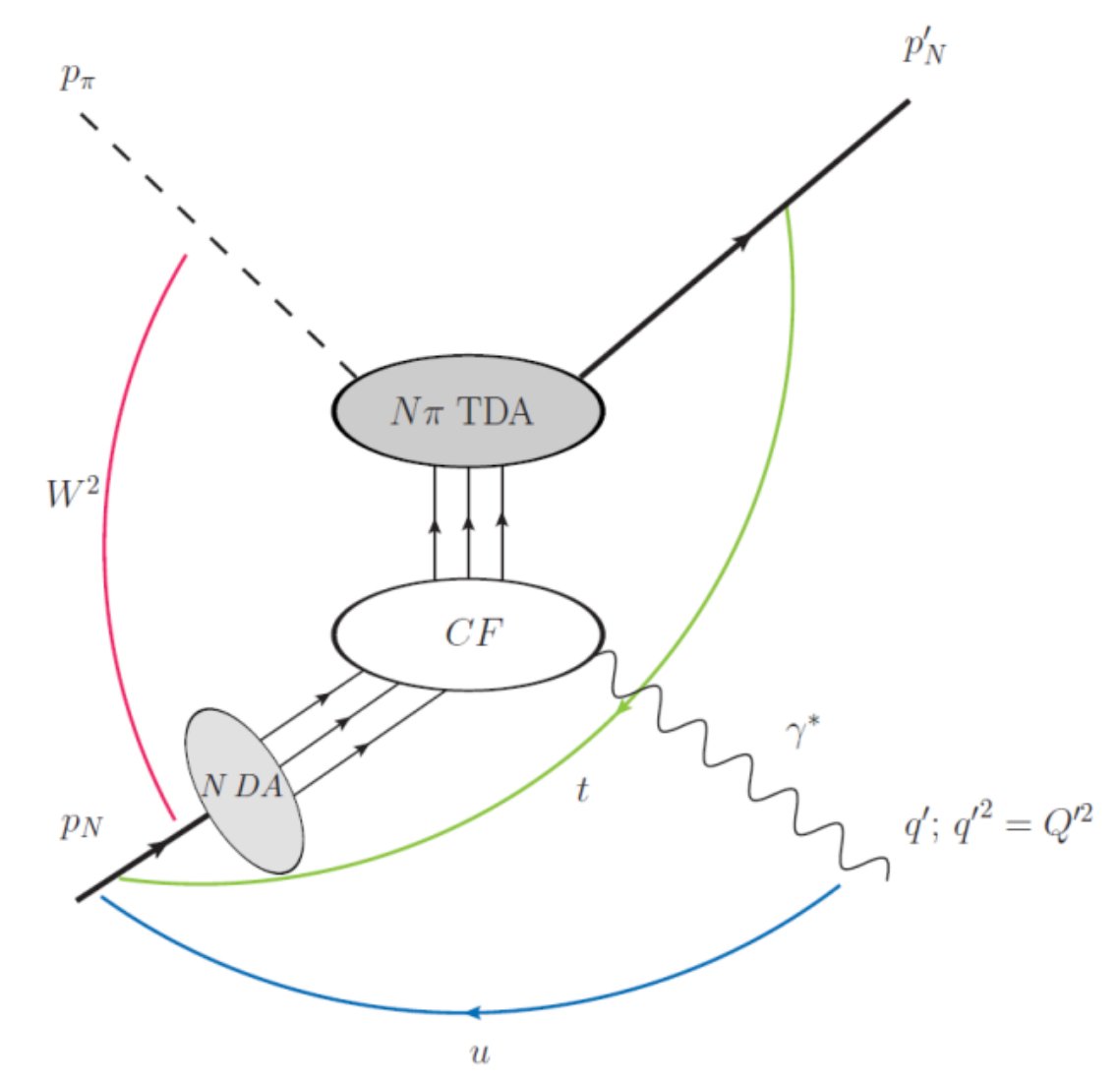}
\ \ \ \ \ \
\includegraphics[width=0.45\textwidth]{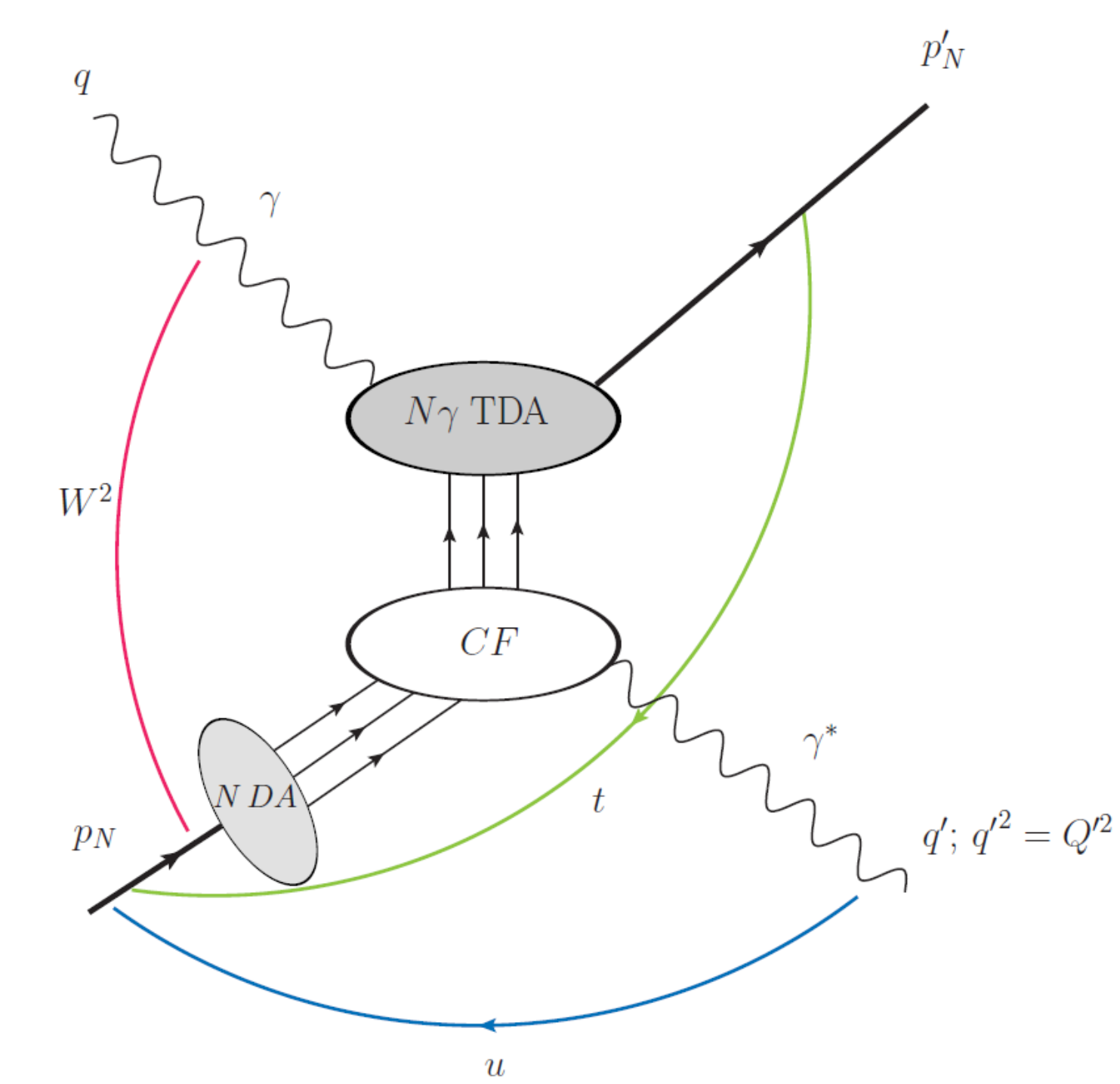}
 \end{center}
    \caption{
     Collinear factorization mechanism for the hard subprocess of near-backward lepton pair production induced by pion beam
     $\pi N \to N' \gamma^*$
      (left panel) and backward TCS  $\gamma N \to N' \gamma^*$ (right panel);
      (large scale is provided by $Q'^2$ and $W^2 \equiv (p_N+q)^2 $; fixed $\tau$ (\ref{Def_tau});   $|u|\equiv |(p_N-q')^2| \sim 0$);
 $N\pi$  ($N\gamma$) TDA stands for the transition distribution amplitudes from a pion-to-nucleon (photon-to-nucleon); $N$ DA stands for the nucleon distribution amplitude. CF denotes the
 coefficient function (hard convolution kernel).}
\label{Fig_TDAfact_gammastar}
\end{figure}

\begin{figure}[H]
 \begin{center}
\includegraphics[width=0.45\textwidth]{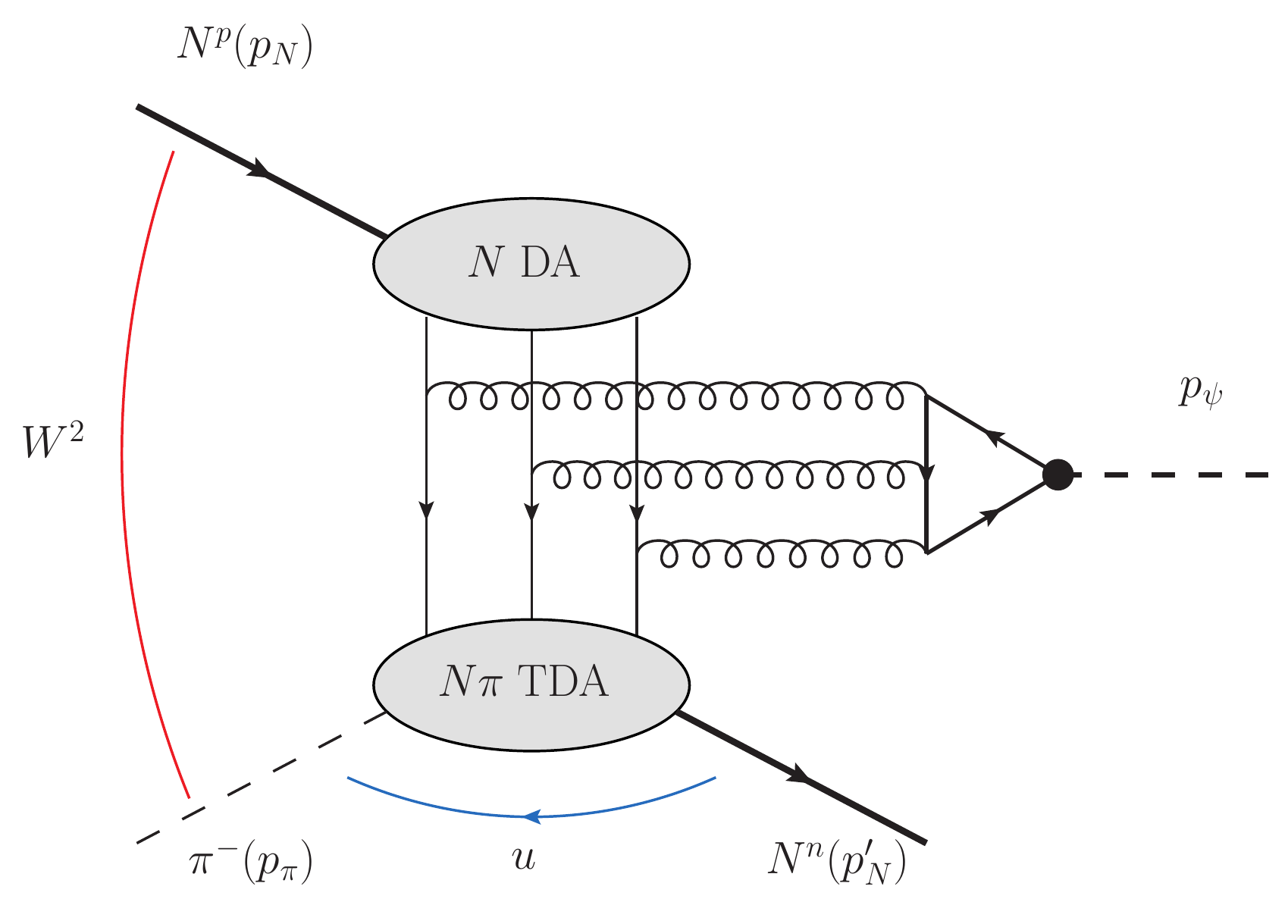}
\ \ \ \ \ \
\includegraphics[width=0.43\textwidth]{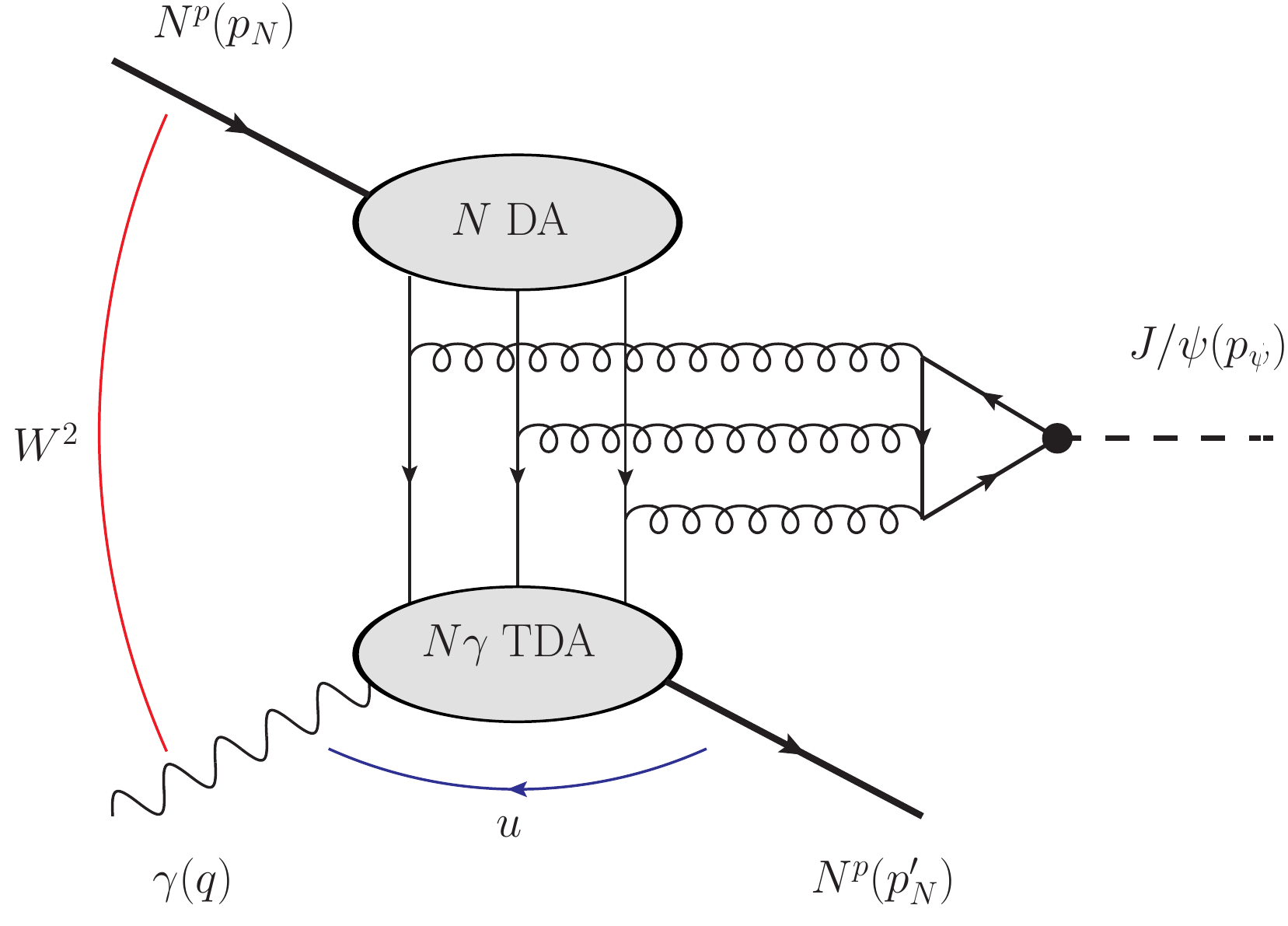}
 \end{center}
    \caption{
    Collinear factorization mechanism of charmonium photoproduction ($\gamma N \to J/\psi N' $)  in the  near-backward  kinematical regime (large scale is provided by $M_\psi^2$ and $W^2 \equiv (p_N+q)^2 $; fixed $\tau$ (\ref{Def_tau});   $|u|\equiv |(p'_N-q)^2| \sim 0$);
 $N\gamma$ TDA stands for the transition distribution amplitudes from a photon-to-a-nucleon; $N$ DA stands for the nucleon distribution amplitude. Black dots denote the non-relativistic light-cone wave function of heavy quarkoinum (\ref{WFNR}). }
\label{Fig_TDAfact_Jpsi}
\end{figure}

In Figs.~\ref{Fig_TDAfact_gammastar}, \ref{Fig_TDAfact_Jpsi} we present the hard scattering mechanisms for the near-backward kinematical regime for hard
reactions (\ref{Reaction_cases})
(see Sec.~\ref{Sec_Kinematics}) involving pion-to-nucleon ($N \pi$), respectively photon-to-nucleon ($N \gamma$),  TDAs and nucleon DAs.

$N \pi$ and $N \gamma $  TDAs are defined as Fourier transforms of matrix elements of non-local three-antiquark light-cone operator between a pion (or a photon) state and
a nucleon state.
For definiteness we consider $\pi^- \to n$ and $\gamma \to p$ $\bar{u} \bar{u} \bar{d}$ TDAs defined with the trilocal light-cone operator%
\footnote{We assume the use of the light-cone gauge $A^+ \equiv 2 (A \cdot n)=0$ and omit the Wilson lines along the light-like path.}
\begin{equation}
\hspace{2em}\widehat{O}_{\rho \tau \chi}^{\bar u \bar u \bar d}(\lambda_1n,\lambda_2n,\lambda_3n) =\varepsilon_{c_1 c_2 c_3}
 \bar u^{c_1}_{\rho}(\lambda_1 n)
\bar u^{c_2}_{\tau}(\lambda_2 n)\bar d^{c_3}_{\chi}(\lambda_3 n).
\end{equation}
Here $c_{1,2,3}$ stand for the color group indices; and $\rho$, $\tau$, $\chi$
denote the Dirac indices.
Other isospin channels can be worked out with help of the isospin symmetry relations worked out in Ref.~\cite{Pire:2011xv}.

The number of  leading twist TDAs matches the number of independent helicity amplitudes
$T_{h_{\bar u} h_{\bar u},h_{\bar d}}^{ h_N }$
and
$T_{h_{\bar u} h_{\bar u},h_{\bar d}}^{ h_\gamma , h_N }$
for $\pi \,  u u d \to N^p$  and $\gamma \,  u u d \to N^p$ process, where
$h_{\bar q}$, $h_\gamma$ and $h_N$
refer to the light-cone helicity of, respectively, quark, initial state photon and the
final state nucleon.

To the leading twist-$3$ accuracy, the parametrization of pion-to-nucleon TDAs
involves $8$ independent $N \pi$ TDAs
\begin{eqnarray}
&&
4 {\cal F} \langle  N^n(p_N,s_N)| \widehat{O}_{\rho \tau \chi}^{\bar u \bar u \bar d}(\lambda_1n,\lambda_2n,\lambda_3n)| \pi^- (p_\pi) \rangle
\nonumber \\ &&
\hspace{2em}
= \delta(x_1+x_2+x_3-2\xi)  \times i \frac{f_N}{f_\pi}  
 \Big[
\sum_{\Upsilon= 1,2 } (v^{N \pi }_\Upsilon)_{\rho \tau, \, \chi} V_{\Upsilon}^{N \pi}(x_1,x_2,x_3, \xi, \Delta^2; \, \mu^2)
\nonumber \\ &&
+\sum_{\Upsilon= 1,2  } (a^{N \pi}_\Upsilon)_{\rho \tau, \, \chi} A_{\Upsilon}^{N \pi}(x_1,x_2,x_3, \xi, \Delta^2; \, \mu^2)
+
\sum_{\Upsilon= 1,2,3,4 } (t^{N \pi}_\Upsilon)_{\rho \tau, \, \chi} T_{\Upsilon}^{N \pi}(x_1,x_2,x_3, \xi, \Delta^2; \, \mu^2)
\Big]\,,
\label{Def_N_pi_TDAs_param}
\end{eqnarray}
where the  Fourier transform operation is defined as
$4 {\cal F} \equiv 4(p \cdot n)^{3} \int\left[\prod_{j=1}^{3} \frac{d \lambda_{j}}{2 \pi}\right] e^{-i \sum_{k=1}^{3} x_{k} \lambda_{k}(p \cdot n)}$;
$f_N=5.0 \times 10^{-3}$~GeV$^2$
is the nucleon light-cone wave function normalization constant
\cite{Chernyak:1987nv};
and
$f_\pi=93~{\rm MeV}$ is the pion weak decay constant.
The explicit expressions for the Dirac structures $\{v^{N \pi },\,a^{N \pi },\, t^{N \pi } \}$ are presented in Appendix~\ref{App_Crossing}.
Each of TDAs is a function of three momentum fraction variables $x_i$; skewness variable $\xi$
(\ref{Def_xi}), $u$-channel invariant momentum transfer $\Delta^2$
and of the factorization scale $\mu^2$.

The parametrization of photon-to-nucleon TDAs
involves $16$ independent $N \gamma$ TDAs
$V_{\Upsilon}^{N \gamma}, \, A_{\Upsilon}^{N \gamma}, \, T_{\Upsilon}^{N \gamma}$:
\begin{eqnarray}
&&
4 {\cal F} \langle  N^p(p_N,s_N)| \widehat{O}_{\rho \tau \chi}^{\bar u \bar u \bar d}(\lambda_1n,\lambda_2n,\lambda_3n)|\gamma(q, \lambda_\gamma) \rangle
\nonumber \\ &&
\hspace{2em}
= \delta(x_1+x_2+x_3-2\xi)  \times m_N 
\Big[
\sum_{\Upsilon= 1 {\cal E}, 1T, \atop 2 {\cal E}, 2T  } (v^{N \gamma }_\Upsilon)_{\rho \tau, \, \chi} V_{\Upsilon}^{N \gamma}(x_1,x_2,x_3, \xi, \Delta^2; \, \mu^2)
\nonumber \\ &&
+\sum_{\Upsilon= 1 {\cal E}, 1T,  \atop 2 {\cal E}, 2T  } (a^{N \gamma}_\Upsilon)_{\rho \tau, \, \chi} A_{\Upsilon}^{N \gamma}(x_1,x_2,x_3, \xi, \Delta^2; \, \mu^2)
+
\sum_{\Upsilon= 1 {\cal E}, 1T,    2{\cal E}, 2T, \atop 3 {\cal E}, 3T,  4 {\cal E}, 4T } (t^{N \gamma}_\Upsilon)_{\rho \tau, \, \chi} T_{\Upsilon}^{N \gamma}(x_1,x_2,x_3, \xi, \Delta^2; \, \mu^2)
\Big].
\nonumber
\label{Def_N_gamma_TDAs_param}
\end{eqnarray}

The crossing relations expressing pion-to-nucleon and photon-to-nucleon through nucleon-to-pion and nucleon-to-photon
TDAs occurring in the description of electroproduction reactions are summarized in Appendix~\ref{App_Crossing}.
We refer the reader to Sec.~4 of the review paper \cite{Pire:2021hbl} for a detailed overview of
symmetry, support and evolution properties of TDAs
and their physical contents and interpretation.

\section{Amplitudes and cross sections of pion beam induced reactions}
\label{Sec_Pi_beam_reactions}
\mbox

In this Section we present a set of formulas for  pion beam induced
near-backward production of a highly virtual lepton pair (or of heavy quarkonium).
For this issue we consider the hard subprocesses
\be
\pi(p_\pi)
+ N(p_N,s_N) \to N(p'_N,s'_N)+
\left.
\begin{cases}{\gamma^*(q',\lambda'_\gamma)
} \\ {J/\psi(p_\psi, \lambda_\psi)}\end{cases}  \!\!\!\!\!\!\!
\right\}.
\ee

Within the $u$-channel factorized description in terms of $N \pi$ TDAs (and nucleon DAs)  to the leading order in
$\alpha_s$,
the amplitude of  the hard $ \pi N   \rightarrow N' \gamma^{*} $ subprocess
$\mathcal{M}^{\lambda'_\gamma}_{s_N s'_N}$
is expressed as
\begin{equation}
\mathcal{M}^{\lambda'_\gamma}_{s_N s'_{N}}(\pi N \to N' \gamma^*)=
\mathcal{C}_\pi
\frac{1}{Q'^4}
\Bigl[
\mathcal{S}_{s_N s'_{N}}^{(1)\, \lambda'_\gamma  }
\mathcal{J}^{(1)}_{\pi N \to N' \gamma^*}(\xi, \Delta^2)
-
\mathcal{S}_{s_N s'_{N}}^{(2) \, \lambda'_\gamma}
\mathcal{J}^{(2)}_{\pi N \to N' \gamma^*}(\xi, \Delta^2)
\Bigr],
\label{Def_ampl_pi_photon}
\end{equation}
with the overall normalization constant $\mathcal{C}_\pi$
\begin{equation}
{\mathcal{C}}_\pi\equiv-i \frac{(4 \pi \alpha_s)^2 \sqrt{4 \pi \alpha_{em}} f_N^2 }{54 f_\pi },
\label{Def_C_PiN}
\end{equation}
where
$\alpha_{em} = \frac{e^2}{4 \pi} \simeq \frac{1}{137}$  is the electromagnetic fine structure constant,
$\alpha_s \simeq 0.3$ is the strong coupling.

The spin structures
$\mathcal{S}^{(k) \, \lambda'_\gamma}_{s_N s'_N}$, $k=1,\,2$
are defined as
\bea
  &&
\mathcal{S}^{(1) \, \lambda'_\gamma}_{s_N s_{{N}}} \equiv
\bar{U}(p'_{N},s'_{N}) \hat{\mathcal{E}}^{*}_\gamma(q',\lambda'_\gamma) \gamma_5 U(p_N,s_N);
\nonumber \\   &&
\mathcal{S}^{(2) \, \lambda'_\gamma}_{s_N s'_N} \equiv
\frac{1}{m_N}
\bar{U}(p'_{N},s'_{N}) \hat{\mathcal{E}}^{*}_\gamma(q', \lambda'_\gamma) \hat{\Delta}_T \gamma_5 U(p_N,s_N),
\label{Def_S_Def_ampl_pi_photon}
\eea
where
${\mathcal{E}}_\gamma(q,\lambda_\gamma)$
stands for the polarization vector of the virtual photon; $U$  are the
nucleon Dirac spinors; and the standard Dirac's ``hat'' notations are adopted.
$\mathcal{J}^{(k)}_{\pi N \to N' \gamma^*}$ $k=1,2$
denote the convolution integrals of
$\pi N$
TDAs and antinucleon DAs with the hard scattering kernels computed from the set of
$21$
relevant scattering diagrams (see~\cite{Lansberg:2007ec}):
\be
&&
{\mathcal{J}}^{(k)}_{\pi N \to N' \gamma^*}
(\xi,\Delta^2) \nn \\ && =
{\int^{1+\xi}_{-1+\xi} }\! \! \!d_3x  \; \delta \left( \sum_{j=1}^3 x_j-2\xi \right)
{\int^{1}_{0} } \! \! \! d_3y \; \delta \left( \sum_{l=1}^3 y_l-1 \right) \;
{\Biggl(2\sum_{\alpha=1}^{7}   R_{\alpha \; \pi N \to N' \gamma^*}^{(k)} +
\sum_{\alpha=8}^{14}   R_{\alpha \; \pi N \to N' \gamma^*}^{(k)} \Biggr)}. \nn \\ &&
\label{Def_JandJprime}
\ee
The integrals  in
$x_i$'s ($y_i$'s) in
(\ref{Def_JandJprime})
stand  over the support of
$N \pi$
TDA (nucleon DA).
Within the $u$-channel factorization regime  of
$\pi N \to N' \gamma^*$
the coefficients
$R_\alpha^{(k)}$
($\alpha=1,\ldots,14$)
correspond to the coefficients
$T_\alpha^{(k)}=D_\alpha \times N_\alpha^{(k)}$,
that can be read off from Table~2 of Ref.~\cite{Pire:2021hbl},
with the replacement of nucleon-to-pion ($\pi N$) TDAs by pion-to-nucleon ($N \pi$) TDAs multiplied by an irrelevant charge conjugation phase factor; and
with the modification $-i 0 \to i 0$ of the regulating prescription in the denominators of hard scattering kernels $D_\alpha$.
The latter changes mirror the difference between the electroproduction hard exclusive reactions with spacelike $\gamma^*$ and 
pion-production
hard exclusive reactions in which $\gamma^*$ is timelike.

Now, the cross section of the near-backward lepton pair production reaction
\be
\pi(p_\pi)+ N(p_N,s_N) \to N'(p'_N,s'_N)+\gamma^*(q', \lambda'_\gamma) \to  N'(p'_N,s'_N) + \ell^+(k_{\ell^+})+ \ell^-(k_{\ell^-})
\label{React_pi_lept}
\ee
can be expressed as
\begin{equation}
d \sigma = \frac{1}{2 (2 \pi)^5
{\Lambda(W^2,m_N^2,m_\pi^2)}}
| \overline{\mathcal{M}_{ \pi N   \rightarrow N' \ell^+ \ell^-   }}| ^2
\frac{d \Omega^{*}_{N'}}{8 W^2 }
{\Lambda(W^2,Q'^2,m_N^2)} \frac{d \Omega_\ell}{8} dQ'^2,
\end{equation}
where
$d\Omega^{*}_{N'} \equiv  d \cos \theta_{N'}^{*} d \varphi_{N'}^{*}$
is the final nucleon solid angle
in the $ \pi N $ CMS.
By $d \Omega_\ell
\equiv
d \cos \theta_\ell d \varphi_\ell$
we denote the produced lepton solid angle in $\ell^+ \ell^-$ CMS (corresponding to the rest frame of the virtual photon). By expressing
$ \cos \theta_{N'}^{*}$
through
$u = (p'_{N}-p_\pi)^2$ as
\begin{equation}
du = \frac{d \cos \theta_{N'}^{*}}{2 W^2}
{\Lambda(W^2,m^2_N,m_\pi^2)}
{\Lambda(W^2,Q^2,m_N^2)}
\end{equation}
and  integrating over the azimuthal angle
$\varphi_{N'}^{*}$
of the produced nucleon and over the azimuthal angle of the lepton
$\varphi_\ell$
the following formula for the unpolarized differential cross section of the reaction
(\ref{React_pi_lept})
is established:
\begin{equation}
\frac{d^3 \sigma}{du d Q^2 d \cos \theta_{\ell}}=
 \frac{\int d \varphi_\ell | \overline{\mathcal{M}_{ \pi N   \rightarrow  N' \ell^+ \ell^- }}| ^2 }
 {\Lambda^2(W^2,m_N^2,m_\pi^2) (2 \pi)^4}.
 \label{Unpol_CS_React_pi_lept}
\end{equation}

The average-squared amplitude
$| \overline{\mathcal{M}_{ \pi N   \rightarrow N' \ell^+ \ell^-   }}| ^2$
is expressed through the helicity amplitudes of the hard subprocess
$\mathcal{M}^{\lambda'_\gamma}_{s_N s'_{{N}}}(\pi N \to N' \gamma^*) $
(\ref{Def_ampl_pi_photon})
\begin{equation}
| \overline{\mathcal{M}_{ \pi N   \rightarrow  N' \ell^+ \ell^-  }}| ^2=
\frac{1}{2}
\sum_{s_N, \, s'_{{N}}, \, \lambda'_\gamma}
\frac{1}{Q^4}
e^2
\mathcal{M}^{\lambda'_\gamma}_{s_N s'_{{N}}}
 {\rm Tr}
\left\{
\hat{k}_{\ell^-} \hat{\mathcal{E}}(q',\lambda'_\gamma) \hat{k}_{\ell^+} \hat{\mathcal{E}}^{*}(q',\lambda'_\gamma)
\right\}
\left( \mathcal{M}^{\lambda'_\gamma}_{s_N s'_{{N}}} \right)^{*},
\label{WF_Cross_sec}
\end{equation}
where the factor $\frac{1}{2}$ corresponds to averaging over polarization
of the initial state nucleon.

Within the factorized description in terms of $N\pi $ TDAs (and nucleon DAs)
to the leading twist accuracy, only the contribution of
transverse polarization of the virtual photon
is relevant.
By computing the leptonic trace in
(\ref{WF_Cross_sec})
in the
$\ell^+ \ell^-$
CMS,
and integrating over the lepton polar angle $\varphi_\ell$,
we obtain:
\begin{equation}
 \int d \varphi_{\ell} \, \vert \overline{\mathcal{M}_{ \pi N   \rightarrow  N' \ell^+ \ell^-  }}\vert ^2
\Big|\mbox{}_{{\rm Leading }\,{\rm  twist}-3}=
\vert \overline{\mathcal{M}_{T}(\pi N \to N' \gamma^*)}\vert ^2 \frac{2 \pi e^2(1+\cos^2 \theta_\ell)}{Q'^2},
\label{MT_squared_React_pi_lept}
\end{equation}
where
\begin{eqnarray}
  &&
| \overline{\mathcal{M}_{T}(\pi N \to N' \gamma^*)}| ^2 = \frac{1}{2} \sum_{s_N, \, s'_{N}, \, {\lambda^{'T}_{\gamma }}}
\mathcal{M}^{\lambda'_\gamma}_{s_N s'_{N}} (\pi N \to N' \gamma^*)
\left(\mathcal{M}^{\lambda'_\gamma}_{s_N s'_{{N}}} (\pi N \to N' \gamma^*) \right)^{*} \nn \\ &&
=\frac{1}{2}
| {\mathcal{C}}_\pi| ^2 \frac{1}{Q'^6} \left(
\frac{2  (1+\xi)}{\xi} \big|  \mathcal{J}^{(1)}_{\pi N \to N' \gamma^*} (\xi,\Delta^2) \big|  ^2
-\frac{2 (1+\xi)}{\xi} \frac{\Delta_T^2}{m_N^2} \big|  \mathcal{J}^{(2)}_{\pi N \to N' \gamma^*} (\xi,\Delta^2) \big|  ^2 +{\mathcal{O}} \left (   1/Q'^2    \right) \, \right).
\nn \\ &&
\label{M_T_piN_squared_React_pi_lept}
\end{eqnarray}
Here $\mathcal{J}^{(1,2)}$ are the integral convolutions defined in
(\ref{Def_JandJprime})
and
$\mathcal{C}_\pi$
is the overall normalization constant (\ref{Def_C_PiN}).

Now, following
Ref.~\cite{Pire:2016gut},
we review the basic  amplitude and cross section formulas for near-backward charmonium
production  in
$\pi N$
collisions  within the TDA framework assuming the collinear factorization
reaction mechanism depicted in the left panel of
Fig.~\ref{Fig_TDAfact_Jpsi}.
The leading order and leading twist amplitude
$\mathcal{M}^{\lambda_\psi}_{s_N s'_N}$
of the hard reaction
$ \pi N   \rightarrow N' J/\psi $
admits the following parametrization
\bea
&&
{\cal M}_{s_N s'_N}^{\lambda_\psi}(\pi N   \rightarrow N' J/\psi )= {\cal C}^\psi_\pi \frac{1}{{\bar M}^5 } \Big[
\mathcal{S}^{(1) \, \lambda_\psi}_{s_N s'_N}
{\cal J}^{(1)}_{\pi N   \rightarrow N' J/\psi '}(\xi,\Delta^2)
-\mathcal{S}^{(2) \, \lambda_\psi}_{s_N s'_N}
{\cal J}^{(2)}_{\pi N \to N' J/\psi}(\xi,\Delta^2)
\Big], \nn \\ &&
\label{Amplitude_pi_Jpsi}
\eea
where  the average mass $\bar{M}$  approximately equals the charmonium mass that is
roughly twice the mass of the charmed quark:
\be
 \bar{M}= 3\, {\rm GeV} \simeq M_\psi \simeq 2m_c.
\ee
The spin structures
$\mathcal{S}^{(k) \, \lambda_\psi}_{s_N s'_N}$, $k=1,\,2$
occurring in (\ref{Amplitude_pi_Jpsi})
are the same as in Eq.~(\ref{Def_S_Def_ampl_pi_photon})
with the virtual photon polarization vector
${\mathcal{E}}^{*}_\gamma(q', \lambda'_\gamma)$
replaced by the charmonium polarization vector
${\mathcal{E}}^{*}_\psi(p_\psi, \lambda_\psi)$.
The explicit expressions for the convolution integrals
${\cal J}^{(1,2)}_{\pi N   \rightarrow N' J/\psi '}(\xi,\Delta^2)$
are presented in Appendix~\ref{App_Jpsi_pion_prod}.
The normalization  constant ${\cal C}_\pi^\psi$ reads
\be
{\cal C}_\pi^\psi
=\left(4 \pi \alpha_s\right)^3 \frac{f^2_N    f_\psi}{f_\pi} \frac{10}{81},
\label{Def_CpiPSI}
\ee
where $f_\psi$ determines the normalization
of  the non-relativistic light-cone wave function of heavy quarkonium
\cite{Chernyak:1983ej}:
\begin{equation}
\Phi_{\rho \tau}(z, p_\psi,\lambda_\psi)= \langle 0|  \bar{c}_\tau(z) c_\rho(-z) |  J/\psi (p_{\psi}, \lambda_\psi) \rangle
=
\frac{1}{4} f_\psi
\left[ 2 m_c \hat{\mathcal{E}}_\psi(p_{\psi}, \lambda_\psi) +\sigma_{ p_{\psi} \nu}   {\mathcal{E}}^\nu_\psi(p_{\psi}, \lambda_\psi)
\right]_{\rho \tau},
\label{WFNR}
\end{equation}
The normalization constant $f_\psi$ can be extracted from the charmonium
leptonic decay width  $\Gamma(J/ \psi \to e^+ e^-)$.
With the values quoted in Ref.~\cite{Workman:2022ynf}, we get $f_\psi= 415.5 \pm 4.9$~MeV.

To work out the cross section formula we
square the amplitude
(\ref{Amplitude_pi_Jpsi})
and average (sum) over spins of initial (final) nucleon.
Staying at the leading twist-$3$ accuracy we account for the
production of transversely  polarized
$J/\psi$. Summing over the transverse polarizations we find
\bea
|\overline{\mathcal{M}_{T}(\pi N   \rightarrow N' J/\psi )}|^2 =
\sum_{\lambda_\psi^T} \left( \frac{1}{2} \sum_{s_N s'_N}
{\cal M}_{s_N s'_N}^{\lambda_{\psi  }}(\pi N   \rightarrow N' J/\psi ) ({\cal M}_{s_N s'_N}^{\lambda_{\psi  }})^*(\pi N   \rightarrow N' J/\psi ) \right).
\nonumber  \\ &&
\eea
The leading twist-3 differential cross section of
$\pi +N   \to J/\psi + N'$
then reads
\bea
&&
\frac{d \sigma}{d u}= \frac{1}{16 \pi \Lambda^2(s,m_N^2,m_\pi^2) } 
|\overline{\mathcal{M}_{T}(\pi N   \rightarrow N' J/\psi )}|^2
\nonumber  \\ &&
=\frac{1}{16 \pi \Lambda^2(s,m_N^2,m_\pi^2)}
\frac{1}{2} |\mathcal{C}_\pi^\psi|^2 \frac{2(1+\xi)}{\xi {\bar{M}}^8}  \left( |\mathcal{J}^{(1)}_{\pi N \to N' J/\psi}(\xi, \Delta^2)|^2 - \frac{\Delta_T^2}{m_N^2} |\mathcal{J}^{(2)}_{\pi N \to N' J/\psi}(\xi, \Delta^2)|^2 \right), \nonumber  \\ &&
\label{CS_def_delta2}
\eea
where
$\xi$
is the $u$-channel skewness variable (\ref{Def_xi})
\be
\xi \simeq  \frac{\bar{M}^2}{2 W^2- \bar{M}^2};
\ee
and $\Lambda(x,y,z)$
is defined in
(\ref{Def_Mandelstam_f}).

\section{Amplitude and cross sections of photoproduction reactions}
\label{Sec_photon_beam_reactions}
\mbox

In this Section we review the near-backward  photoproduction off nucleon of a highly virtual lepton pair or of heavy quarkonium.
The hard subprocesses of these two reactions
\be
\gamma(q, \lambda_\gamma)
+ N(p_N,s_N) \to N(p'_N,s'_N)+
\left.
\begin{cases}{\gamma^*(q',\lambda'_\gamma)
} \\ {J/\psi(p_\psi, \lambda_\psi)}\end{cases}  \!\!\!\!\!\!\!
\right\}
\ee
are considered within the collinear factorization framework. The corresponding hard scattering
mechanisms are depicted  in the right panels of Figs.~\ref{Fig_TDAfact_gammastar}  and \ref{Fig_TDAfact_Jpsi} respectively.

For the backward Timelike Compton Scattering (TCS) reaction
\be
\gamma(q, \lambda_\gamma) + N(p_N,s_N) \to N'(p'_N,s'_N) + \gamma^*(q', \lambda'_\gamma)
\label{HardReact_bTCS}
\ee
we closely follow the exposition of Ref.~\cite{Pire:2022fbi}.
The  helicity amplitudes  ${\cal M}^{\lambda_\gamma \lambda'_\gamma}_{s_N s'_N}$ of
(\ref{HardReact_bTCS})
involve $4$ independent tensor structures%
\footnote{The indexes $k=1,3,4,5$ were chosen to match the notations established for the description of near-backward vector meson electroproduction in Ref.~\cite{Pire:2015kxa}.}
\be
{\cal M}_{s_N s'_N}^{\lambda_\gamma \lambda'_\gamma}(\gamma N \to N' \gamma^*)=
{\cal C}_V
\frac{1}{Q'^4}
\sum_{k=1,3,4,5}
{\cal S}_{s_N s'_N}^{(k) \, \lambda_\gamma \lambda'_\gamma} {\cal J}_{ \gamma N \to N' \gamma^* }^{(k)} (\xi, \, \Delta^2).
\label{Hel_ampl_def_photo_gamma}
\ee
with the overall normalization constant $\mathcal{C}_V$
\begin{equation}
{\mathcal{C}}_V\equiv-i \frac{(4 \pi \alpha_s)^2 \sqrt{4 \pi \alpha_{em}} f_N m_N }{54}.
\label{Def_C_V}
\end{equation}

There is one tensor structure independent of $\Delta_T$:
\be
&&
{\cal S}_{s_N s'_N}^{(1) \, \lambda_\gamma \lambda'_\gamma}=
\bar{U}(p'_N,s'_N) \hat{\cal E}^*(q',\lambda'_\gamma) \hat{\cal E}(q, \lambda_\gamma) U(p_N,s_N)\,,
\label{S_photon_Dindep}
\ee
and three
$\Delta_T$-dependent tensor structures:
\be
&&
{\cal S}_{s_N s'_N}^{(3) \, \lambda_\gamma \lambda'_\gamma}= \frac{1}{m_N} ({\cal E}(q, \lambda_\gamma) \cdot \Delta_T) \, \bar{U}(p'_N,s'_N) \hat{{\cal E}}^*(q',\lambda'_\gamma)  U(p_N,s_N);
\nonumber \\ &&
{\cal S}_{s_N s'_N}^{(4) \, \lambda_\gamma \lambda'_\gamma}=\frac{1}{m_N^2} ({\cal E}(q, \lambda_\gamma) \cdot \Delta_T) \, \bar{U}(p'_N,s'_N) \hat{{\cal E}}^*(q',\lambda') \hat{\Delta}_T U(p_N,s_N);
\nonumber \\
&&
{\cal S}_{s_N s'_N}^{(5)\,\lambda_\gamma \lambda'_\gamma}= \bar{U}(p'_N,s'_N) \hat{{\cal E}}^*(q',\lambda'_\gamma) \hat{\cal E}^*(q, \lambda_\gamma) \hat{\Delta}_T U(p_N,s_N)\,.
\label{S_photon_Ddep}
\ee
The explicit expressions for the convolution integrals ${\cal J}_{ \gamma N \to N' \gamma^* }^{(k)} (\xi, \, \Delta^2)$ can be read off from the expressions for the
set of 21 relevant scattering diagrams summarized in Table~1 of \cite{Pire:2022fbi}.

The cross section of the near-backward lepton pair production reaction
\be
\gamma(q, \lambda_\gamma) + N(p_N,s_N) \to  N'(p'_N,s'_N) +\gamma^*(q', \lambda'_\gamma) \to N'(p'_N,s'_N) + \ell^+(k_{\ell^+})+ \ell^-(k_{\ell^-})
\label{HardReact_bTCSleptons}
\ee
integrated over the final nucleon azimuthal angle $\varphi_{N'}^*$ and lepton azimuthal angle $\varphi_\ell$ reads
\be
\frac{d^3 \sigma}{du d Q'^2 d \cos \theta_{\ell}}=
\frac{\int d \varphi_\ell \overline{|{\cal M}_{ \gamma N \to N' \ell^+ \ell^-}|^2} }{64
(s-m_N^2)^2
(2 \pi)^4}\,,
\label{CS_main_formula}
\ee
with  the average-squared amplitude
$\overline{|{\cal M}_{ \gamma N \to N' \ell^+ \ell^-}|^2}$
expressed as
\be
&&
\overline{|{\cal M}_{ \gamma N \to N' \ell^+ \ell^-}|^2} =
\frac{1}{4}
\sum_{s_N s'_N {\lambda_\gamma  \lambda'_\gamma}} \frac{1}{Q'^4} e^2
{\cal M}^{\lambda_\gamma \lambda'_\gamma}_{s_N s'_N}(\gamma N \to N' \gamma^*) \nn \\ && \times
{\rm Tr}
\left\{
\hat{k}_{\ell^-} {\cal E}(q', \lambda'_\gamma) \hat{k}_{\ell^+} {\cal E}^*(q', \lambda'_\gamma)
\right\}
\left( {\cal M}^{\lambda_\gamma \lambda'_\gamma}_{s_N s'_N}(\gamma N \to N' \gamma^*) \right)^* . \nn \\ &&
\label{Average_sq_Ampl}
\ee

To the leading twist-3 accuracy, the averaged-squared amplitude
(\ref{Average_sq_Ampl})
integrated
over the lepton azimuthal angle
$\varphi_\ell$
reads
\be
\int d \varphi_\ell \overline{|{\cal M}_{ \gamma N \to N' \ell^+ \ell^-}|^2} \Big|_{\text{Leading twist-3}}=
|\overline{{\cal M}_T(\gamma N \to N' \gamma^*)}|^2 \frac{2 \pi e^2 (1+  \cos^2 \theta_\ell)}{Q'^2},
\label{M2_leading_tw}
\ee
where
\be
&&
|\overline{{\cal M}_T(\gamma N \to N' \gamma^*)}|^2 =
\frac{1}{4}
\sum_{s_N s'_N {\lambda  \lambda'}}
{\cal M}^{\lambda \lambda'}_{s_N s'_N}
\left( {\cal M}^{\lambda \lambda'}_{s_N s'_N} \right)^* 
=\frac{1}{4} |{\cal C}_V|^2 \frac{1}{Q'^6}
\frac{2(1+\xi)}{\xi}
\Big[
2|{\cal J}^{(1)}_{ \, \gamma N \to N' \gamma^*}|^2 \nn \\ && + \frac{\Delta_T^2}{m_N^2} \Big\{
-|{\cal J}^{(3)}_{ \, \gamma N \to N' \gamma^*}|^2+ \frac{\Delta_T^2}{m_N^2} |{\cal J}^{(4)}_{ \, \gamma N \to N' \gamma^*}|^2+
\left( {\cal J}^{(4)}_{ \, \gamma N \to N' \gamma^*} {\cal J}^{(1)*}_{ \, \gamma N \to N' \gamma^*}+{\cal J}^{(4)*}_{ \, \gamma N \to N' \gamma^*} {\cal J}^{(1)}_{ \, \gamma N \to N' \gamma^*}\right) \nn \\ &&
-2 |{\cal J}^{(5)}_{ \, \gamma N \to N' \gamma^*}|^2 -
\left( {\cal J}^{(5)}_{ \, \gamma N \to N' \gamma^*} {\cal J}^{(3)*}_{ \, \gamma N \to N' \gamma^*}+{\cal J}^{(5)*}_{ \, \gamma N \to N' \gamma^*} {\cal J}^{(3)}_{ \, \gamma N \to N' \gamma^*}\right)
\Big\} + {\cal O}(1/Q'^2)
\Big];
\label{TransAmpl_squared_Cross}
\ee
and the $(1 + \cos^2 \theta_\ell)$  dependence reflects the dominance of transversely polarized virtual photon.

In a similar way, the  helicity amplitudes  ${\cal M}^{\lambda_\gamma \lambda_\psi}_{s_N s'_N}(\gamma N \to N' J/\psi)$ of the hard  reaction
\be
\gamma(q, \lambda_\gamma) + N(p_N,s_N) \to  N'(p'_N,s'_N) + J/\psi(p_\psi, \lambda_\psi)
\label{HardReactJpsiphoton}
\ee
involve the same $4$ independent tensor structures (\ref{S_photon_Dindep}), (\ref{S_photon_Ddep}) with virtual
photon polarization vector
${\cal E}(q', \lambda'_\gamma)$
replaced by that of the heavy quarkonium ${\cal E}(p_\psi, \lambda_\psi)$:
\be
{\cal M}_{s_N s'_N}^{\lambda_\gamma \lambda_\psi}(\gamma N \to N' J/\psi)=
{\cal C}^\psi_V
\frac{1}{Q'^4}
\sum_{k=1,3,4,5}
{\cal S}_{s_N s'_N}^{(k) \, \lambda_\gamma \lambda_\psi} {\cal J}_{ \gamma N \to N' J/\psi }^{(k)} (\xi, \, \Delta^2).
\label{Hel_ampl_def_photo_Jpsi}
\ee
The normalization  constant ${\cal C}^\psi_V$ reads
\be
{\cal C}^\psi_V
=\left(4 \pi \alpha_s\right)^3  f_N  m_N  f_\psi  \frac{10}{81}.
\label{Def_CVpsi}
\ee

To the leading order in $\alpha_s$, the hard amplitude of chramonium photoproduction is calculated from the $3$ Feynman diagrams
analogous to the case of charmonium production with a pion beam.
The explicit expressions for the integral convolutions ${\cal J}_{ \gamma N \to N' J/\psi }^{(k)}$ are presented in Appendix~\ref{App_B2}.
The averaged-squared amplitude (\ref{Hel_ampl_def_photo_Jpsi}), to the leading twist-3 accuracy, is expressed as
\be
&&
|\overline{{\cal M}_T(\gamma N \to N' J/\psi)}|^2 =
\frac{1}{4}
\sum_{s_N s'_N {\lambda_\gamma  \lambda_{\psi }}}
{\cal M}^{\lambda_\gamma  \lambda_{\psi}}_{s_N s'_N}
\left( {\cal M}^{\lambda_\gamma \lambda_\psi}_{s_N s'_N} \right)^* 
=\frac{1}{4} |{\cal C}_\psi^V|^2 \frac{1}{\bar{M}^8}
\frac{2(1+\xi)}{\xi}
\Big[
2|{\cal J}^{(1)}_{\, \gamma N \to N' J/\psi}|^2 \nn \\ && + \frac{\Delta_T^2}{m_N^2} \Big\{
-|{\cal J}^{(3)}_{\, \gamma N \to N' J/\psi}|^2+ \frac{\Delta_T^2}{m_N^2} |{\cal J}^{(4)}_{\, \gamma N \to N' J/\psi}|^2+
\left( {\cal J}^{(4)}_{\, \gamma N \to N' J/\psi} {\cal J}^{(1)*}_{\, \gamma N \to N' J/\psi}+{\cal J}^{(4)*}_{\, \gamma N \to N' J/\psi} {\cal J}^{(1)}_{\, \gamma N \to N' J/\psi}\right) \nn \\ &&
-2 |{\cal J}^{(5)}_{\, \gamma N \to N' J/\psi}|^2 -
\left( {\cal J}^{(5)}_{\, \gamma N \to N' J/\psi} {\cal J}^{(3)*}_{\, \gamma N \to N' J/\psi}+{\cal J}^{(5)*}_{\, \gamma N \to N' J/\psi} {\cal J}^{(3)}_{\, \gamma N \to N' J/\psi}\right)
\Big\} + {\cal O}(1/Q'^2)
\Big];
\label{TransAmplJpsi_squared_Cross}
\ee
and the differential cross section of the reaction
(\ref{HardReactJpsiphoton}) reads
\be
\frac{d \sigma}{d u}=\frac{1}{16 \pi \Lambda^2\left(W^2, m_N^2, 0\right)}\left|\overline{\mathcal{M}_T(\gamma N \to N' J/\psi)}\right|^2.
\ee


\section{Estimates of pion-beam-induced near-backward lepton pair and charmonium production cross sections}

\begin{figure}[H]
\begin{center}
\includegraphics[width=0.33\textwidth]{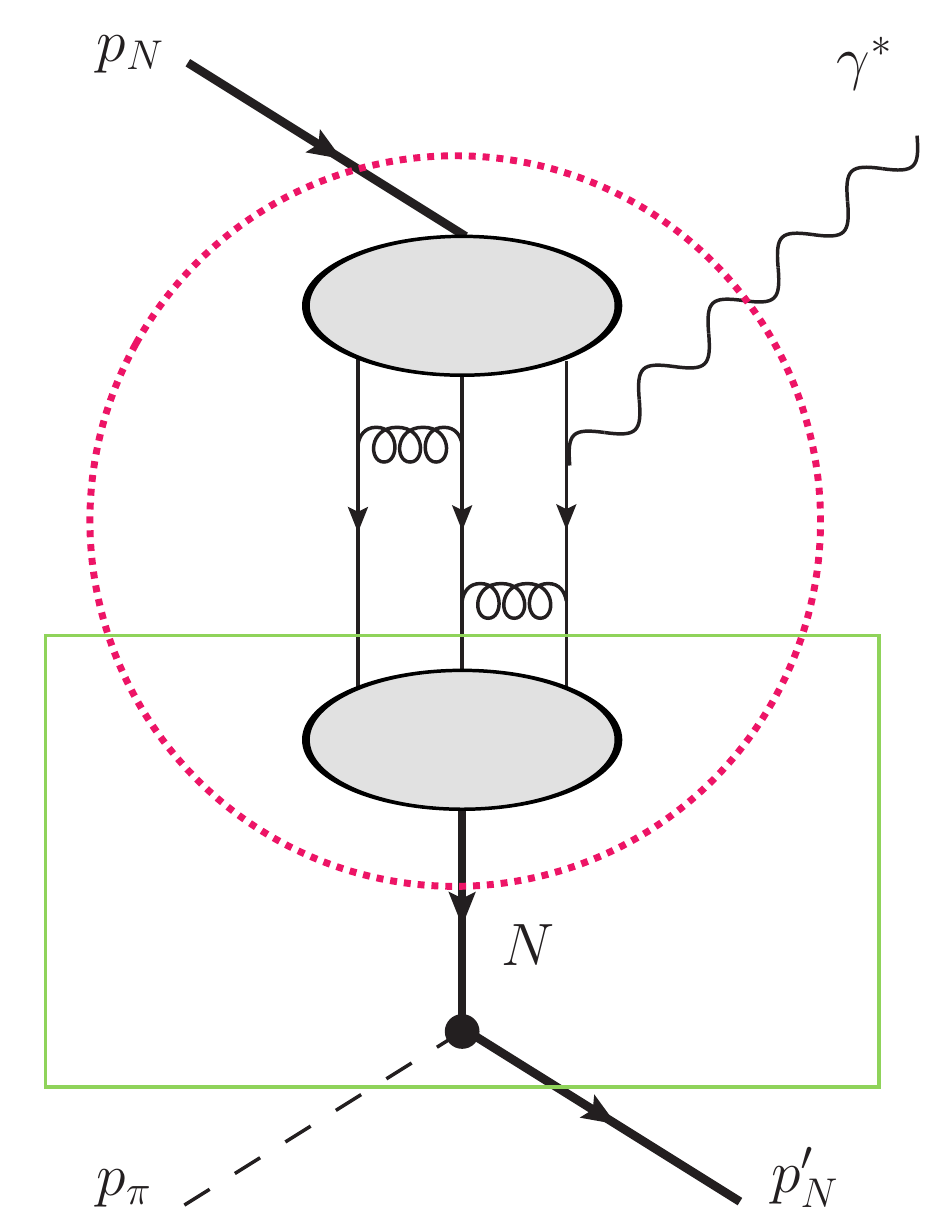}
\includegraphics[width=0.4\textwidth]{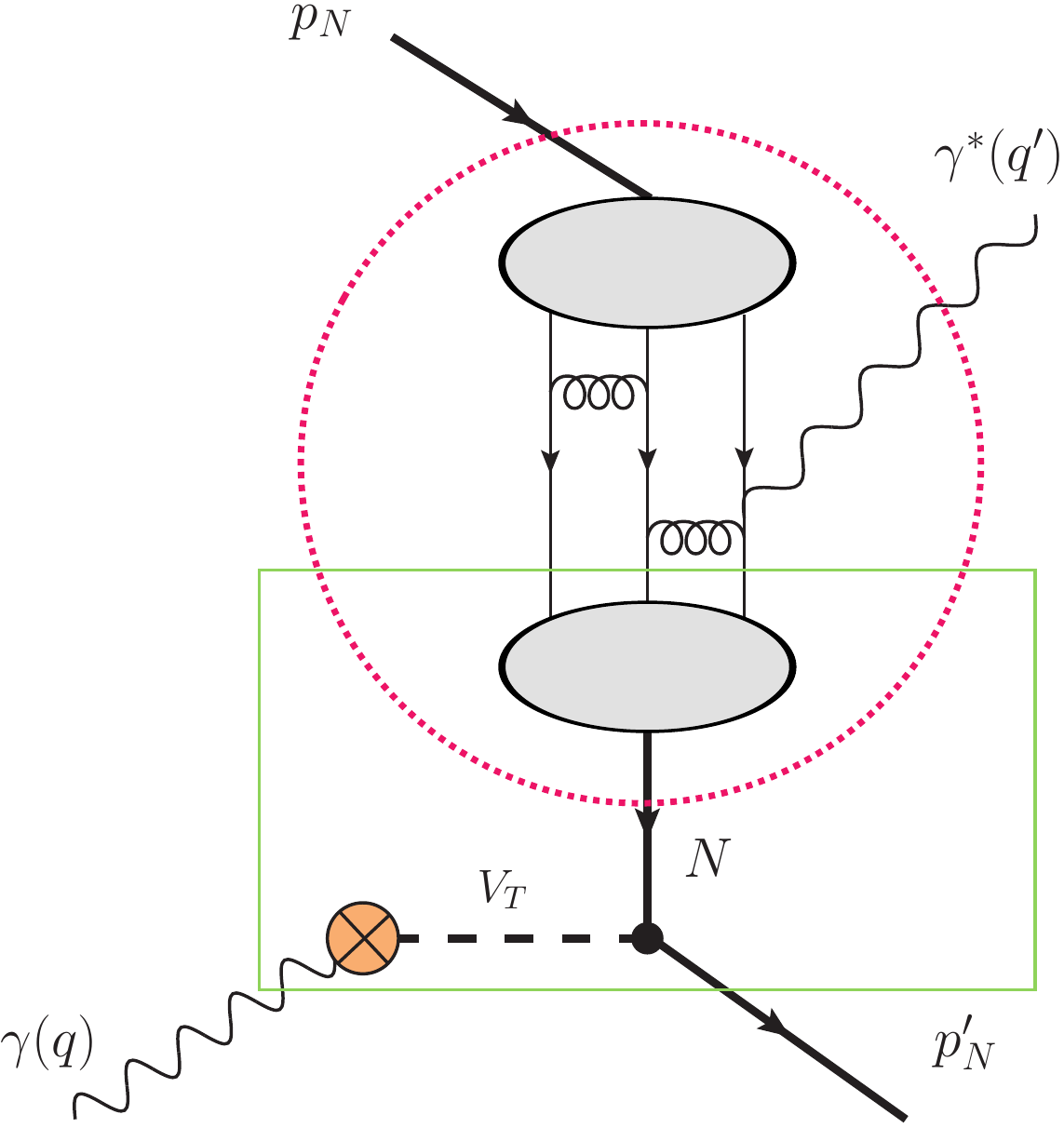}
\end{center}
     \caption{{\bf Left panel:} Cross channel nucleon exchange amplitude graph for $\pi N \to N' \gamma^*$ in perturbative QCD; {\bf Right panel:} Cross channel nucleon exchange amplitude graph for $ \gamma N \to N' \gamma^*$ within the VMD framework in perturbative QCD; dashed circles contain a typical LO graph for the nucleon electromagnetic form factor in perturbative QCD; the rectangles contain  the cross channel nucleon contribution into $N \pi$ and $N\gamma$ TDAs. The crossed circle depicts the $\gamma$ to transversely polarized vector meson vertex.}
\label{Fig_TDA_VMD}
\end{figure}

In this Section we present our estimates of pion-beam-induced near-backward lepton pair and charmonium production cross sections
for the kinematical conditions of J-PARC within the cross channel nucleon exchange model for $N \pi$ TDAs (see left panel of Fig.~\ref{Fig_TDA_VMD}).
The explicit expressions for nucleon-to-pion ($\pi N$) TDAs with the cross channel nucleon exchange model
are summarized in Sec.~5.1 of Ref.~\cite{Pire:2021hbl}.
$N \pi$ TDAs are expressed thorough $\pi N$ TDAs using the crossing relations
(\ref{Crossing_piN_TDAs}).

The integral convolutions
$\mathcal{I}^{(k)}_{\, \pi N \to N'
\genfrac\{\}{0pt}{2}{\gamma^*}{J/\psi}
}$, $k=1,\,2$,
introduced in Eqs.~(\ref{Def_ampl_pi_photon}), (\ref{Amplitude_pi_Jpsi}),
within the cross channel nucleon exchange model read
\be
&&
\mathcal{I}^{(1)}_{\, \pi N \to N'
\genfrac\{\}{0pt}{2}{\gamma^*}{J/\psi}
}(\xi, \Delta^2) \Big|_{N(940)}=- \sqrt{2}
\genfrac\{\}{0pt}{0}{{\cal I}_0}{M_0}
\frac{  f_\pi \,   g_{\pi NN}  m_N (1+\xi) } {   (\Delta^2-m_N^2) (1-\xi )}; \nn \\ &&
\mathcal{I}^{(2)}_{\, \pi N \to N'
\genfrac\{\}{0pt}{2}{\gamma^*}{J/\psi}
}(\xi, \Delta^2) \Big|_{N(940)}=- \sqrt{2}
\genfrac\{\}{0pt}{0}{{\cal I}_0}{M_0}
\frac{  f_\pi \,   g_{\pi NN}  m_N   } {   (\Delta^2-m_N^2)  }.
\label{ConvInt_pion_ex_model}
\ee
Here
\bi
\item $ g_{\pi NN}  \simeq 13$ stands for the pion-nucleon dimensionless coupling constant;
\item $\sqrt{2}$ is the isospin factor for the 
$\pi^- p \to n \genfrac\{\}{0pt}{2}{\gamma^*}{J/\psi}$ channel;
\item   ${\cal I}_0$  is the  constant occurring in the
leading order perturbative QCD description of proton electromagnetic form factor $F_1^p(Q^2)$
\cite{Chernyak:1987nv}:
\be
Q^4 F_1^p(Q^2)= \frac{(4 \pi \alpha_s)^2 f_N^2}{54}{\cal I}_0.
\label{pQCDF1}
\ee
Thus, the cross section of
 $\pi N \to N' \ell^+ \ell^-$
within
the   cross channel nucleon exchange model for $N \pi$ TDAs
turns out to be proportional to the square of
the perturbative QCD nucleon electromagnetic form factor. COZ input nucleon DA \cite{Chernyak:1987nv}
provides ${\cal I}_0 \Big|_{\rm COZ}=1.45 \cdot 10^5$;
an input DA with a shape close to the asymptotic form
({\it e.g.} Bolz-Kroll
\cite{Bolz:1996sw}, 
Braun-Lenz-Wittmann NLO
\cite{Braun:2006hz},
or that computed from the chiral soliton model 
\cite{Kim:2021zbz})
results
in a negligibly small value of  ${\cal I}_0$ and is thus unable to describe current experimental data on the nucleon electromagnetic form factor staying at the leading twist accuracy.
The use of the CZ-type DA solutions  can be seen as a way to partially take into account the contribution of the soft spectator mechanism . The regularization of the potential end point singularities then require further theoretical efforts, see {\it e.g.} the discussion on the pQCD description of
$\gamma^* \gamma$ form factors in Ref.~\cite{Musatov:1997pu}.

\item  $M_0$ is a well known convolution of nucleon DAs with hard
scattering kernel (\ref{Def_M0})
occurring in the $J/\psi \to \bar{p} p$ decay amplitude
\be
\mathcal{M}_{J / \psi \rightarrow \bar{p} p}=\left(4 \pi \alpha_s\right)^3 \frac{f_N^2 f_\psi}{\bar{M}^5} \frac{10}{81} \bar{U} \hat{\mathcal{E}}_\psi  V M_0.
\label{Jpsi_decay_Chernyak}
\ee
Therefore, the cross section of  near-backward $J/\psi$  production within the  cross channel nucleon exchange model for $N \pi$ TDAs turns out to be proportional
to  the
$J/\psi \to \bar{p} p$
decay width within the pQCD approach
\cite{Chernyak:1987nv}:
\be
\Gamma_{J/\psi \to p \bar{p}}
= (\pi \alpha_s)^6 \frac{1280 f_\psi^2 f_N^4 }{243 \pi {\bar{M}^9}} |M_0|^2.
\label{Charm_dec_width}
\ee
For the COZ input nucleon DA, $M_0 \Big|_{\rm COZ} \simeq 0.79 \cdot 10^4$. 
Note a very strong dependence of the decay width on $\alpha_s$. In the analysis of
Ref.~\cite{Pire:2016gut} the value
of $\alpha_s$ was adjusted to reproduce the experimental value $\Gamma_{J/\psi \to p \bar{p}}$
for a given input nucleon DA. In this paper we keep the compromise value of the strong coupling $\alpha_s=0.3$.
A discussion on the sensitivity of the result on $\alpha_s$ and on the form of the nucleon DA used as the phenomenological input
for our model can be found in Ref.~\cite{Pire:2013jva}.
\ei

\begin{figure}[H]
 \begin{center}
\includegraphics[width=0.7\textwidth]{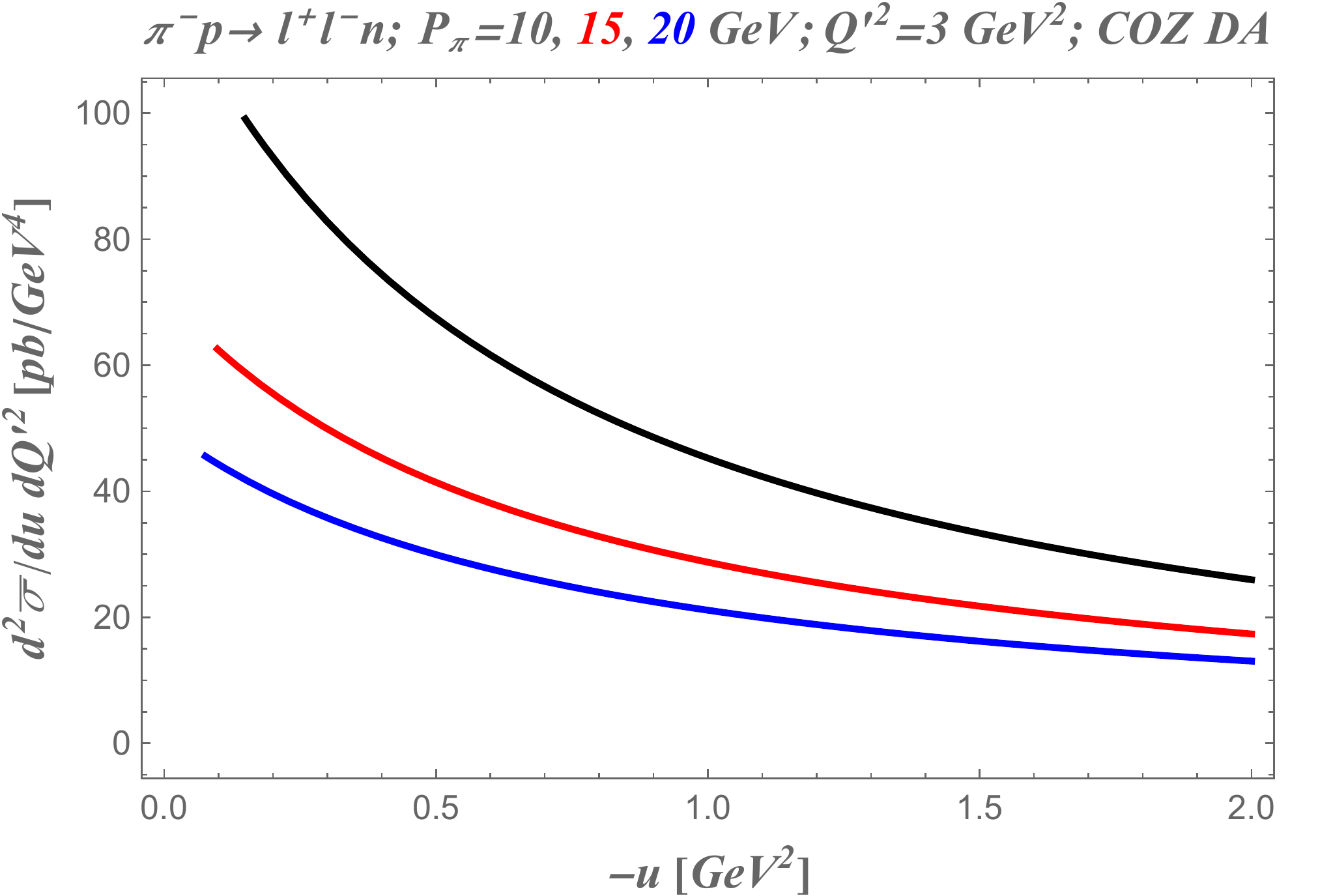}
 \end{center}
    \caption{ The integrated cross section for backward lepton pair production
$
\frac{d^2 \bar{\sigma}}{d \Delta^2 dQ'^2}
$
as a function of $-u$ from the threshold value $-u_0$ (corresponding to the exactly
backward production) up to $-u=2$~GeV$^2$
for the three values of pion beam momentum $P_\pi$ from the range of the J-PARC experiment: $P_\pi= 10$ GeV;   $P_\pi= 15$ GeV;  $P_\pi= 20$ GeV (from top to bottom);
the invariant mass of the lepton pair $Q'^2=3$~GeV$^2$;
 the COZ
\cite{Chernyak:1987nv}
solution for the leading twist nucleon DA
is used as the phenomenological input.
    }
\label{Fig_CSgammastar_pionbeam}
\end{figure}

In Fig.~\ref{Fig_CSgammastar_pionbeam}, within the cross channel nucleon exchange model for $\pi N$ TDAs
(\ref{ConvInt_pion_ex_model}), we present
the integrated cross section (\ref{Unpol_CS_React_pi_lept}) for backward lepton pair production
\be
\frac{d^2 \bar{\sigma}}{d u dQ'^2} = \int d \cos \theta_\ell  \frac{d^3  {\sigma}}{d u dQ'^2 d \cos \theta_\ell}
\ee
as a function of $-u$ from the threshold value $-u_0$ (corresponding to the exactly
backward production) up to $-u=2$~GeV$^2$
for the three values of pion beam momentum $P_\pi$ from the range of the J-PARC experiment
\cite{PhysRevD.93.114034} :
$P_\pi= 10$ GeV;   $P_\pi= 15$ GeV;  $P_\pi= 20$ GeV.
The invariant mass of the lepton pair  is set to $Q'^2=3$~GeV$^2$;
 the COZ
\cite{Chernyak:1987nv}
solution for the leading twist nucleon DA
is used as the phenomenological input.

\begin{figure}[H]
 \begin{center}
\includegraphics[width=0.7\textwidth]{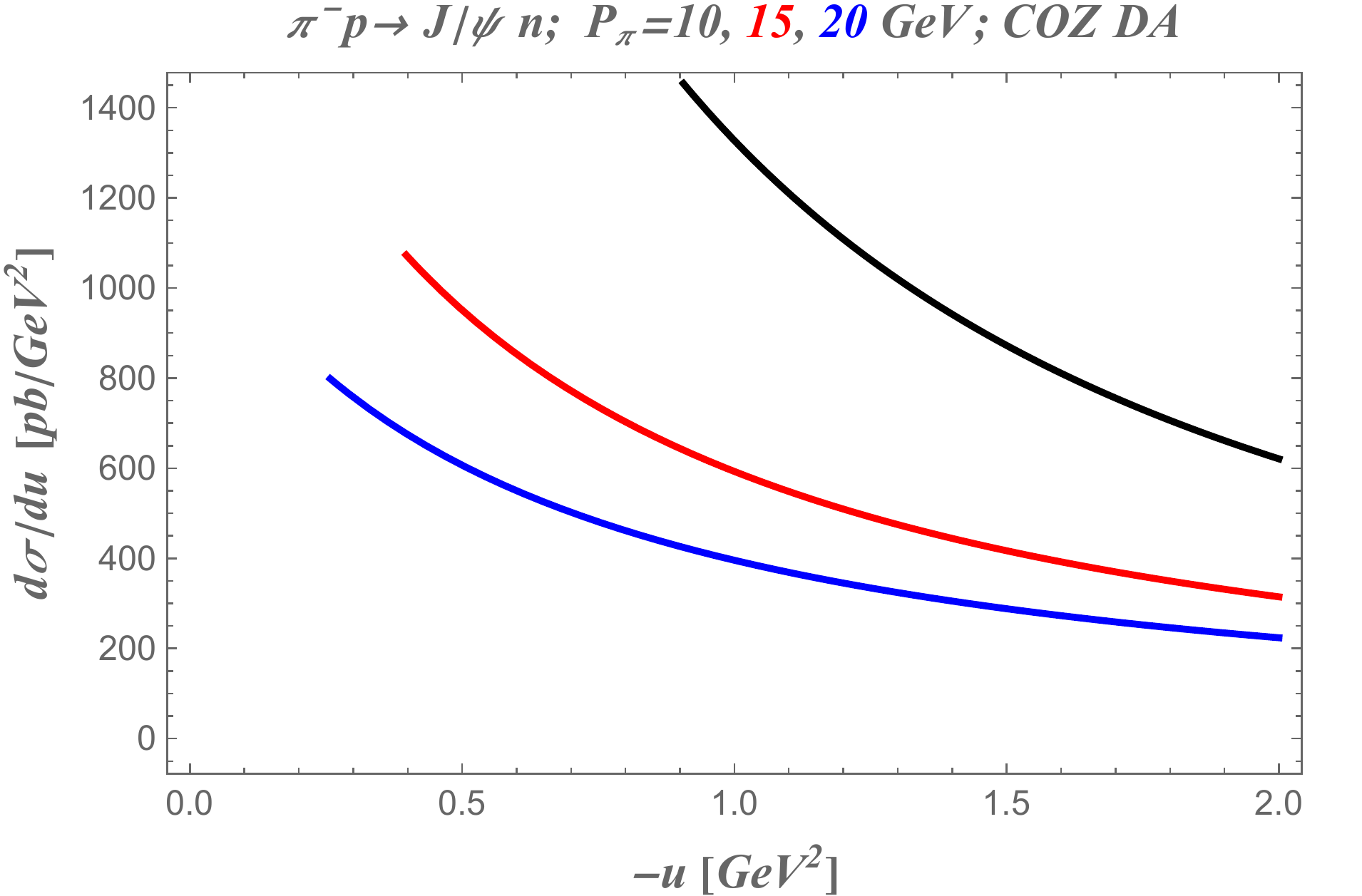}
 \end{center}
    \caption{ The  cross section for backward $J/\psi$ production
$
\frac{d \bar{\sigma}}{d \Delta^2}
$
as a function of $-u$ from the threshold value $-u_0$ (corresponding to the exactly
backward production) up to $-u=2$~GeV$^2$
for the three values of pion beam momentum $P_\pi$ from the range of the J-PARC experiment $P_\pi= 10$ GeV;   $P_\pi= 15$ GeV (from top to bottom);  $P_\pi= 20$ GeV;
COZ
\cite{Chernyak:1987nv}
solution for the leading twist nucleon DA
is used as the phenomenological input.
    }
\label{Fig_CSJpsi_pionbeam}
\end{figure}

In Figure~\ref{Fig_CSJpsi_pionbeam} we show
the  cross section for backward $J/\psi$ production
$
\frac{d \bar{\sigma}}{d \Delta^2}
$
(\ref{CS_def_delta2})
as a function of $-u$ from the threshold value $-u_0$ (corresponding to the exactly
backward production) up to $-u=2$~GeV$^2$ within the cross channel nucleon exchange model for $N \pi$ TDAs
(\ref{ConvInt_pion_ex_model})
for the same three values of pion beam momentum $P_\pi$ :
$P_\pi= 10$ GeV;   $P_\pi= 15$ GeV;  $P_\pi= 20$ GeV;
again using the COZ solution for the leading twist nucleon DA as the phenomenological input.

The presented cross sections estimates give hope of experimental
accessibility of the reactions at J-PARC. Dedicated feasibility studies,
similar to that  performed for accessing
of $\pi N$
TDAs
at \={P}ANDA
\cite{PANDA:2014qiz,PANDA:2016scz},
extending the analysis of  
\cite{Sawada:2016mao} 
to the near-backward kinematical regime are highly 
demanded to carry out a final conclusion on feasibility 
and prepare a detailed experimental proposal for J-PARC.

\section{Data driven model estimates  for near-backward $J/\psi$ photoproduction and for  TCS cross-sections }
\label{NEWSec_Model_estimates}
\mbox

In this Section we employ the data on   $J/\psi$
photoproduction  over the full near-threshold
kinematic region
recently presented by the  GlueX collaboration in JLab  \cite{GlueX:2019mkq,Adhikari:2023fcr}
to constrain the normalization and $u$-dependence of $N\gamma$ TDAs. 
With help of the resulting $N\gamma$ TDA models we present the cross section estimates for
the near-backward lepton pair  photoproduction, both  in the 
$J/\psi$ 
resonance region and in the continuum (TCS),
for the kinematic conditions corresponding to the recent analysis of the GlueX collaboration .
The relevant values of the LAB frame photon energy $E_\gamma$, the center-of-mass invariant energy $W=\sqrt{m_N^2+2m_N E_\gamma}$ and
 the threshold values of the invariants $t$ and $u$ corresponding to exactly backward scattering are summarized in Table~\ref{Table_1}.

\begin{table}[H]
\begin{center}
\caption{Kinematical range of the GlueX $J/\psi$ photoproduction experiment \cite{Adhikari:2023fcr}.}\label{Table_1}%
\begin{tabular}{@{}llll@{}}
\toprule
$\ \ E_\gamma$ [GeV] \ \ & \ \ $W$ [GeV] \ \ & \ \ $-t_1$ [GeV$^2$] \ \ & \ \ $-u_0$ [GeV$^2$] \ \ \\
8.2 $-$  9.28   & 4.04 $-$ 4.28   & 3.99 $-$ 6.58 & 1.54 $-$ 0.98 \\
9.28 $-$  10.36  & 4.28 $-$ 4.51   & 6.58 $-$ 8.84  & 0.98 $-$ 0.75 \\
10.36 $-$  11.44  & 4.51 $-$ 4.73  & 8.84 $-$ 11.01  & 0.75 $-$ 0.61   \\
\botrule
\end{tabular}
\end{center}
\end{table}

A sensible estimate of the scattering amplitudes for near-backward photoproduction reactions
is more difficult to justify than for the pion-beam-induced reactions. Indeed,   
the normalization of photon-to-nucleon TDAs is not constrained, contrarily to pion-to-nucleon  TDAs which are naturally normalized in the limit $\xi \to 1$ thanks to the soft pion chiral limit \cite{Lansberg:2011aa}. 

An estimate 
based on the hypothesis of the applicability of Vector Meson Dominance (VMD)
\cite{Hakioglu:1991pn,Schildknecht:2005xr}
to such reactions, was suggested in Ref.~\cite{Pire:2022fbi}. The nucleon-to-photon TDAs were related to the corresponding TDAs for the nucleon-to-transversely-polarized-vector-mesons, as:
\be
&&
\{V, \, A\}_{\Upsilon}^{\gamma N}=\frac{e}{f_{\rho}} \{V, \, A\}_{\Upsilon}^{\rho N}+\frac{e}{f_{\omega}} \{V, \, A\}_{\Upsilon}^{\omega N}+\frac{e}{f_{\phi}} \{V, \, A\}_{\Upsilon}^{\phi N}, \ \ \text{with} \ \
\Upsilon= 1{\cal E}, \, 1T, \, 2{\cal E}, \, 2T;
\nn \\
&&
T_{\Upsilon}^{\gamma N}=\frac{e}{f_{\rho}} T_{\Upsilon}^{\rho N}+\frac{e}{f_{\omega}} T_{\Upsilon}^{\omega N}+\frac{e}{f_{\phi}} T_{\Upsilon}^{\phi N}, \ \ \text{with} \ \
\Upsilon= 1{\cal E}, \, 1T, \, 2{\cal E}, \, 2T, \, 3{\cal E}, \, 3T, 4{\cal E}; \, 4T,
\label{VMD_based_gN}
\ee
where $e$ is the electron charge and vector-meson-to-photon couplings
$f_{\rho,\omega,\phi}$
 are estimated from
$\Gamma_{V \to e^+e^-}$ decay widths.
Such a model is very constrained since  data on vector meson backward electroproduction \cite{JeffersonLabFp:2019gpp} exist at comparable energy and skewness. It turns out that  this naive model leads to an unmeasurably small  cross-section for near-backward $J/\psi$ photoproduction. The very fact that the GlueX collaboration detects some backward scattered $J/\psi$ mesons leads us - perhaps not surprisingly \cite{Lee:2022ymp} - to disregard the VMD-based approach.

Let us thus try to define the various steps of a phenomenological program which may lead to a sensible extraction of at least some features of the photon to nucleon TDAs before some of their properties are discovered in non-perturbative studies such as those using lattice QCD or QCD sum-rules \cite{Haisch:2021nos} techniques.

Since the scattering amplitudes come from  the convolution of TDAs with coefficient functions integrated over the momentum fraction variables $x_i$, the dependence of TDAs on  $x_i$ is deeply buried in observables and thus may be the least accessible feature  from phenomenological studies. Such a difficulty already exists in the GPD case where it is recognized that the ($x=\pm \xi$) restricted domain dominance of the lowest order DVCS and TCS amplitudes calls for the study of complementary processes \cite{Qiu:2022pla, Deja:2023ahc}.

On the contrary, the overall normalization and the $u$-dependence of the TDAs should be easier to access. A possible strategy is thus emerging. Firstly, we extract from our study of the transversely polarized vector meson backward electroproduction case in a simplified  nucleon exchange model (\ref{VMD_based_gN})  that the 
 integral convolutions
$\mathcal{I}^{(k)}_{\, \gamma N \to N'
\genfrac\{\}{0pt}{2}{\gamma^*}{J/\psi}
}$, $k=1,\,3,\,4,\,5$,
written in Eqs.~(\ref{Hel_ampl_def_photo_gamma}), (\ref{Hel_ampl_def_photo_Jpsi}),
turn out to be  linear combinations of
$\mathcal{I}^{(k)}$s, $k=1,\,3,\,4,\,5$, which are proportional to the constant ${\cal I}_0$ ({\it resp.} $M_0$) coming from the ($x_1,x_2,x_3$) integration in the calculation of the nucleon form factor ({\it resp.} the $J/\psi \to \bar{p} p$ decay amplitude (\ref{Jpsi_decay_Chernyak})):
\be
&&
\mathcal{I}^{(1)}_{\, \gamma N \to N'
\genfrac\{\}{0pt}{2}{\gamma^*}{J/\psi}
}(\xi, \Delta^2) \Big|_{N(940)}=-
 \frac{e K_{1 {\cal E}}(-\xi, \Delta^2)}{2\xi}
\genfrac\{\}{0pt}{0}{{\cal I}_0}{M_0}
; \nn \\ &&
\mathcal{I}^{(3)}_{\,  \gamma N \to N' \genfrac\{\}{0pt}{2}{\gamma^*}{J/\psi}}(\xi, \Delta^2) \Big|_{N(940)}= -\frac{eK_{1 T}(-\xi, \Delta^2)+eK_{2 {\cal E}}(-\xi, \Delta^2)}{2\xi}
\genfrac\{\}{0pt}{0}{{\cal I}_0}{M_0} ; \nn \\ &&
\mathcal{I}^{(4)}_{\,  \gamma N \to N' \genfrac\{\}{0pt}{2}{\gamma^*}{J/\psi}}(\xi, \Delta^2) \Big|_{N(940)}=- \frac{eK_{2 T}(-\xi, \Delta^2)}{2\xi}
 \genfrac\{\}{0pt}{0}{{\cal I}_0}{M_0} ; \nn \\ &&
\mathcal{I}^{(5)}_{\,  \gamma N \to N' \genfrac\{\}{0pt}{2}{\gamma^*}{J/\psi}}(\xi, \Delta^2) \Big|_{N(940)}=  \frac{eK_{2 {\cal E}}(-\xi, \Delta^2)}{2\xi}\genfrac\{\}{0pt}{0}{{\cal I}_0}{M_0} \,. 
\label{Conv_I_Jpsi}
\ee
Here we assume  the dominance of the vector coupling  of transversely polarized vector
meson to nucleons
${\cal L}_{VNN}^{\rm eff.}=\bar{N} G^V_{VNN} \gamma^\mu V_\mu N$; and
the  explicit expressions for the functions
$K_{1 {\cal E}}$,
$K_{2 {\cal E}}$,
$K_{1 T}$,
$K_{2 T}$,
read as suggested from \cite{Pire:2015kxa}:
\be
&&
K_{1 {\mathcal{E}}}( \xi, \Delta^2)= f_N G (\Delta^2)  \frac{2\xi(1-\xi)}{1+\xi} ; \nn \\ &&
K_{2 {\mathcal{E}}}( \xi, \Delta^2)=f_N G (\Delta^2) (-2 \xi) ;
\nn \\ &&
K_{1 T}( \xi, \Delta^2)=f_N G (\Delta^2) \frac{2 \xi  (1+3 \xi)}{1-\xi}   ;
\nn \\ &&
K_{2 T}( \xi, \Delta^2)= 0.
\label{Def_K_factors}
\ee
The kinematical factors occurring from
Eqs.~(\ref{Def_K_factors}) bring additional dependence on $Q'^2$ and $W^2$
through the skewness variable $\xi$ (\ref{Xi_collinear}). The dependence on the invariant 
$u$-channel momentum transfer 
$\Delta^2 \equiv u$
is implemented through $G(\Delta^2)$. 


Having taken into account the various kinematical factors, which are the same in the present processes and in our previous study of backward vector meson electroproduction
\cite{Pire:2015kxa},
we are now in a position to show how one can fix the normalization and $\Delta^2$-dependence of the photon to nucleon TDAs from the $J/\psi$ photoproduction data and work out a set of predictions for the future photoproduction experiments. 
To achieve this goal we propose two educated guesses for the $\Delta^2$-dependence, the first one reminiscent of the cross channel nucleon exchange model:
\be
G(\Delta^2) = G^{(1)}(\Delta^2)=  \frac{C^{(1)}}{\Delta^2-m_N^2}.
\label{Dipole_type_ANZ}
\ee
and the second one
\be
G(\Delta^2) = G^{(2)}(\Delta^2)=\frac{C^{(2)}e^{\alpha \Delta^2}} {\Delta^2-m_N^2}.
\label{exponential}
\ee
with positive intercept $\alpha$ letting the amplitude fall off with 
$-u$ faster than the original dipole-type formula (\ref{Dipole_type_ANZ}). 
In both cases, we let the overall normalization 
$C^{(1,\,2)}$
as an adjustable parameter to be fixed from
the data.

\begin{figure}[H]
 \begin{center}
\includegraphics[width=0.49\textwidth]{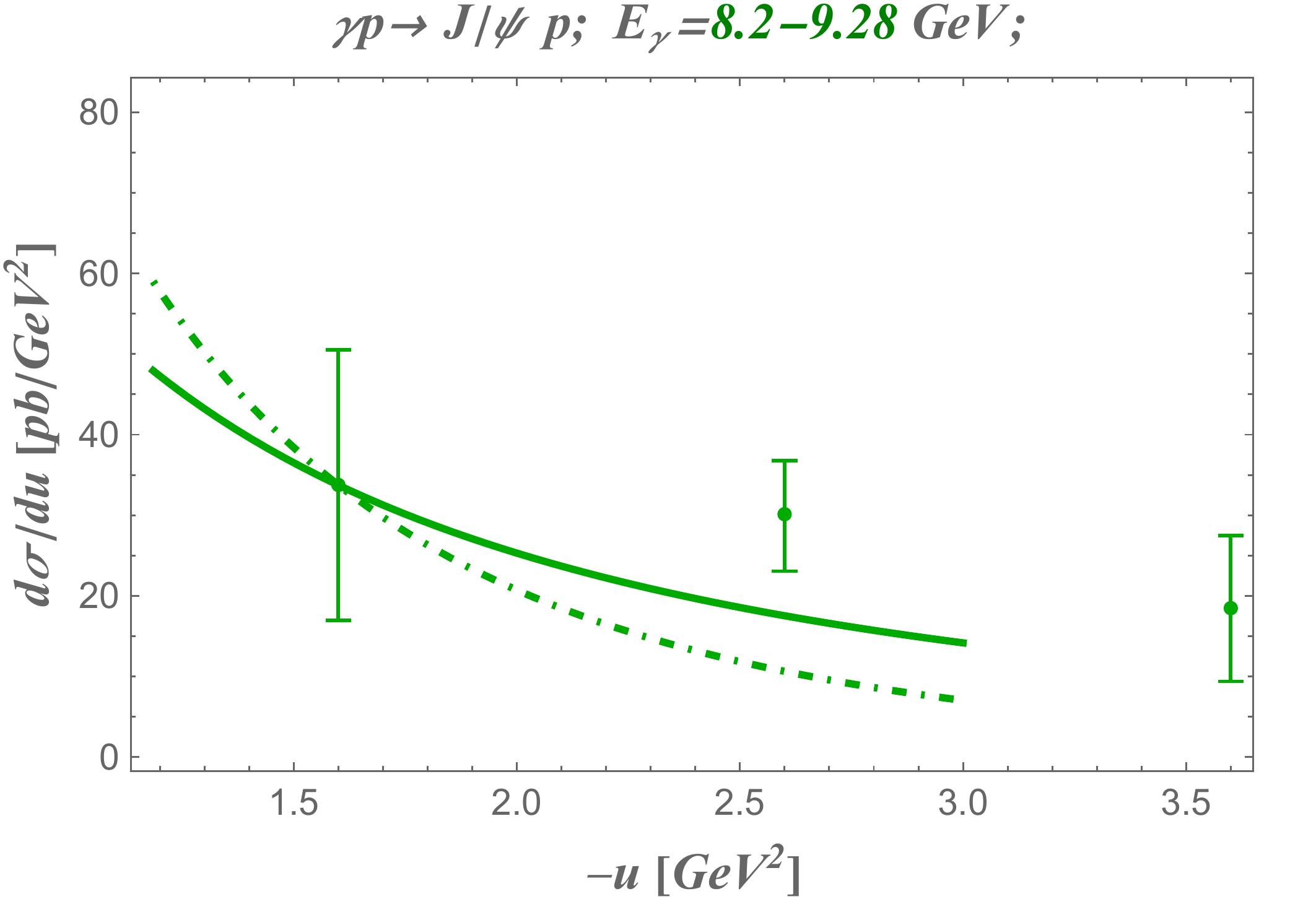}
\includegraphics[width=0.49\textwidth]{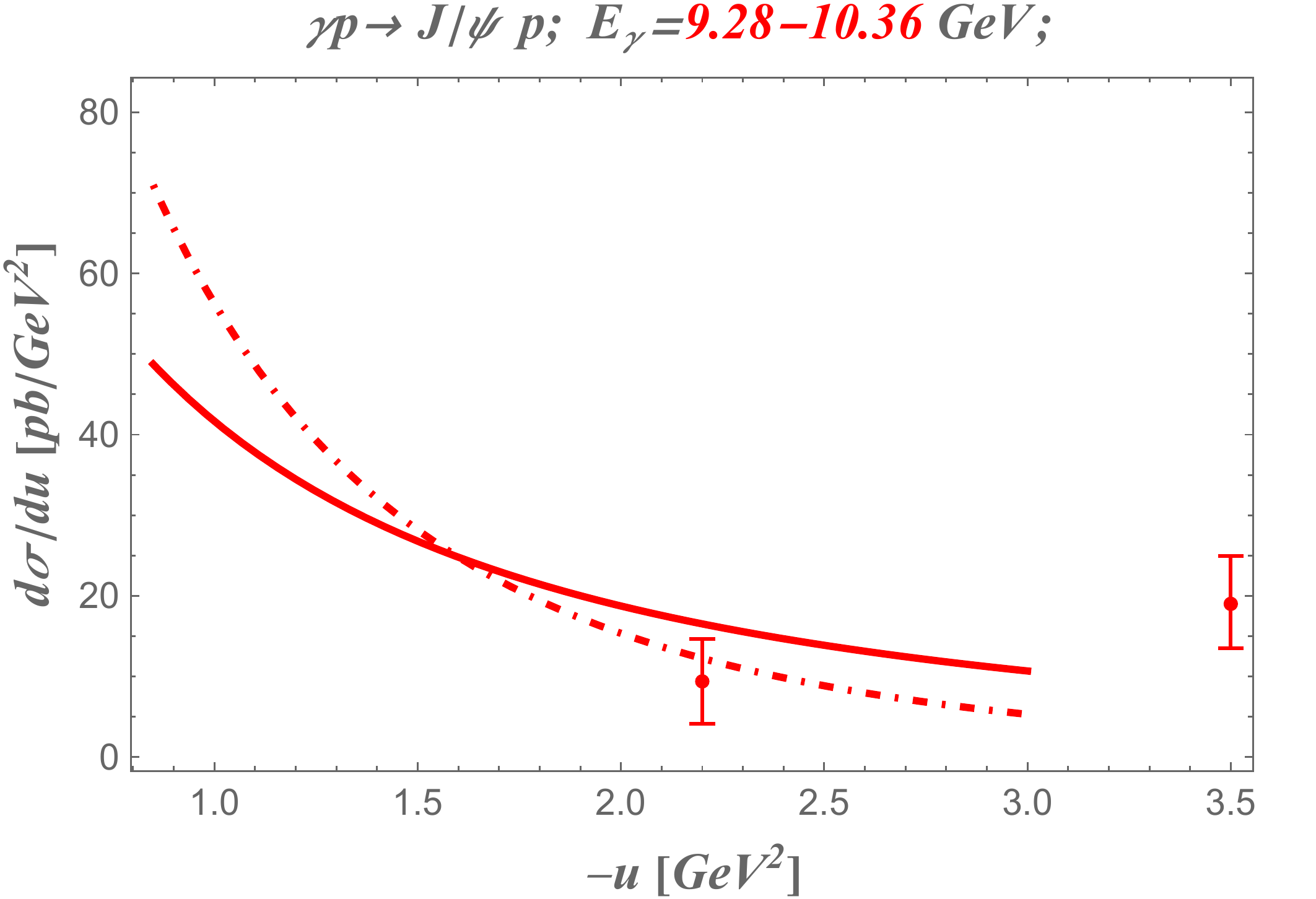}
\includegraphics[width=0.49\textwidth]{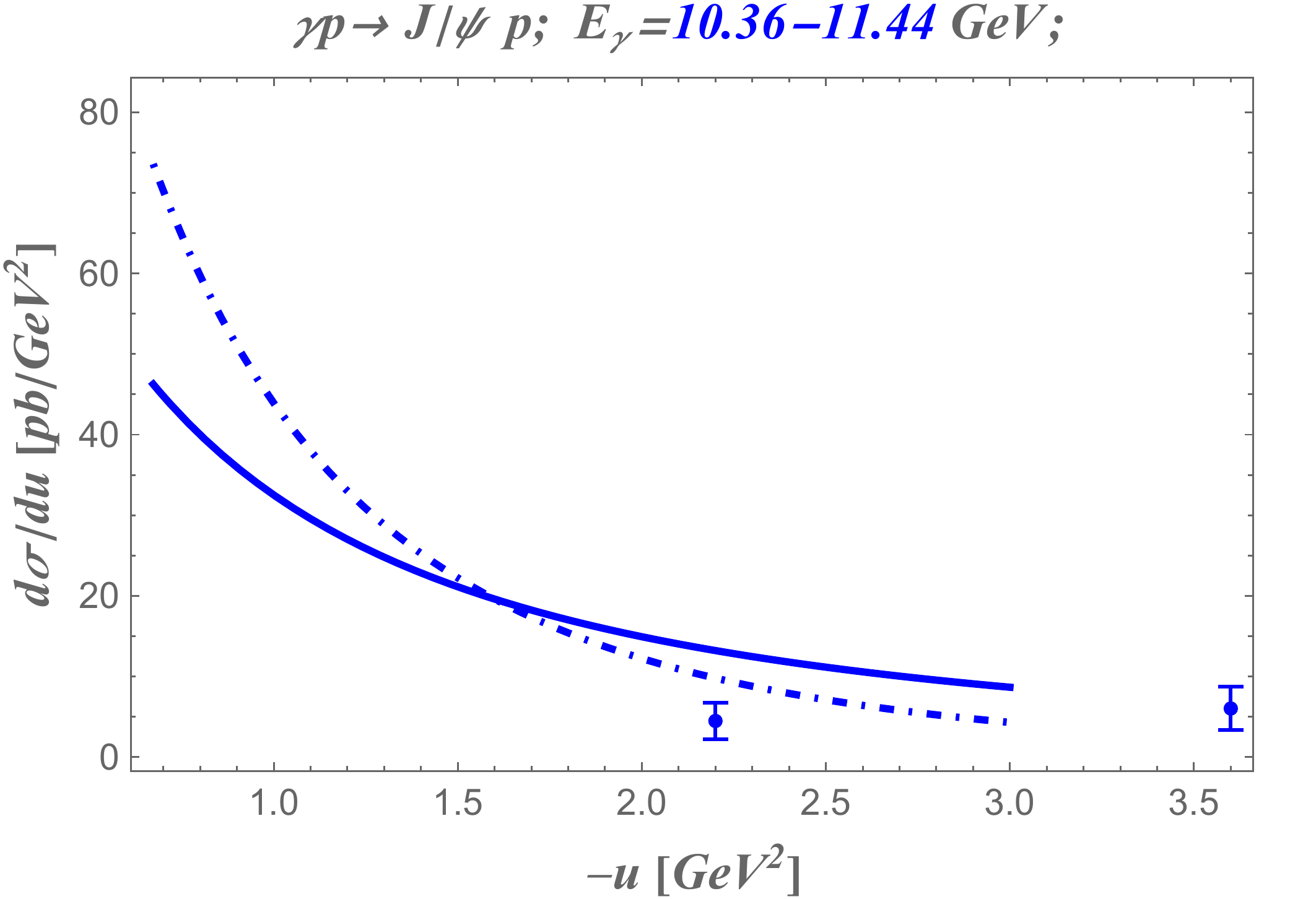}
 \end{center}
    \caption{ Solid lines: the cross section for backward $J/\psi$ photoproduction
$
\frac{d \sigma}{d u}
$
as a function of $-u$ from the threshold value $-u_0$ (corresponding to the exactly
backward production, see Table~\ref{Table_1}) up to $-u=3$~GeV$^2$
for the three central values of $E_\gamma$ from the range of the GlueX experiment 
$\langle E_\gamma \rangle= 8.74$ GeV;  $\langle E_\gamma \rangle= 9.82$; and $\langle E_\gamma \rangle= 10.9$ GeV
in
the model 
(\ref{Conv_I_Jpsi}) with the dipole-type $u$-dependence (\ref{Dipole_type_ANZ}).
Dot-dashed lines show the cross sections within the TDA model 
 with a modified normalization and $u$-dependence
(\ref{exponential})
with
$\alpha=0.25$~GeV$^{-2}$. COZ
\cite{Chernyak:1987nv}
solution for the leading twist nucleon DA
is used as the phenomenological input. With error bars we show the points closest to the backward threshold from the experimental data presented in Fig.~13 of Ref.~\cite{Adhikari:2023fcr}. Note that for the $\langle E_\gamma \rangle= 9.82$; and $\langle E_\gamma \rangle= 10.9$ GeV  bins no data point approaches close enough to threshold values of $-u=-u_0$. 
    }
\label{Fig_CS_JPsiPhoto}
\end{figure}

In Fig.~\ref{Fig_CS_JPsiPhoto}, 
the solid lines  show the predictions of the model 
(\ref{Conv_I_Jpsi}) with the dipole-type $u$-dependence (\ref{Dipole_type_ANZ})
 for the three central values of $E_\gamma$ from the range of the GlueX experiment
\cite{Adhikari:2023fcr}
$\langle E_\gamma \rangle= 8.74$ GeV;  $\langle E_\gamma \rangle= 9.82$; and $\langle E_\gamma \rangle= 10.9$ GeV.
With error bars we show the points closest to the backward threshold from the experimental data presented in Fig.~13 of \cite{Adhikari:2023fcr}.  Note that for the $\langle E_\gamma \rangle= 9.82$; and $\langle E_\gamma \rangle= 10.9$ GeV  bins no data point approach close enough to threshold values of $-u=-u_0$. 
We use the value of $M_0$ computed with the COZ
\cite{Chernyak:1987nv}
solution%
\footnote{DA solutions with a shape close to the asymptotic form 
\cite{Braun:2006hz}
are known to
largely underestimate the value of $J/\psi \to N\bar{N}$ decay width  with the leading twist-$3$ 
pQCD description since the $M_0$ constant appears to be too small for the compromise
value of $\alpha_s \simeq 0.3$, see discussion in Ref.~\cite{Stefanis:1997zyh}. 
}
for the leading twist nucleon DA
as the phenomenological input and fit the value of the normalization constant
$C^{(1)}=10.9$ from the experimental cross section value corresponding to
the experimental point with lowest $-u$ ({\it i.e.} largest $-t$) of the
$ \langle E_\gamma \rangle=8.74$~GeV bin (see the first panel of Fig.~\ref{Fig_CS_JPsiPhoto}).
This allows to present our estimates of the backward peak for the
$\langle E_\gamma \rangle= 9.82$; and $\langle E_\gamma \rangle= 10.9$ GeV bins. 
Note that the GlueX data points for these bins stay too far from the backward threshold.
Getting more data close to the backward threshold require augmenting the GlueX luminosity 
for the corresponding values of $E_\gamma$. The suggested $17$~GeV electron beam upgrade 
(see discussion in \cite{LubomirTrento})
will provide the necessary increase in statistics to challenge the manifestation of 
the backward peak for higher bins in $ E_\gamma$. The estimated luminosity of the GlueX
with  $17$~GeV electron beam results in large expected counting rates (several thousands of evens)
in the vicinity of the backward peak. 

With the dot-dashed lines we
show the predictions of the TDA model  with the normalization and $u$-dependence
(\ref{exponential})
with the value of the intercept set to  $\alpha=0.25$~GeV$^{-2}$. The normalization constant $C^{(2)}=16.3$
is fixed analogously to the previous case. The exponential form of $u$-dependence results in a  narrower 
cross section backward peak. 

Note that with our choice of the normalization constants $C^{(1,\,2)} \simeq G_V^{\omega NN}$,
see {\it e.g.} \cite{Machleidt:2000ge}, the magnitude of the corresponding photon-to-nucleon TDAs
turn out to be roughly of the same order as those of the  nucleon-to-vector meson TDAs (multiplied by the charge factor $e$):
\be
{\rm TDA}_{N \gamma}(x,\xi,t) \sim e {\rm TDA}_{V N}(x,-\xi,t); 
\ee
the normalization of the latter TDAs was found consistent with the experiment \cite{JeffersonLabFp:2019gpp}.

In order to address the universality 
of $N \gamma$ TDAs through the TCS cross section measurements 
we also present our estimates of near-backward TCS cross section within the 
$N \gamma$ TDA models  
(\ref{Conv_I_Jpsi})
with the normalization  
and $u$-dependence
(\ref{Dipole_type_ANZ}),
(\ref{exponential})
chosen to fit the GlueX $J/\psi$ photoproduction cross section  for $E_\gamma=8.74$ GeV in the near-backward region 
and $\alpha=0.25$~GeV$^{-2}$.

\begin{figure}[H]
 \begin{center}
\includegraphics[width=0.49\textwidth]{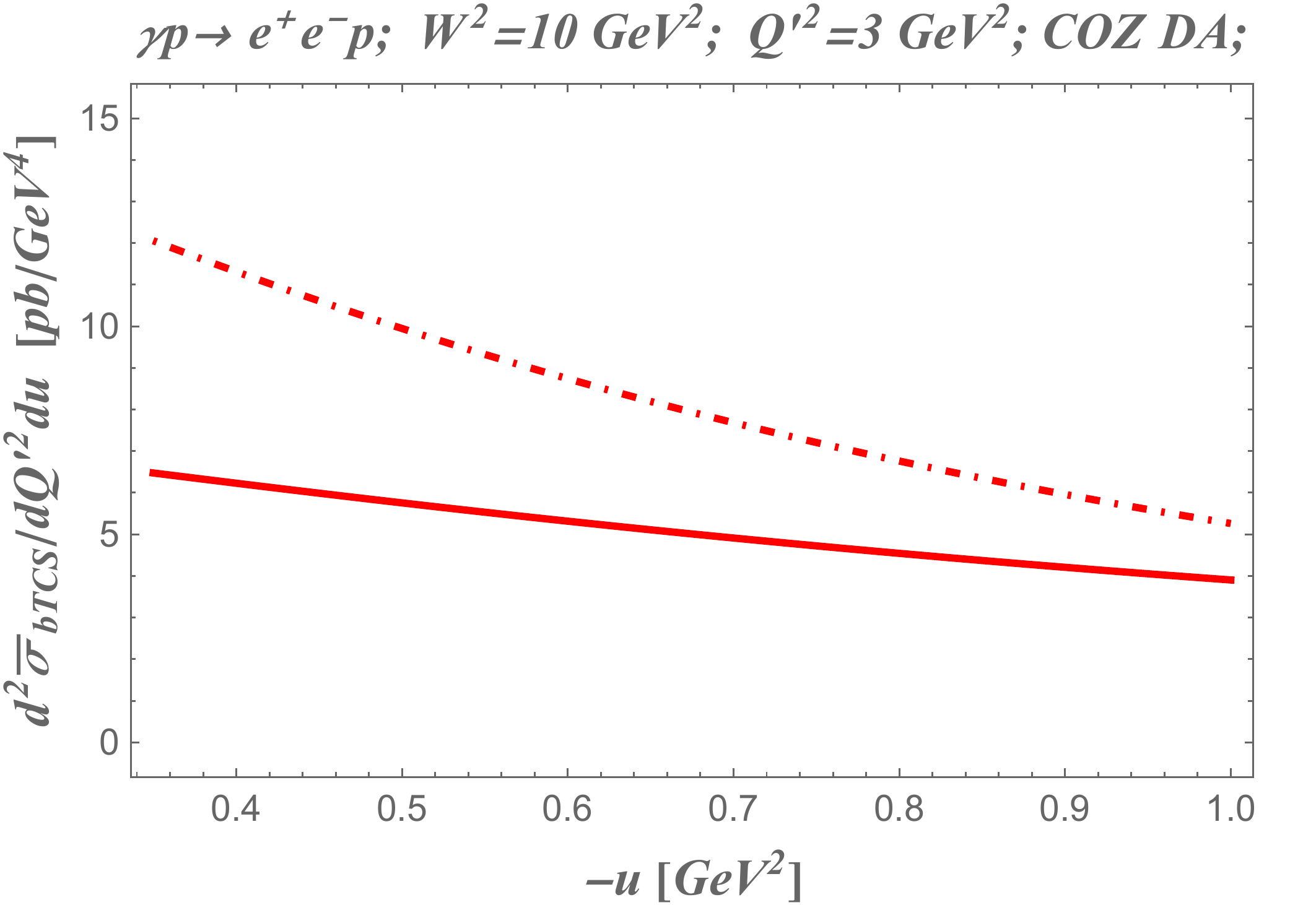}
\includegraphics[width=0.49\textwidth]{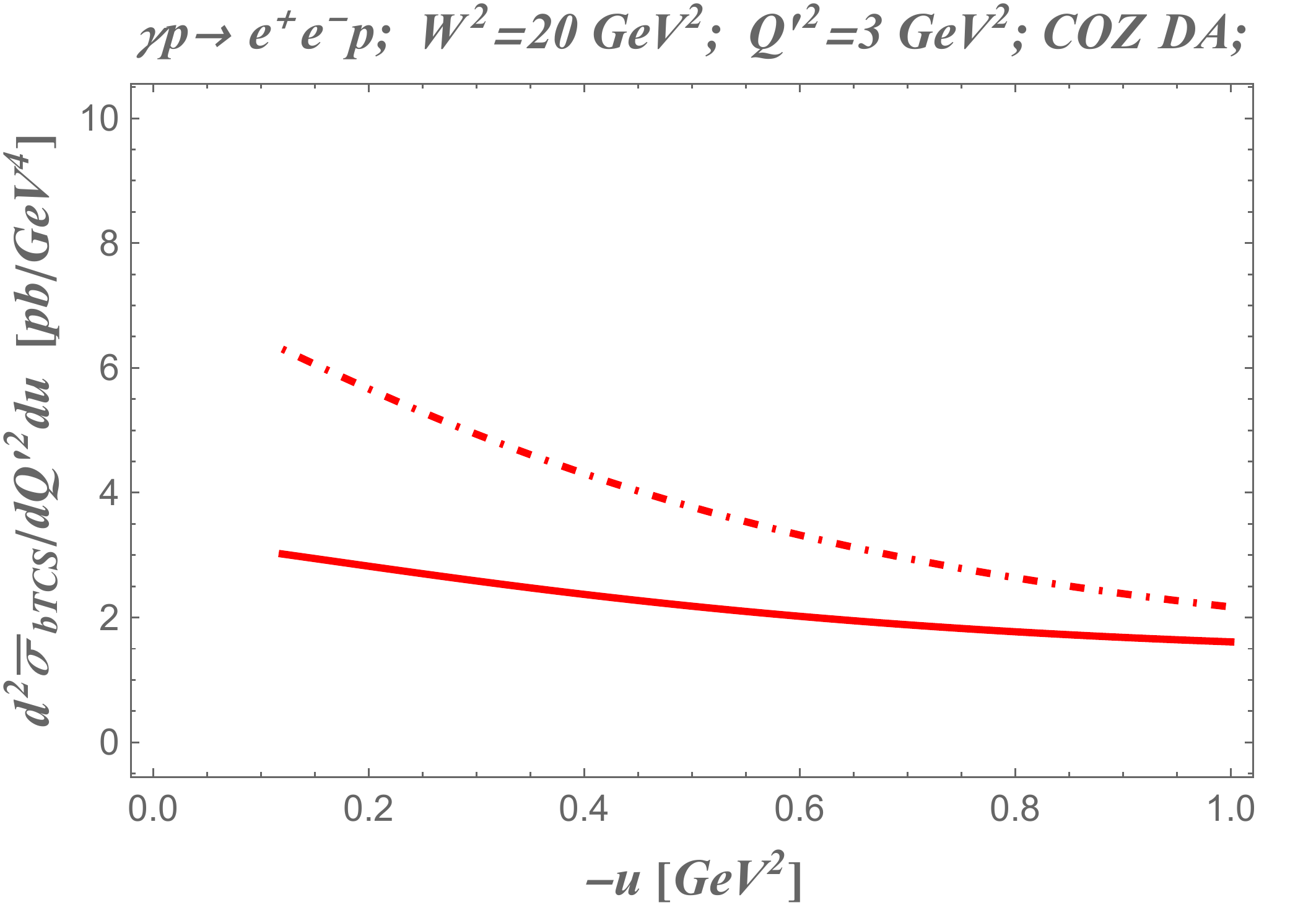}
\includegraphics[width=0.49\textwidth]{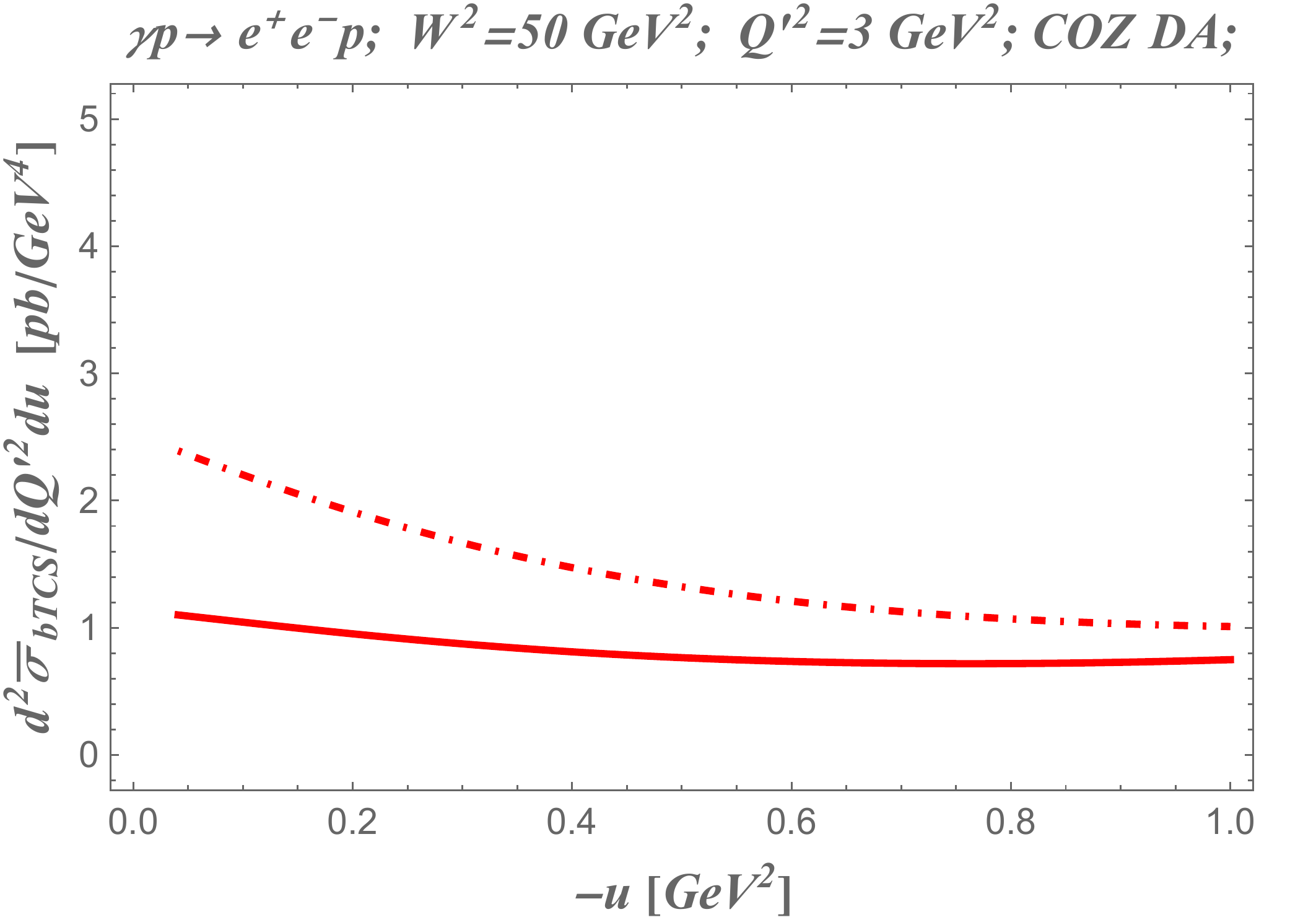}
\includegraphics[width=0.49\textwidth]{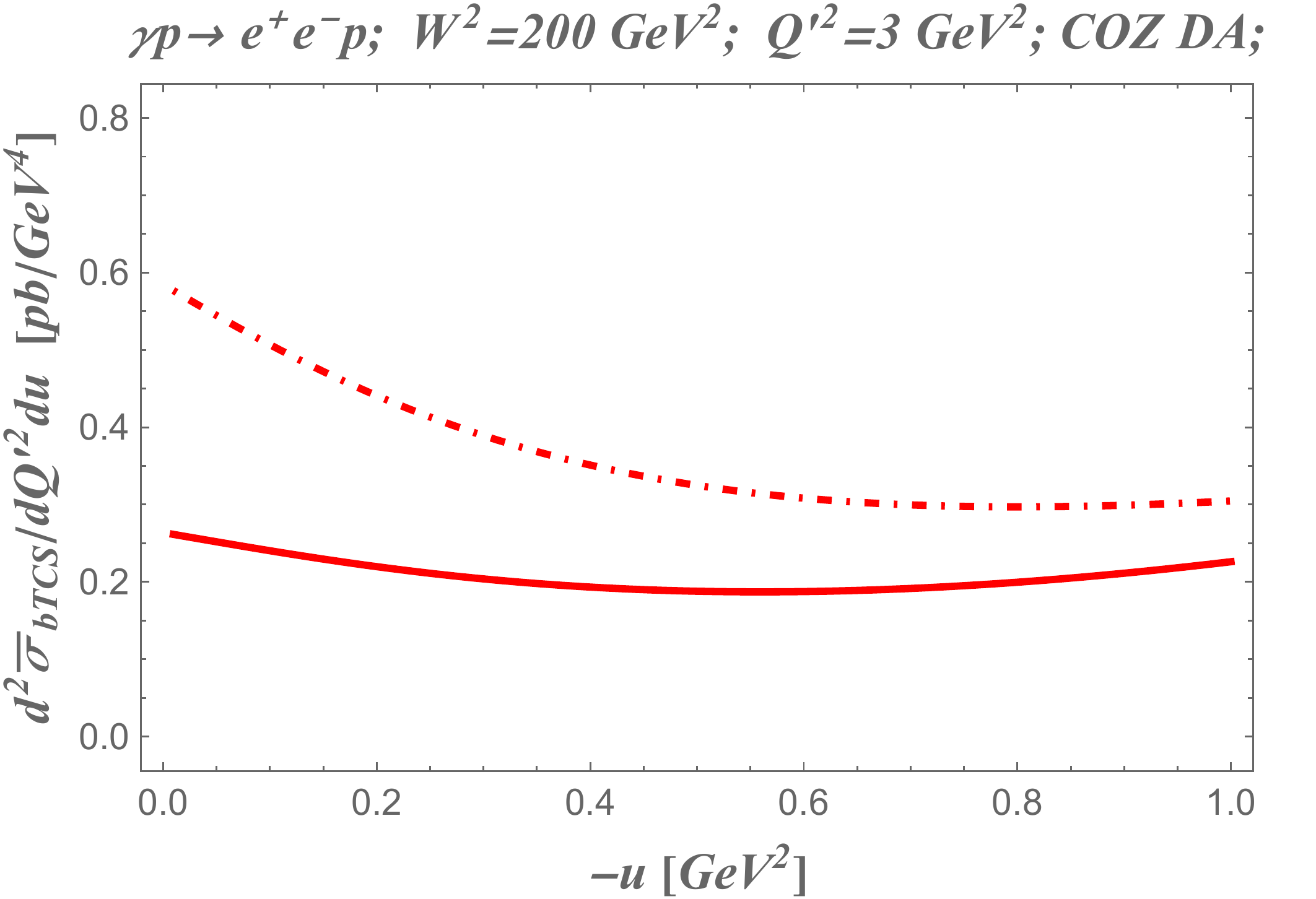}
 \end{center}
    \caption{  The near-backward  $\gamma p \to p e^+ e^-$ scattering cross section
(\ref{CS_TCS_integrated}) 
for several values of $W^2$; $Q'^2=3$ GeV$^2$; as a function of $-u$ from the minimal value $-u_0$ up to $1$ GeV$^2$ in the $\gamma N$ TDA model (\ref{Conv_I_Jpsi}).
Solid lines show the  cross section estimates within the 
TDA model with the dipole-type $u$-dependence
(\ref{Dipole_type_ANZ}); dot-dashed lines: the cross sections from the TDA model with modified
 $u$-dependence (\ref{exponential}) with $\alpha = 0.25$~GeV$^{-2}$. Normalization is fixed from the 
 $J/\psi$ photoproduction data, as explained above. COZ solution is employed as input phenomenological
solution for nucleon DAs.  
    }
\label{Fig_CS_bTCS}
\end{figure}

In Fig.~\ref{Fig_CS_bTCS} we show the cross section (\ref{CS_main_formula}) of the near-backward $\gamma p \to p e^+ e^-$ 
integrated over the lepton polar angle
\be
\frac{d^2 \bar{\sigma}}{ du d Q'^2} \equiv  \int_0^\pi d \cos \theta_\ell \frac{d^3   {\sigma}}{ du d Q'^2 d \cos \theta_\ell }
\label{CS_TCS_integrated}
\ee
as a function of $-u$ from $-u_0$ up to 1~GeV$^2$. 
We present the cross section for several values of $W^2$ corresponding to the kinematical range of JLab@12 GeV and JLab@24
and the future EIC and EIcC.
 The invariant mass squared of the lepton pair is set to
$Q'^2=3$ GeV$^2$. We plot the cross sections as  functions of $-u$ from the minimal value $-u_0$ up to $1$ GeV$^2$.
Solid lines  show the cross sections within the  $\gamma N$ TDA model with the dipole-type $u$-dependence (\ref{Dipole_type_ANZ})  
 with  overall normalization $C^{(1)}$
 adjusted to the Gluex data for $J\psi$ photoproduction cross section in the near-backward region for
$\langle E_\gamma \rangle = 8.74$ GeV. 
 Dot-dashed lines show the  cross section estimates with the TDA model  with a modified
 $u$-dependence (\ref{exponential})
 with  $\alpha = 0.25$
 and the normalization $C^{(2)}$ chosen to match the GlueX $J/\psi$ photoproduction data. 
The COZ solution is employed as input phenomenological
solution for nucleon DAs.

The Bethe-Heitler background cross section for TCS reaction has been estimated in Ref.~\cite{Pire:2022fbi} in the near backward region 
and found to be negligibly small apart from the very narrow peaks in the vicinity of $\theta_\ell=0$. Its contribution into the
integrated cross section (\ref{CS_TCS_integrated}) can, therefore, be safely neglected. 

Let us stress that it is essential to firmly establish the physical normalization of 
 $N \gamma$  TDAs and to
develop a reliable framework to model them in the complete domain of their definition. A possible approach can be the calculations performed within the light-cone quark model of Ref.~\cite{Pasquini:2009ki} or the QCD sum-rules techniques \cite{Haisch:2021nos}. Lattice QCD calculations might also help to get some constraints on these TDAs.

The essence of our implementation of the $N\gamma$ TDA framework is addressing the universality 
of manifestation of the backward peak  from the collinear factorization mechanism involving 
$N\gamma$ TDAs. We fit our TDA model to match the backward peak revealed for $J/\psi$
photoproduction by the GlueX collaboration in the lowest bin in $E_\gamma$. 
This requires that the normalization of $N \gamma$ TDAs be roughly that of 
nucleon-to-vector meson TDAs, the latter being to some expend tested experimentally.  
This implies that the backward peak must be also manifest in other  $E_\gamma$ bins  for $J/\psi$
photoproduction by the GlueX  as well as in the backward TCS reaction.

The suggested $17$~GeV upgrade of the GlueX experiment will provide enough statistics to confirm or reject the manifestation of the backward peak required by the universality of $N\gamma$ TDAs.
This will bring strong arguments to make a decision on the validity of the collinear factorization framework for backward photoproduction reactions and will allow a more detailed comparison with predictions from alternative reaction mechanisms \cite{Yu:2018ydp,Strakovsky:2022jnx,Strakovsky:2023kqu}.

Also, the magnitude of the near-backward TCS cross section from the $\gamma N$ TDA models 
(\ref{Conv_I_Jpsi})
with $u$-dependence 
(\ref{Dipole_type_ANZ}), (\ref{exponential}); and
 normalization adjusted to the GlueX $J/\psi$ photoproduction data in the near-backward region is
considerably larger than the predictions made from the pure VMD-based cross channel nucleon exchange model
\cite{Pire:2022fbi}.
An observation of a sizable backward peak for TCS cross section would provide a crucial test for the verification of the universality of TDAs and hence of the consistency of the QCD approach to backward hard exclusive reactions.

\section{Conclusions}
\mbox

This short review of the applications of the TDA concept in exclusive reactions initiated by a photon or a $\pi$-meson beam illustrates the ability of the QCD collinear factorization approach to describe high invariant mass dilepton
and heavy quarkonium production amplitudes in terms of meson-to-nucleon and photon-to-nucleon TDAs in a new kinematical window, complementary to the forward kinematics, where GPDs play a leading role. 
In particular, the comparison between backward timelike virtual photon and backward charmonium production, {\it i.e.} dilepton production at or off the resonance peak at the $J/\psi$ mass, will offer a clear cut proof of the validity of a partonic interpretation of backward exclusive scattering. It also would be instructive to compare the 
predictions of QCD collinear factorization approach with that of the 
Regge theory framework
\cite{Yu:2018ydp,Strakovsky:2022jnx}
within the near-backward kinematics. 

Let us stress that working out predictions that may allow to really distinguish between the Regge-based approach and the collinear factorization approach in the backward regime is a tricky task.
Recent Regge analysis of backward meson electroproduction in Ref.~\cite{Laget:2021qwq} demonstrates
that by including the contribution of box diagrams in the backward kinematics a consistent
description of the JLab experimental data can be achieved based on the Regge approach.
The  two description can be seen as sort of dual in the kinematical domain in which factorization can be justified.

We did not cover the  deep electroproduction processes~\cite{Lansberg:2011aa, Pire:2015kxa} nor the antiproton - nucleon annihilation case~\cite{Lansberg:2012ha}  that we studied earlier in a quite detailed way. We believe that near future studies at JLab as well as at future experimental facilities \cite{AbdulKhalek:2021gbh, Anderle:2021wcy, PANDA:2009yku} will test the universality of TDAs extracted from various processes.  Anyhow, the study of hard exclusive reactions in the backward region will provide very interesting data that should drastically improve our understanding of the baryonic structure.

\section*{Acknowledgements}
We acknowledge useful and inspiring discussions with Shunzo Kumano, Bill Li, Zein-Eddine Meziani and Lubomir Pentchev. We also thank both anonymous referees for their careful reports that helped us to improve our paper.

This work was supported in part by the European Union's Horizon 2020 research and innovation programme under Grant Agreement No. 824093 and by the LABEX P2IO.
The work of K.S. is supported by the Foundation for
the Advancement of Theoretical Physics and Mathematics ``BASIS'' and by
the National Research Foundation
of Korea (NRF) under Grants No. NRF-2020R1A2C1007597 and No.
NRF-2018R1A6A1A06024970 (Basic Science Research Program).
The work of  A.S. is supported by the Foundation for
the Advancement of Theoretical Physics and Mathematics ``BASIS''.
The work of L.S. is supported by the grant 2019/33/B/ST2/02588 of the National Science Center in Poland.

\setcounter{section}{0}
\setcounter{equation}{0}
\renewcommand{\thesection}{\Alph{section}}
\renewcommand{\theequation}{\thesection\arabic{equation}}

\section{Crossing $\pi \to N$ TDAs to $N \to \pi$ TDAs}
\label{App_Crossing}
\mbox

The study of electroproduction processes~\cite{Lansberg:2007ec} lead us to   parameterize the nucleon-to-pion
($ \pi N $)
TDAs defined through the Fourier transform of the
$\pi N$
matrix element of the trilinear quark operator on the light cone.
The parametrization involves
eight
invariant functions each being the function of three
longitudinal momentum fractions, skewness variable,
momentum transfer squared as well as of the factorization scale.

Let us consider the neutron-to-$\pi^-$ $uud$ TDA. We make use of the parametrization of
Ref.~\cite{Lansberg:2007ec},
where only three invariant functions turn out to be relevant in the
$\Delta_T=0$
limit:
\bea
&&
4 (p \cdot n)^3 \int \left[ \prod_{j=1}^3 \frac{d \lambda_j}{2 \pi}\right]
e^{i \sum_{k=1}^3 \tilde{x}_k \lambda_k (p \cdot n)}
 \langle     \pi^-(p_\pi)|\,  \varepsilon_{c_1 c_2 c_3} u^{c_1}_{\rho}(\lambda_1 n)
u^{c_2}_{\tau}(\lambda_2 n)d^{c_3}_{\chi}(\lambda_3 n)
\,|n(p_N,s_N) \rangle
\nonumber \\ &&
= \delta(\tilde{x}_1+\tilde{x}_2+\tilde{x}_3-2 \tilde{\xi}) i \frac{f_N}{f_\pi}\Big[  V^{   \pi^- n}_{1}(\tilde{x}_{1,2,3}, \tilde{\xi} ,\tilde{\Delta}^2)  (  \hat{p} C)_{\rho \tau}(U^+)_{\chi}
\nonumber \\ &&
+A^{  \pi^- n}_{1}(\tilde{x}_{1,2,3}, \tilde{\xi},\tilde{\Delta}^2)  (  \hat{p} \gamma^5 C)_{\rho \tau}(\gamma^5 U^+ )_{\chi}
 +T^{   \pi^- n}_{1}(\tilde{x}_{1,2,3}, \tilde{\xi},\tilde{\Delta}^2)  (\sigma_{p\mu} C)_{\rho \tau }(\gamma^\mu U^+ )_{\chi}
 \nonumber \\ &&
 + m_N^{-1} V^{   \pi^- n }_{2} (\tilde{x}_{1,2,3}, \tilde{\xi},\tilde{\Delta}^2)
 ( \hat{ p}  C)_{\rho \tau}( \hat{\tilde{\Delta}}_T U^+)_{\chi}
+ m_N^{-1}
 A^{   \pi^- n}_{2}(\tilde{x}_{1,2,3}, \tilde{\xi},\tilde{\Delta}^2)  ( \hat{ p}  \gamma^5 C)_{\rho\tau}(\gamma^5  \hat{\tilde{\Delta}}_T  U^+)_{\chi}
  \nonumber \\ &&
+ m_N^{-1} T^{    \pi^- n}_{2} (\tilde{x}_{1,2,3}, \tilde{\xi} ,\tilde{\Delta}^2) ( \sigma_{p \tilde{\Delta}_T} C)_{\rho \tau} (U^+)_{\chi}
+  m_N ^{-1}T^{     \pi^- n }_{3} (\tilde{x}_{1,2,3}, \tilde{\xi},\tilde{\Delta}^2) ( \sigma_{p\mu} C)_{\rho \tau} (\sigma^{\mu \tilde{\Delta}_T}
 U^+)_{\chi}
 \nonumber \\ &&
+ m_N^{-2} T^{  \pi^- n}_{4} (\tilde{x}_{1,2,3}, \tilde{\xi},\tilde{\Delta}^2)  (\sigma_{p \tilde{\Delta}_T} C)_{\rho \tau}
(\hat{ \tilde{\Delta}}_T U^+)_{\chi} \Big]
 \nonumber \\ &&
\equiv
\delta(\tilde{x}_1+\tilde{x}_2+\tilde{x}_3-2 \tilde{\xi}) i \frac{f_N}{f_\pi}
\sum_{\rm  Dirac \atop structures} s^{ \pi N}_{\rho \tau, \, \chi} H_s^{     \pi^- n }( \tilde{x}_1,\tilde{x}_2,\tilde{x}_3, \tilde{\xi}, \tilde{\Delta}^2).
 \label{Old_param_TDAs}
\eea
We adopt Dirac's ``hat'' notation $\hat{v} \equiv v_\mu \gamma^\mu$;
$\sigma^{\mu\nu}= \frac{1}{2} [\gamma^\mu, \gamma^\nu]$; $\sigma^{v \mu} \equiv v_\lambda \sigma^{\lambda \mu}$;
$C$
is the charge conjugation matrix and
$U^+= \hat{p} \hat{n} \, U(p_N,s_N)$
is the large component of the nucleon spinor; and employ compact notations    for the set of the relevant Dirac structures
$s^{   \pi  N} \equiv \{ a^{  \pi N}_{1,2} , \, a^{  \pi N}_{1,2}, \, t^{  \pi N}_{1,2,3,4} \}$.

Note that the
$ \pi N$
TDA
(\ref{Old_param_TDAs})
is defined with respect to the natural kinematical variables of $\gamma^* N \to \pi N'$  reaction.
Namely the cross channel momentum transfer is
$\tilde{\Delta}=p_\pi-p_N$
and the skewness parameter
$\tilde{\xi}$
is defined from the longitudinal momentum transfer
between pion and nucleon
$$
\tilde{\xi}
 \equiv  -\frac{( p_\pi-p_N) \cdot n}{(p_\pi+p_N) \cdot n}
$$
({\it i.e.} it differs by the sign from the definition
(\ref{Def_xi})
natural for the reactions
(\ref{Reaction_cases}) in the near-backward regime).

\begin{figure}[H]
 \begin{center}
\epsfig{figure= 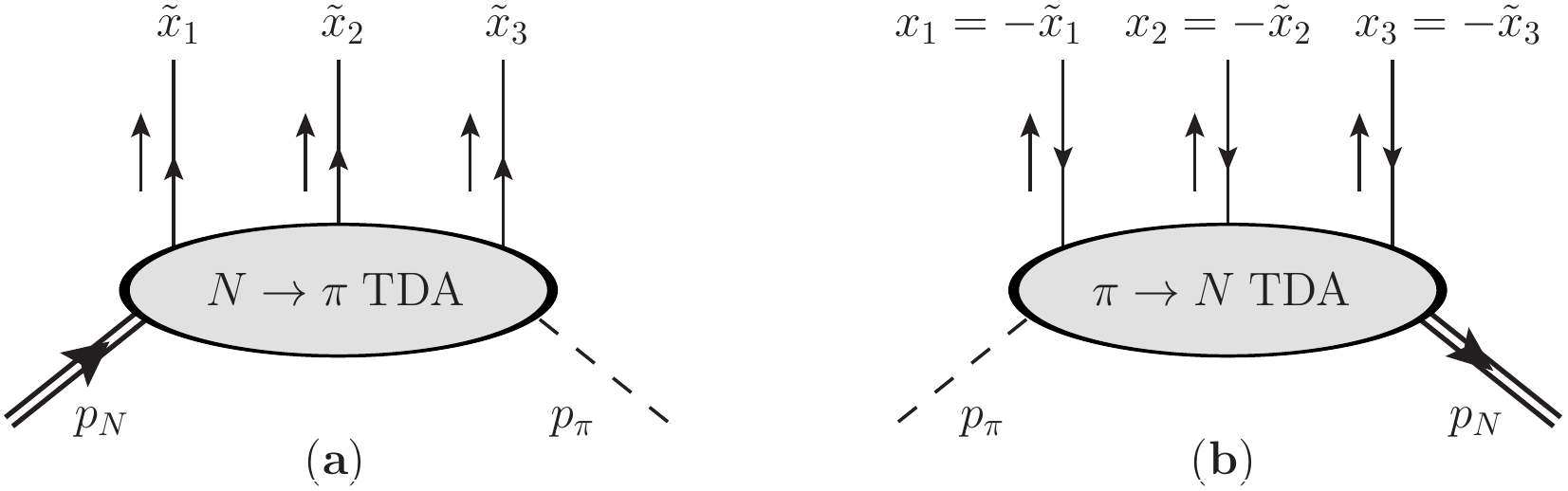, height=4.0cm}
   \end{center}
     \caption{Small arrows show the direction of the longitudinal momentum flow in the ERBL-like
      regime for
{\bf (a)}: The longitudinal momentum flow for nucleon-to-pion
($\pi N$) TDAs
defined in
(\ref{Old_param_TDAs}).
The longitudinal momentum transfer is $(p_\pi-p_N) \cdot n \equiv \tilde{\Delta} \cdot n$.
{\bf (b)}: The longitudinal momentum flow for pion-to-nucleon
($N \pi$) TDAs defined in (\ref{Def_N_pi_TDAs_param}).
The longitudinal momentum transfer is $(p_N-p_\pi) \cdot n \equiv  \Delta \cdot n$.
Arrows on the nucleon and quark (antiquark) lines show the direction of flow of the baryonic charge.}
\label{Fig_Flow}
\end{figure}


In order to express
pion-to-nucleon
($ N \pi $)
TDAs through
($\pi N  $)
TDAs occurring in
(\ref{Old_param_TDAs})
 we apply the Dirac conjugation
(complex conjugation and convolution with
$\gamma_0$ matrices
in the appropriate spinor indices)
for both sides of eq.~(\ref{Old_param_TDAs}) and
compare the result to the definition of  $\pi  N$ TDAs:
\be
&&
-4 (p \cdot n)^3 \int \left[ \prod_{j=1}^3 \frac{d \lambda_j}{2 \pi}\right]
e^{-i \sum_{k=1}^3 \tilde{x}_k \lambda_k (p \cdot n)}
 \langle    n(p_N,s_N) |\,  \varepsilon_{c_1 c_2 c_3} \bar{u}^{c_1}_{\rho}(\lambda_1 n)
\bar{u}^{c_2}_{\tau}(\lambda_2 n) \bar{d}^{c_3}_{\chi}(\lambda_3 n)
\,| \pi^-(p_\pi) \rangle
\nonumber \\ &&
=
-\delta(\tilde{x}_1+\tilde{x}_2+\tilde{x}_3-2 \tilde{\xi}) i \frac{f_N}{f_\pi}
\sum_s   \underbrace{(\gamma_0^T)_{\tau \tau'} \left[ s_{\rho' \tau', \chi'}^{  \pi N} \right]^\dag
(\gamma_0)_{\rho'\rho }
(\gamma_0)_{\chi' \chi}}_{s_{\rho \tau, \chi}^{  N \pi }} H_s^{  \pi N}
(\tilde{x}_{1},\tilde{x}_{2},\tilde{x}_{3}, \tilde{\xi} ,\Delta^2).
\label{DiracConjTDA}
\ee

For the relevant Dirac structures occurring in the parametrization (\ref{Def_N_pi_TDAs_param}) we get
\be
&&
(v_1^{N \pi})_{\rho \tau, \chi}= (C \hat{p})_{\rho \tau} \bar{U}^+_\chi;
\nonumber \\ &&
(a_1^{N \pi})_{\rho \tau, \chi}= (C \hat{p} \gamma_5)_{\rho \tau} \left(\bar{U}^+ \gamma_5 \right)_\chi;
\nonumber \\ &&
(t_1^{N \pi})_{\rho \tau, \chi}= -(C \sigma_{p \mu})_{\rho \tau} \left(\bar{U}^+ \gamma_\mu \right)_\chi;
\nonumber \\ &&
(v_2^{N \pi})_{\rho \tau, \chi}= (C \hat{p})_{\rho \tau}  \left( \hat{\tilde{\Delta}}_T\bar{U}^+ \right)_\chi=-
 (C \hat{p})_{\rho \tau}  \left( \hat{ \Delta}_T\bar{U}^+ \right)_\chi;
\nonumber \\ &&
(a_2^{N \pi})_{\rho \tau, \chi}= (C \hat{p} \gamma_5)_{\rho \tau} \left( \bar{U}^+ \hat{\tilde{\Delta}}_T \gamma_5 \right)_\chi=-
 (C \hat{p} \gamma_5)_{\rho \tau} \left( \bar{U}^+  \hat{\Delta}_T \gamma_5 \right)_\chi
\nonumber \\ &&
(t_2^{\pi \to N)})_{\rho \tau, \chi}= -(C \sigma_{p \tilde{\Delta}_T})_{\rho \tau} \left(\bar{U}^+ \right)_\chi=(C \sigma_{p  \Delta_T})_{\rho \tau} \left(\bar{U}^+ \right)_\chi;
\nonumber \\ &&
(t_3^{N \pi})_{\rho \tau, \chi}= (C \sigma_{p \mu})_{\rho \tau}
\left(\bar{U}^+ \sigma_{  \mu \tilde{\Delta}_T} \right)_\chi=
- (C \sigma_{p \mu})_{\rho \tau}
\left(\bar{U}^+ \sigma_{  \mu  \Delta_T} \right)_\chi;
\nonumber \\ &&
(t_4^{N \pi})_{\rho \tau, \chi}= -(C \sigma_{p \tilde{\Delta}_T})_{\rho \tau}
\left(\bar{U}^+  \hat{\tilde{\Delta}}_T\right)_\chi=
-(C \sigma_{p \Delta_T})_{\rho \tau}
\left(\bar{U}^+  \hat{\Delta}_T \right)_\chi\,,
\ee
where we switch to the definition of momentum transfer
natural for the timelike reactions:
$\tilde{\Delta} \to -\Delta$.
$\bar{U}^+\equiv \bar{U}(p_N) \hat{n} \hat{p}$
stands for the large component of the
$\bar{U}(p_N)$
Dirac spinor.

The flow of the longitudinal momentum for
$N \to \pi$
TDAs defined as in eq.~(\ref{Old_param_TDAs}) and
$\pi \to N$
TDAs   is presented on Fig.~\ref{Fig_Flow}.
By switching to the variables
$\xi=-\tilde{\xi}$
and
$x_i=-\tilde{x}_i$
natural for the timelike reactions
and
$\tilde{\Delta}^2 \to \Delta^2$
we conclude that
\be
&&
\left\{
V_{1,\, 2}^{n \pi^- }, \,
A_{1,\, 2}^{n \pi^- }, \,
T_{1,\, 2, \, 3, \,4}^{n \pi^- }
\right\}
(x_{1,2,3}, \xi, \Delta^2)
=
\left\{
V_{1,\, 2}^{    \pi^- n}, \,
A_{1,\, 2}^{  \pi^- n}, \,
T_{1,\, 2, \, 3, \,4}^{  \pi^- n}
\right\}
(-x_{1,2,3}, -\xi, \Delta^2).
\label{Crossing_piN_TDAs}
\ee

The set of the Dirac structures for photon-to-nucleon ($N \gamma$) TDAs can be
established according to the pattern of Eq.~(\ref{DiracConjTDA}) employing the set of the
Dirac structures for photon-to-nucleon TDAs summarized in Appendix of Ref.~\cite{Pire:2022fbi}.
The set of relations between nucleon-to-photon and photon-to-nucleon TDAs
is fully analogous to  Eq.~(\ref{Crossing_piN_TDAs}).

\setcounter{equation}{0}
\section{Backward charmonium production amplitudes}
\label{App_B}
\subsection{Backward $J/\psi$ pion-production amplitude}
\label{App_Jpsi_pion_prod}

\begin{figure}[H]
\begin{center}
\includegraphics[width=0.3\textwidth]{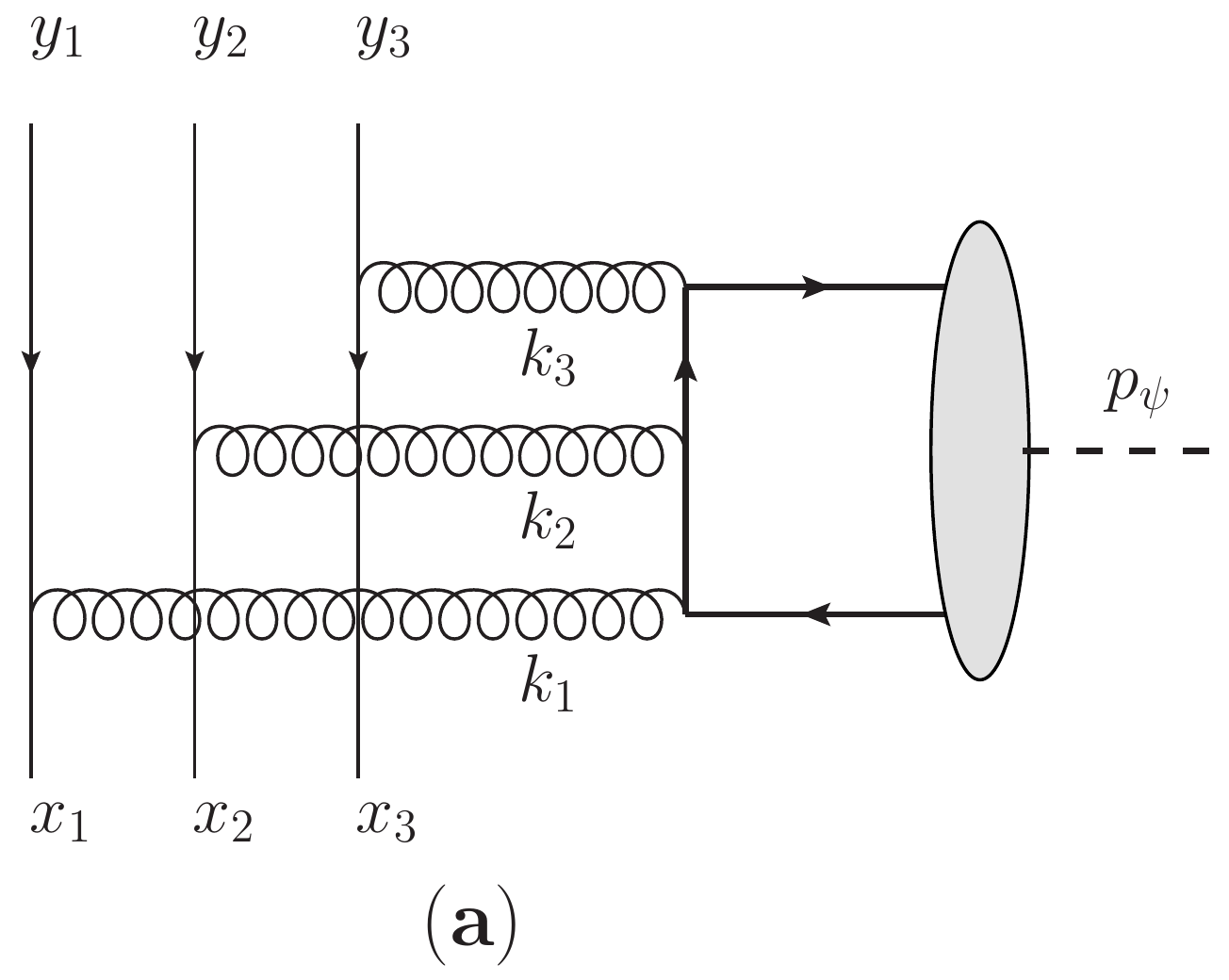}
\includegraphics[width=0.3\textwidth]{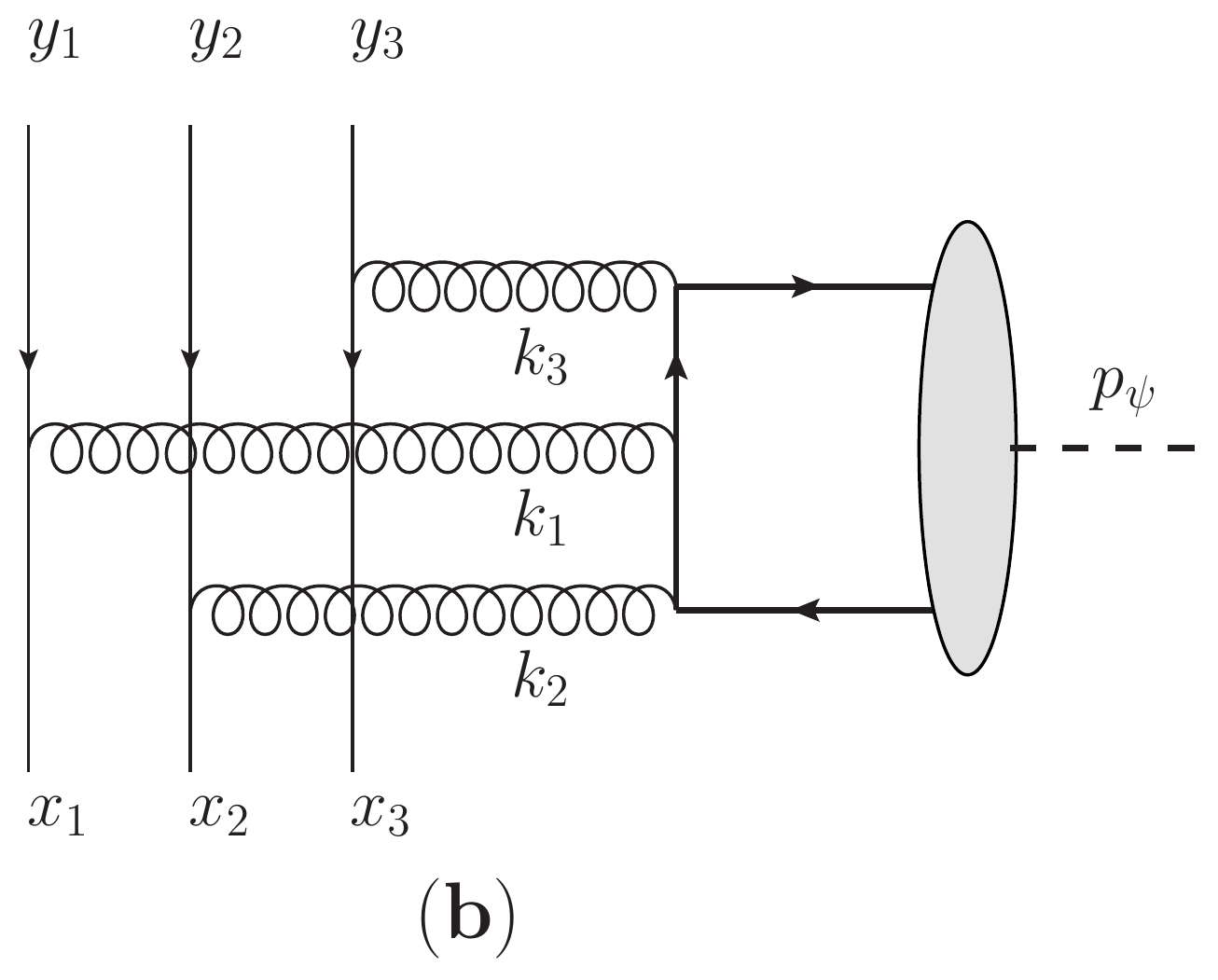}
\includegraphics[width=0.3\textwidth]{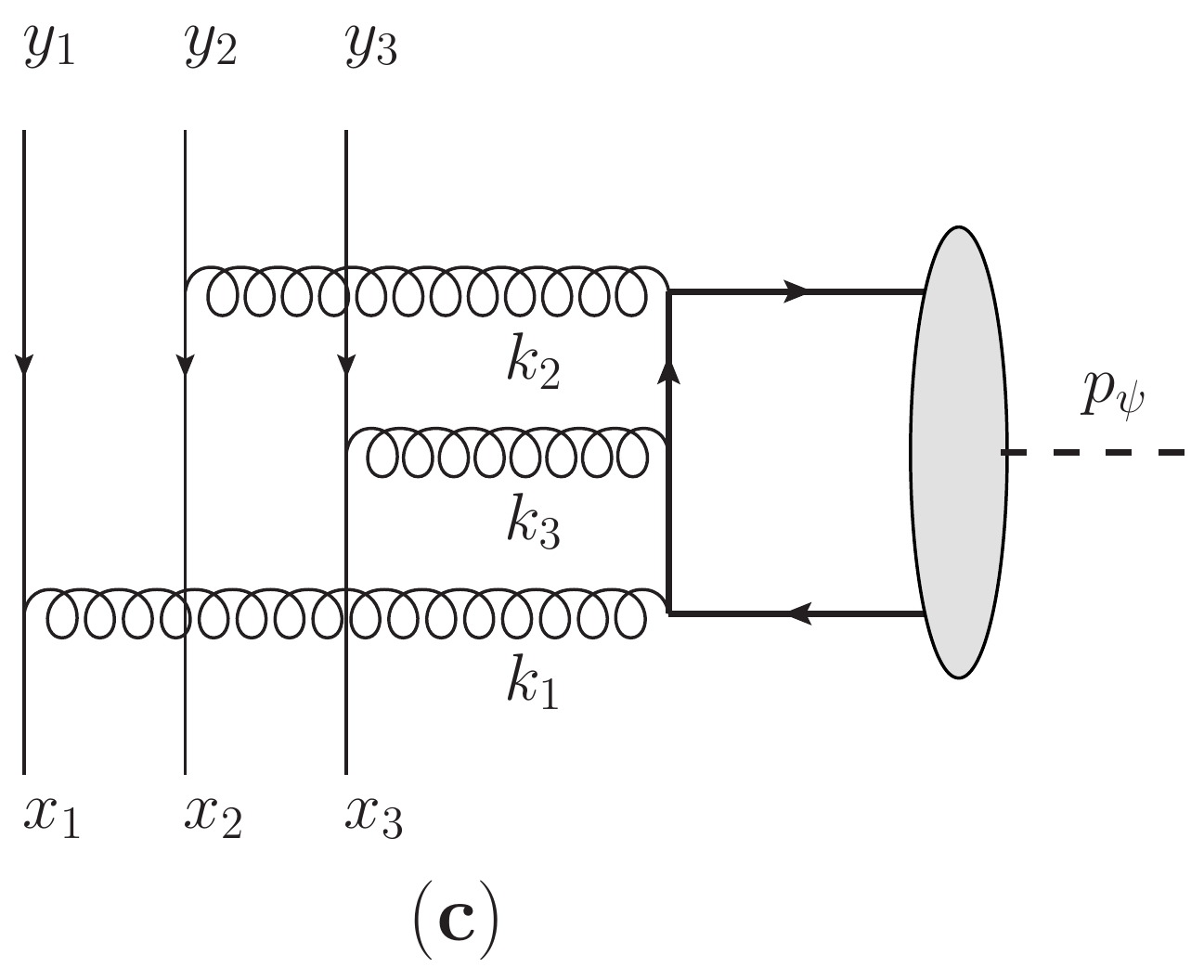}
\end{center}
     \caption{Leading order diagrams for the backward $J/\psi ~\pi$-production and photoproduction off nucleons. $x_{1,2,3}$  and $y_{1,2,3}$ stand for the longitudinal momentum fraction variables
     of, respectively, TDAs and nucleon DAs. }
\label{Fig_diagrams_Jpsi}
\end{figure}

The amplitude of backward charmonium production  (\ref{Amplitude_pi_Jpsi})
was computed in \cite{Pire:2016gut} from the $3$ leading twist-$3$ diagrams
presented in Fig.~\ref{Fig_diagrams_Jpsi}. These diagrams  are analogous to those familiar from the calculation of the
charmonium decay width (see {\it e.g.} Ref.~\cite{Chernyak:1987nv}) dominated by the transverse  polarization of the charmonium.

Here we quote the result
for the integral convolutions ${\mathcal{J}}^{(1,2)}_{\pi N \to N' J/\psi}(\xi,\Delta^2)$
of hard kernels with $N \pi$ TDAs, nucleon DAs and  non-relativistic light-cone wave function of $J/\psi$:
\begin{eqnarray}
 &&
 {\mathcal{J}}^{(1)}_{\pi N \to N' J/\psi}(\xi,\Delta^2)
= { {\int^{1+\xi}_{-1+\xi} }\! \! \!
d_3 x
\, \delta \left(\sum_{j=1}^3 x_j-2\xi\right)
}
\;
{{\int^{1}_{0} }\! \! \! d_3y
\,
\delta\left(\sum_{l=1}^3 y_l-1\right) }
\nonumber
\\ &&
\left\{ \frac{\xi ^3 (x_1 y_3+x_3 y_1) (V_1^{N \pi}-A_1^{N \pi})
(V^{p}-A^{p}) }{  y_1 y_2 y_3 (x_1+i0) (x_2+i0) (x_3+i0)   (x_1 (2
   y_1-1)-2 \xi  y_1+i0) (x_3 (2 y_3-1)-2 \xi  y_3+i0)} \right.
\nonumber \\ &&
+\left. \frac{ \xi ^3 (x_1 y_2+x_2 y_1) (2 T_1^{N \pi}
+
\frac{\Delta_T^2}{m_N^2}T_4^{N \pi}
)
T^{p}}{  y_1 y_2 y_3 (x_1+i0) (x_2+i0) (x_3+i0)
(x_1 (2y_1-1)-2 \xi  y_1+i0) (x_2 (2 y_2-1)-2 \xi  y_2+i0)}  \right\}; \nn \\ &&
   \label{Amplitude_result_I}
\end{eqnarray}
\begin{eqnarray}
 &&
 {\mathcal{J}}^{(2)}_{\pi N \to N' J/\psi}(\xi,\Delta^2)
= { {\int^{1+\xi}_{-1+\xi} }\! \! \!
d_3 x
\, \delta \left(\sum_{j=1}^3 x_j-2\xi\right)
}
\;
{{\int^{1}_{0} }\! \! \! d_3y
\,
\delta\left(\sum_{l=1}^3 y_l-1\right) } \nonumber \\ &&
\left\{ \frac{\xi ^3 (x_1 y_3+x_3 y_1) (V_2^{N \pi}-A_2^{N \pi})
(V^{p}-A^{p}) }{  y_1 y_2 y_3 (x_1+i0) (x_2+i0) (x_3+i0)   (x_1 (2
   y_1-1)-2 \xi  y_1+i0) (x_3 (2 y_3-1)-2 \xi  y_3+i0)} \right.
\nonumber \\ &&
+\left. \frac{ \xi ^3 (x_1 y_2+x_2 y_1) (T_2^{N \pi}+
T_3^{N \pi}) T^{p} }{  y_1 y_2 y_3 (x_1+i0) (x_2+i0) (x_3+i0)
(x_1 (2   y_1-1)-2 \xi  y_1+i0) (x_2 (2 y_2-1)-2 \xi  y_2+i0)}  \right\}. \nn \\ &&
   \label{Amplitude_result_Ip}
\end{eqnarray}

The expressions (\ref{Amplitude_result_I}), (\ref{Amplitude_result_Ip}) have structure similar to the well known
convolution of nucleon DAs with hard
scattering kernel
occurring in the $J/\psi \to \bar{p} p$ decay amplitude (\ref{Jpsi_decay_Chernyak})
\cite{Chernyak:1987nv}:
\be
\begin{aligned}
&M_0=\int_0^1 d_3 x \delta\left(\sum_{j=1}^3 x_j-1\right) \int_0^1 d_3 y \delta\left(\sum_{k=1}^3 y_k-1\right) \\
&\left\{\frac{y_1 x_3\left(V^p\left(x_{1,2,3}\right)-A^p\left(x_{1,2,3}\right)\right)\left(V^p\left(y_{1,2,3}\right)-A^p\left(y_{1,2,3}\right)\right)}{y_1 y_2 y_3 x_1 x_2 x_3\left(1-\left(2 x_1-1\right)\left(2 y_1-1\right)\right)\left(1-\left(2 x_3-1\right)\left(2 y_3-1\right)\right)}\right. \\
&\left.+\frac{2 y_1 x_2 T^p\left(x_{1,2,3}\right) T^p\left(y_{1,2,3}\right)}{y_1 y_2 y_3 x_1 x_2 x_3\left(1-\left(2 x_1-1\right)\left(2 y_1-1\right)\right)\left(1-\left(2 x_2-1\right)\left(2 y_2-1\right)\right)}\right\} .
\end{aligned}
\label{Def_M0}
\ee

\subsection{Backward $J/\psi$ photoproduction amplitude}
\label{App_B2}

The amplitude of backward $J/\psi$ photoproduction off nucleons
(\ref{HardReactJpsiphoton})
is calculated  from the same $3$ diagrams presented in
 Fig.~\ref{Fig_diagrams_Jpsi}. It yields the following result for the convolutions
$ {{\mathcal{J}}}^{(1,3,4,5)}_{\gamma  N \to N' J/\psi}(\xi,\Delta^2)$

\begin{eqnarray}
 &&
 {\mathcal{J}}^{(1)}_{\gamma  N \to N' J/\psi}(\xi,\Delta^2)
= { {\int^{1+\xi}_{-1+\xi} }\! \! \!
d_3 x
\, \delta \left(\sum_{j=1}^3 x_j-2\xi\right)
}
\;
{{\int^{1}_{0} }\! \! \! d_3y
\,
\delta\left(\sum_{l=1}^3 y_l-1\right) }
\nonumber
\\ &&
\left\{- \frac{\xi ^3 (x_1 y_3+x_3 y_1) (V_{1 {\cal E}}^{N \gamma}-A_{1 {\cal E}}^{N \gamma})
(V^{p}-A^{p}) }{  y_1 y_2 y_3 (x_1+i0) (x_2+i0) (x_3+i0)   (x_1 (2
   y_1-1)-2 \xi  y_1+i0) (x_3 (2 y_3-1)-2 \xi  y_3+i0)} \right.
\nonumber \\ &&
+\left. \frac{ \xi ^3 (x_1 y_2+x_2 y_1) ( T_{1 {\cal E}}^{N \gamma}+T_{2 {\cal E}}^{N \gamma}
)
T^{p}}{  y_1 y_2 y_3 (x_1+i0) (x_2+i0) (x_3+i0)
(x_1 (2y_1-1)-2 \xi  y_1+i0) (x_2 (2 y_2-1)-2 \xi  y_2+i0)}  \right\}; \nn \\ &&
   \label{Amplitude_result_J1gammaJpsi}
\end{eqnarray}
\begin{eqnarray}
 &&
 {\mathcal{J}}^{(3)}_{\gamma  N \to N' J/\psi}(\xi,\Delta^2)
= { {\int^{1+\xi}_{-1+\xi} }\! \! \!
d_3 x
\, \delta \left(\sum_{j=1}^3 x_j-2\xi\right)
}
\;
{{\int^{1}_{0} }\! \! \! d_3y
\,
\delta\left(\sum_{l=1}^3 y_l-1\right) } \nonumber \\ &&
\left\{ -\frac{\xi ^3 (x_1 y_3+x_3 y_1) (V_{1 T}^{N \gamma}-A_{1 T}^{N \gamma}+V_{2 {\cal E}}^{N \gamma}-A_{2 {\cal E}}^{N \gamma})
(V^{p}-A^{p}) }{  y_1 y_2 y_3 (x_1+i0) (x_2+i0) (x_3+i0)   (x_1 (2
   y_1-1)-2 \xi  y_1+i0) (x_3 (2 y_3-1)-2 \xi  y_3+i0)} \right.
\nonumber \\ &&
+\left. \frac{ \xi ^3 (x_1 y_2+x_2 y_1) ( 2 T_{3 {\cal E}}^{N \gamma}+2 T_{1 T}^{N \gamma} + \frac{\Delta_T^2}{m_N^2} 2 T_{4 T}^{N \gamma} ) T^{p} }{  y_1 y_2 y_3 (x_1+i0) (x_2+i0) (x_3+i0)
(x_1 (2   y_1-1)-2 \xi  y_1+i0) (x_2 (2 y_2-1)-2 \xi  y_2+i0)}  \right\}; \nn \\ &&
   \label{Amplitude_result_J3gammaJpsi}
\end{eqnarray}

\begin{eqnarray}
 &&
 {\mathcal{J}}^{(4)}_{\gamma  N \to N' J/\psi}(\xi,\Delta^2)
= { {\int^{1+\xi}_{-1+\xi} }\! \! \!
d_3 x
\, \delta \left(\sum_{j=1}^3 x_j-2\xi\right)
}
\;
{{\int^{1}_{0} }\! \! \! d_3y
\,
\delta\left(\sum_{l=1}^3 y_l-1\right) } \nonumber \\ &&
\left\{ -\frac{\xi ^3 (x_1 y_3+x_3 y_1) (V_{2 T}^{N \gamma}-A_{2 T}^{N \gamma})
(V^{p}-A^{p}) }{  y_1 y_2 y_3 (x_1+i0) (x_2+i0) (x_3+i0)   (x_1 (2
   y_1-1)-2 \xi  y_1+i0) (x_3 (2 y_3-1)-2 \xi  y_3+i0)} \right.
\nonumber \\ &&
+\left. \frac{ \xi ^3 (x_1 y_2+x_2 y_1) (  T_{2T}^{N \gamma}+  T_{3 T}^{N \gamma}  ) T^{p} }{  y_1 y_2 y_3 (x_1+i0) (x_2+i0) (x_3+i0)
(x_1 (2   y_1-1)-2 \xi  y_1+i0) (x_2 (2 y_2-1)-2 \xi  y_2+i0)}  \right\}; \nn \\ &&
   \label{Amplitude_result_J4gammaJpsi}
\end{eqnarray}

\begin{eqnarray}
 &&
{\mathcal{J}}^{(5)}_{\gamma  N \to N' J/\psi}(\xi,\Delta^2)
= { {\int^{1+\xi}_{-1+\xi} }\! \! \!
d_3 x
\, \delta \left(\sum_{j=1}^3 x_j-2\xi\right)
}
\;
{{\int^{1}_{0} }\! \! \! d_3y
\,
\delta\left(\sum_{l=1}^3 y_l-1\right) } \nonumber \\ &&
\left\{ \frac{\xi ^3 (x_1 y_3+x_3 y_1) (V_{2 {\cal E}}^{N \gamma}-A_{2 {\cal E}}^{N \gamma})
(V^{p}-A^{p}) }{  y_1 y_2 y_3 (x_1+i0) (x_2+i0) (x_3+i0)   (x_1 (2
   y_1-1)-2 \xi  y_1+i0) (x_3 (2 y_3-1)-2 \xi  y_3+i0)} \right.
\nonumber \\ &&
-\left. \frac{ \xi ^3 (x_1 y_2+x_2 y_1) (  T_{3 {\cal E}}^{N \gamma}-  T_{4 {\cal E}}^{N \gamma}  ) T^{p} }{  y_1 y_2 y_3 (x_1+i0) (x_2+i0) (x_3+i0)
(x_1 (2   y_1-1)-2 \xi  y_1+i0) (x_2 (2 y_2-1)-2 \xi  y_2+i0)}  \right\}. \nn \\ &&
   \label{Amplitude_result_J5gammaJpsi}
\end{eqnarray}

\bibliography{AAPPS_TDA}

\begin{thebibliography}{50}%
\makeatletter
\providecommand \@ifxundefined [1]{%
 \@ifx{#1\undefined}
}%
\providecommand \@ifnum [1]{%
 \ifnum #1\expandafter \@firstoftwo
 \else \expandafter \@secondoftwo
 \fi
}%
\providecommand \@ifx [1]{%
 \ifx #1\expandafter \@firstoftwo
 \else \expandafter \@secondoftwo
 \fi
}%
\providecommand \natexlab [1]{#1}%
\providecommand \enquote  [1]{``#1''}%
\providecommand \bibnamefont  [1]{#1}%
\providecommand \bibfnamefont [1]{#1}%
\providecommand \citenamefont [1]{#1}%
\providecommand \href@noop [0]{\@secondoftwo}%
\providecommand \href [0]{\begingroup \@sanitize@url \@href}%
\providecommand \@href[1]{\@@startlink{#1}\@@href}%
\providecommand \@@href[1]{\endgroup#1\@@endlink}%
\providecommand \@sanitize@url [0]{\catcode `\\12\catcode `\$12\catcode
  `\&12\catcode `\#12\catcode `\^12\catcode `\_12\catcode `\%12\relax}%
\providecommand \@@startlink[1]{}%
\providecommand \@@endlink[0]{}%
\providecommand \url  [0]{\begingroup\@sanitize@url \@url }%
\providecommand \@url [1]{\endgroup\@href {#1}{\urlprefix }}%
\providecommand \urlprefix  [0]{URL }%
\providecommand \Eprint [0]{\href }%
\providecommand \doibase [0]{https://doi.org/}%
\providecommand \selectlanguage [0]{\@gobble}%
\providecommand \bibinfo  [0]{\@secondoftwo}%
\providecommand \bibfield  [0]{\@secondoftwo}%
\providecommand \translation [1]{[#1]}%
\providecommand \BibitemOpen [0]{}%
\providecommand \bibitemStop [0]{}%
\providecommand \bibitemNoStop [0]{.\EOS\space}%
\providecommand \EOS [0]{\spacefactor3000\relax}%
\providecommand \BibitemShut  [1]{\csname bibitem#1\endcsname}%
\let\auto@bib@innerbib\@empty
\bibitem [{\citenamefont {M\"uller}\ \emph {et~al.}(1994)\citenamefont
  {M\"uller}, \citenamefont {Robaschik}, \citenamefont {Geyer}, \citenamefont
  {Dittes},\ and\ \citenamefont {Ho\v{r}ej\v{s}i}}]{Muller:1994ses}%
  \BibitemOpen
  \bibfield  {author} {\bibinfo {author} {\bibfnamefont {D.}~\bibnamefont
  {M\"uller}}, \bibinfo {author} {\bibfnamefont {D.}~\bibnamefont {Robaschik}},
  \bibinfo {author} {\bibfnamefont {B.}~\bibnamefont {Geyer}}, \bibinfo
  {author} {\bibfnamefont {F.~M.}\ \bibnamefont {Dittes}},\ and\ \bibinfo
  {author} {\bibfnamefont {J.}~\bibnamefont {Ho\v{r}ej\v{s}i}},\ }\bibfield
  {title} {\bibinfo {title} {{Wave functions, evolution equations and evolution
  kernels from light ray operators of QCD}},\ }\href
  {https://doi.org/10.1002/prop.2190420202} {\bibfield  {journal} {\bibinfo
  {journal} {Fortsch. Phys.}\ }\textbf {\bibinfo {volume} {42}},\ \bibinfo
  {pages} {101} (\bibinfo {year} {1994})},\ \Eprint
  {https://arxiv.org/abs/hep-ph/9812448} {arXiv:hep-ph/9812448} \BibitemShut
  {NoStop}%
\bibitem [{\citenamefont {Berger}\ \emph {et~al.}(2002)\citenamefont {Berger},
  \citenamefont {Diehl},\ and\ \citenamefont {Pire}}]{Berger:2001xd}%
  \BibitemOpen
  \bibfield  {author} {\bibinfo {author} {\bibfnamefont {E.~R.}\ \bibnamefont
  {Berger}}, \bibinfo {author} {\bibfnamefont {M.}~\bibnamefont {Diehl}},\ and\
  \bibinfo {author} {\bibfnamefont {B.}~\bibnamefont {Pire}},\ }\bibfield
  {title} {\bibinfo {title} {{Time - like Compton scattering: Exclusive
  photoproduction of lepton pairs}},\ }\href
  {https://doi.org/10.1007/s100520200917} {\bibfield  {journal} {\bibinfo
  {journal} {Eur. Phys. J. C}\ }\textbf {\bibinfo {volume} {23}},\ \bibinfo
  {pages} {675} (\bibinfo {year} {2002})},\ \Eprint
  {https://arxiv.org/abs/hep-ph/0110062} {arXiv:hep-ph/0110062} \BibitemShut
  {NoStop}%
\bibitem [{\citenamefont {Berger}\ \emph {et~al.}(2001)\citenamefont {Berger},
  \citenamefont {Diehl},\ and\ \citenamefont {Pire}}]{Berger:2001zn}%
  \BibitemOpen
  \bibfield  {author} {\bibinfo {author} {\bibfnamefont {E.~R.}\ \bibnamefont
  {Berger}}, \bibinfo {author} {\bibfnamefont {M.}~\bibnamefont {Diehl}},\ and\
  \bibinfo {author} {\bibfnamefont {B.}~\bibnamefont {Pire}},\ }\bibfield
  {title} {\bibinfo {title} {{Probing generalized parton distributions in $\pi
  N \to \ell^+ \ell^- N$}},\ }\href
  {https://doi.org/10.1016/S0370-2693(01)01345-4} {\bibfield  {journal}
  {\bibinfo  {journal} {Phys. Lett. B}\ }\textbf {\bibinfo {volume} {523}},\
  \bibinfo {pages} {265} (\bibinfo {year} {2001})},\ \Eprint
  {https://arxiv.org/abs/hep-ph/0110080} {arXiv:hep-ph/0110080} \BibitemShut
  {NoStop}%
\bibitem [{\citenamefont {Sawada}\ \emph
  {et~al.}(2016{\natexlab{a}})\citenamefont {Sawada}, \citenamefont {Chang},
  \citenamefont {Kumano}, \citenamefont {Peng}, \citenamefont {Sawada},\ and\
  \citenamefont {Tanaka}}]{Sawada:2016mao}%
  \BibitemOpen
  \bibfield  {author} {\bibinfo {author} {\bibfnamefont {T.}~\bibnamefont
  {Sawada}}, \bibinfo {author} {\bibfnamefont {W.-C.}\ \bibnamefont {Chang}},
  \bibinfo {author} {\bibfnamefont {S.}~\bibnamefont {Kumano}}, \bibinfo
  {author} {\bibfnamefont {J.-C.}\ \bibnamefont {Peng}}, \bibinfo {author}
  {\bibfnamefont {S.}~\bibnamefont {Sawada}},\ and\ \bibinfo {author}
  {\bibfnamefont {K.}~\bibnamefont {Tanaka}},\ }\bibfield  {title} {\bibinfo
  {title} {{Accessing proton generalized parton distributions and pion
  distribution amplitudes with the exclusive pion-induced Drell-Yan process at
  J-PARC}},\ }\href {https://doi.org/10.1103/PhysRevD.93.114034} {\bibfield
  {journal} {\bibinfo  {journal} {Phys. Rev. D}\ }\textbf {\bibinfo {volume}
  {93}},\ \bibinfo {pages} {114034} (\bibinfo {year} {2016}{\natexlab{a}})},\
  \Eprint {https://arxiv.org/abs/1605.00364} {arXiv:1605.00364 [nucl-ex]}
  \BibitemShut {NoStop}%
\bibitem [{\citenamefont {M\"uller}\ \emph {et~al.}(2012)\citenamefont
  {M\"uller}, \citenamefont {Pire}, \citenamefont {Szymanowski},\ and\
  \citenamefont {Wagner}}]{Mueller:2012sma}%
  \BibitemOpen
  \bibfield  {author} {\bibinfo {author} {\bibfnamefont {D.}~\bibnamefont
  {M\"uller}}, \bibinfo {author} {\bibfnamefont {B.}~\bibnamefont {Pire}},
  \bibinfo {author} {\bibfnamefont {L.}~\bibnamefont {Szymanowski}},\ and\
  \bibinfo {author} {\bibfnamefont {J.}~\bibnamefont {Wagner}},\ }\bibfield
  {title} {\bibinfo {title} {{On timelike and spacelike hard exclusive
  reactions}},\ }\href {https://doi.org/10.1103/PhysRevD.86.031502} {\bibfield
  {journal} {\bibinfo  {journal} {Phys. Rev. D}\ }\textbf {\bibinfo {volume}
  {86}},\ \bibinfo {pages} {031502} (\bibinfo {year} {2012})},\ \Eprint
  {https://arxiv.org/abs/1203.4392} {arXiv:1203.4392 [hep-ph]} \BibitemShut
  {NoStop}%
\bibitem [{\citenamefont {Chatagnon}\ \emph {et~al.}(2021)\citenamefont
  {Chatagnon} \emph {et~al.}}]{CLAS:2021lky}%
  \BibitemOpen
  \bibfield  {author} {\bibinfo {author} {\bibfnamefont {P.}~\bibnamefont
  {Chatagnon}} \emph {et~al.} (\bibinfo {collaboration} {CLAS}),\ }\bibfield
  {title} {\bibinfo {title} {{First Measurement of Timelike Compton
  Scattering}},\ }\href {https://doi.org/10.1103/PhysRevLett.127.262501}
  {\bibfield  {journal} {\bibinfo  {journal} {Phys. Rev. Lett.}\ }\textbf
  {\bibinfo {volume} {127}},\ \bibinfo {pages} {262501} (\bibinfo {year}
  {2021})},\ \Eprint {https://arxiv.org/abs/2108.11746} {arXiv:2108.11746
  [hep-ex]} \BibitemShut {NoStop}%
\bibitem [{\citenamefont {Ali}\ \emph {et~al.}(2019)\citenamefont {Ali} \emph
  {et~al.}}]{GlueX:2019mkq}%
  \BibitemOpen
  \bibfield  {author} {\bibinfo {author} {\bibfnamefont {A.}~\bibnamefont
  {Ali}} \emph {et~al.} (\bibinfo {collaboration} {GlueX}),\ }\bibfield
  {title} {\bibinfo {title} {{First Measurement of Near-Threshold
  J/\ensuremath{\psi} Exclusive Photoproduction off the Proton}},\ }\href
  {https://doi.org/10.1103/PhysRevLett.123.072001} {\bibfield  {journal}
  {\bibinfo  {journal} {Phys. Rev. Lett.}\ }\textbf {\bibinfo {volume} {123}},\
  \bibinfo {pages} {072001} (\bibinfo {year} {2019})},\ \Eprint
  {https://arxiv.org/abs/1905.10811} {arXiv:1905.10811 [nucl-ex]} \BibitemShut
  {NoStop}%
\bibitem [{\citenamefont {Lee}\ \emph {et~al.}(2022)\citenamefont {Lee},
  \citenamefont {Sakinah},\ and\ \citenamefont {Oh}}]{Lee:2022ymp}%
  \BibitemOpen
  \bibfield  {author} {\bibinfo {author} {\bibfnamefont {T.~S.~H.}\
  \bibnamefont {Lee}}, \bibinfo {author} {\bibfnamefont {S.}~\bibnamefont
  {Sakinah}},\ and\ \bibinfo {author} {\bibfnamefont {Y.}~\bibnamefont {Oh}},\
  }\bibfield  {title} {\bibinfo {title} {{Models of $J/\varPsi $
  photo-production reactions on the nucleon}},\ }\href
  {https://doi.org/10.1140/epja/s10050-022-00901-9} {\bibfield  {journal}
  {\bibinfo  {journal} {Eur. Phys. J. A}\ }\textbf {\bibinfo {volume} {58}},\
  \bibinfo {pages} {252} (\bibinfo {year} {2022})},\ \Eprint
  {https://arxiv.org/abs/2210.02154} {arXiv:2210.02154 [hep-ph]} \BibitemShut
  {NoStop}%
\bibitem [{\citenamefont {Sun}\ \emph {et~al.}(2022)\citenamefont {Sun},
  \citenamefont {Tong},\ and\ \citenamefont {Yuan}}]{Sun:2021pyw}%
  \BibitemOpen
  \bibfield  {author} {\bibinfo {author} {\bibfnamefont {P.}~\bibnamefont
  {Sun}}, \bibinfo {author} {\bibfnamefont {X.-B.}\ \bibnamefont {Tong}},\ and\
  \bibinfo {author} {\bibfnamefont {F.}~\bibnamefont {Yuan}},\ }\bibfield
  {title} {\bibinfo {title} {{Near threshold heavy quarkonium photoproduction
  at large momentum transfer}},\ }\href
  {https://doi.org/10.1103/PhysRevD.105.054032} {\bibfield  {journal} {\bibinfo
   {journal} {Phys. Rev. D}\ }\textbf {\bibinfo {volume} {105}},\ \bibinfo
  {pages} {054032} (\bibinfo {year} {2022})},\ \Eprint
  {https://arxiv.org/abs/2111.07034} {arXiv:2111.07034 [hep-ph]} \BibitemShut
  {NoStop}%
\bibitem [{\citenamefont {Pire}\ \emph {et~al.}(2017)\citenamefont {Pire},
  \citenamefont {Semenov-Tian-Shansky},\ and\ \citenamefont
  {Szymanowski}}]{Pire:2016gut}%
  \BibitemOpen
  \bibfield  {author} {\bibinfo {author} {\bibfnamefont {B.}~\bibnamefont
  {Pire}}, \bibinfo {author} {\bibfnamefont {K.}~\bibnamefont
  {Semenov-Tian-Shansky}},\ and\ \bibinfo {author} {\bibfnamefont
  {L.}~\bibnamefont {Szymanowski}},\ }\bibfield  {title} {\bibinfo {title}
  {{Backward charmonium production in $\pi N$ collisions}},\ }\href
  {https://doi.org/10.1103/PhysRevD.95.034021} {\bibfield  {journal} {\bibinfo
  {journal} {Phys. Rev. D}\ }\textbf {\bibinfo {volume} {95}},\ \bibinfo
  {pages} {034021} (\bibinfo {year} {2017})},\ \Eprint
  {https://arxiv.org/abs/1611.07234} {arXiv:1611.07234 [hep-ph]} \BibitemShut
  {NoStop}%
\bibitem [{\citenamefont {Pire}\ \emph {et~al.}(2022)\citenamefont {Pire},
  \citenamefont {Semenov-Tian-Shansky}, \citenamefont {Shaikhutdinova},\ and\
  \citenamefont {Szymanowski}}]{Pire:2022fbi}%
  \BibitemOpen
  \bibfield  {author} {\bibinfo {author} {\bibfnamefont {B.}~\bibnamefont
  {Pire}}, \bibinfo {author} {\bibfnamefont {K.~M.}\ \bibnamefont
  {Semenov-Tian-Shansky}}, \bibinfo {author} {\bibfnamefont {A.~A.}\
  \bibnamefont {Shaikhutdinova}},\ and\ \bibinfo {author} {\bibfnamefont
  {L.}~\bibnamefont {Szymanowski}},\ }\bibfield  {title} {\bibinfo {title}
  {{Backward timelike Compton scattering to decipher the photon content of the
  nucleon}},\ }\href {https://doi.org/10.1140/epjc/s10052-022-10587-4}
  {\bibfield  {journal} {\bibinfo  {journal} {Eur. Phys. J. C}\ }\textbf
  {\bibinfo {volume} {82}},\ \bibinfo {pages} {656} (\bibinfo {year} {2022})},\
  \Eprint {https://arxiv.org/abs/2201.12853} {arXiv:2201.12853 [hep-ph]}
  \BibitemShut {NoStop}%
\bibitem [{\citenamefont {Pire}\ \emph {et~al.}(2021)\citenamefont {Pire},
  \citenamefont {Semenov-Tian-Shansky},\ and\ \citenamefont
  {Szymanowski}}]{Pire:2021hbl}%
  \BibitemOpen
  \bibfield  {author} {\bibinfo {author} {\bibfnamefont {B.}~\bibnamefont
  {Pire}}, \bibinfo {author} {\bibfnamefont {K.}~\bibnamefont
  {Semenov-Tian-Shansky}},\ and\ \bibinfo {author} {\bibfnamefont
  {L.}~\bibnamefont {Szymanowski}},\ }\bibfield  {title} {\bibinfo {title}
  {{Transition distribution amplitudes and hard exclusive reactions with baryon
  number transfer}},\ }\href {https://doi.org/10.1016/j.physrep.2021.09.002}
  {\bibfield  {journal} {\bibinfo  {journal} {Phys. Rept.}\ }\textbf {\bibinfo
  {volume} {940}},\ \bibinfo {pages} {2185} (\bibinfo {year} {2021})},\ \Eprint
  {https://arxiv.org/abs/2103.01079} {arXiv:2103.01079 [hep-ph]} \BibitemShut
  {NoStop}%
\bibitem [{\citenamefont {Gayoso}\ \emph {et~al.}(2021)\citenamefont {Gayoso}
  \emph {et~al.}}]{Gayoso:2021rzj}%
  \BibitemOpen
  \bibfield  {author} {\bibinfo {author} {\bibfnamefont {C.~A.}\ \bibnamefont
  {Gayoso}} \emph {et~al.},\ }\bibfield  {title} {\bibinfo {title} {{Progress
  and opportunities in backward angle ($u$-channel) physics}},\ }\href
  {https://doi.org/10.1140/epja/s10050-021-00625-2} {\bibfield  {journal}
  {\bibinfo  {journal} {Eur. Phys. J. A}\ }\textbf {\bibinfo {volume} {57}},\
  \bibinfo {pages} {342} (\bibinfo {year} {2021})},\ \Eprint
  {https://arxiv.org/abs/2107.06748} {arXiv:2107.06748 [hep-ph]} \BibitemShut
  {NoStop}%
\bibitem [{\citenamefont {Li}\ \emph {et~al.}(2020)\citenamefont {Li} \emph
  {et~al.}}]{Li:2020nsk}%
  \BibitemOpen
  \bibfield  {author} {\bibinfo {author} {\bibfnamefont {W.~B.}\ \bibnamefont
  {Li}} \emph {et~al.},\ }\bibfield  {title} {\bibinfo {title} {{Backward-angle
  Exclusive $\pi^0$ Production above the Resonance Region}},\ }\Eprint
  {https://arxiv.org/abs/2008.10768} {arXiv:2008.10768 [nucl-ex]}  (\bibinfo
  {year} {2020})\BibitemShut {NoStop}%
\bibitem [{\citenamefont {Aoki}\ \emph {et~al.}(2021)\citenamefont {Aoki} \emph
  {et~al.}}]{Aoki:2021cqa}%
  \BibitemOpen
  \bibfield  {author} {\bibinfo {author} {\bibfnamefont {K.}~\bibnamefont
  {Aoki}} \emph {et~al.},\ }\bibfield  {title} {\bibinfo {title} {{Extension of
  the J-PARC Hadron Experimental Facility: Third White Paper}},\ }\Eprint
  {https://arxiv.org/abs/2110.04462} {arXiv:2110.04462 [nucl-ex]}  (\bibinfo
  {year} {2021})\BibitemShut {NoStop}%
\bibitem [{\citenamefont {Abdul~Khalek}\ \emph {et~al.}(2022)\citenamefont
  {Abdul~Khalek} \emph {et~al.}}]{AbdulKhalek:2021gbh}%
  \BibitemOpen
  \bibfield  {author} {\bibinfo {author} {\bibfnamefont {R.}~\bibnamefont
  {Abdul~Khalek}} \emph {et~al.},\ }\bibfield  {title} {\bibinfo {title}
  {{Science Requirements and Detector Concepts for the Electron-Ion Collider}:
  {EIC Yellow Report}},\ }\href
  {https://doi.org/10.1016/j.nuclphysa.2022.122447} {\bibfield  {journal}
  {\bibinfo  {journal} {Nucl. Phys. A}\ }\textbf {\bibinfo {volume} {1026}},\
  \bibinfo {pages} {122447} (\bibinfo {year} {2022})},\ \Eprint
  {https://arxiv.org/abs/2103.05419} {arXiv:2103.05419 [physics.ins-det]}
  \BibitemShut {NoStop}%
\bibitem [{\citenamefont {Burkert}\ \emph {et~al.}(2022)\citenamefont {Burkert}
  \emph {et~al.}}]{Burkert:2022hjz}%
  \BibitemOpen
  \bibfield  {author} {\bibinfo {author} {\bibfnamefont {V.}~\bibnamefont
  {Burkert}} \emph {et~al.},\ }\bibfield  {title} {\bibinfo {title} {{Precision
  Studies of QCD in the Low Energy Domain of the EIC}},\ }\Eprint
  {https://arxiv.org/abs/2211.15746} {arXiv:2211.15746 [nucl-ex]}  (\bibinfo
  {year} {2022})\BibitemShut {NoStop}%
\bibitem [{\citenamefont {Anderle}\ \emph {et~al.}(2021)\citenamefont {Anderle}
  \emph {et~al.}}]{Anderle:2021wcy}%
  \BibitemOpen
  \bibfield  {author} {\bibinfo {author} {\bibfnamefont {D.~P.}\ \bibnamefont
  {Anderle}} \emph {et~al.},\ }\bibfield  {title} {\bibinfo {title}
  {{Electron-ion collider in China}},\ }\href
  {https://doi.org/10.1007/s11467-021-1062-0} {\bibfield  {journal} {\bibinfo
  {journal} {Front. Phys. (Beijing)}\ }\textbf {\bibinfo {volume} {16}},\
  \bibinfo {pages} {64701} (\bibinfo {year} {2021})},\ \Eprint
  {https://arxiv.org/abs/2102.09222} {arXiv:2102.09222 [nucl-ex]} \BibitemShut
  {NoStop}%
\bibitem [{\citenamefont {Adhikari}\ \emph {et~al.}(2023)\citenamefont
  {Adhikari} \emph {et~al.}}]{Adhikari:2023fcr}%
  \BibitemOpen
  \bibfield  {author} {\bibinfo {author} {\bibfnamefont {S.}~\bibnamefont
  {Adhikari}} \emph {et~al.},\ }\href@noop {} {\bibinfo {title} {{Measurement
  of the J/$\psi $ photoproduction cross section over the full near-threshold
  kinematic region}}} (\bibinfo {year} {2023}),\ \Eprint
  {https://arxiv.org/abs/2304.03845} {arXiv:2304.03845 [nucl-ex]} \BibitemShut
  {NoStop}%
\bibitem [{\citenamefont {Pire}\ \emph {et~al.}(2011)\citenamefont {Pire},
  \citenamefont {Semenov-Tian-Shansky},\ and\ \citenamefont
  {Szymanowski}}]{Pire:2011xv}%
  \BibitemOpen
  \bibfield  {author} {\bibinfo {author} {\bibfnamefont {B.}~\bibnamefont
  {Pire}}, \bibinfo {author} {\bibfnamefont {K.}~\bibnamefont
  {Semenov-Tian-Shansky}},\ and\ \bibinfo {author} {\bibfnamefont
  {L.}~\bibnamefont {Szymanowski}},\ }\bibfield  {title} {\bibinfo {title} {{
  $\pi N$ transition distribution amplitudes: their symmetries and constraints
  from chiral dynamics}},\ }\href {https://doi.org/10.1103/PhysRevD.84.074014}
  {\bibfield  {journal} {\bibinfo  {journal} {Phys. Rev. D}\ }\textbf {\bibinfo
  {volume} {84}},\ \bibinfo {pages} {074014} (\bibinfo {year} {2011})},\
  \Eprint {https://arxiv.org/abs/1106.1851} {arXiv:1106.1851 [hep-ph]}
  \BibitemShut {NoStop}%
\bibitem [{\citenamefont {Chernyak}\ \emph {et~al.}(1989)\citenamefont
  {Chernyak}, \citenamefont {Ogloblin},\ and\ \citenamefont
  {Zhitnitsky}}]{Chernyak:1987nv}%
  \BibitemOpen
  \bibfield  {author} {\bibinfo {author} {\bibfnamefont {V.~L.}\ \bibnamefont
  {Chernyak}}, \bibinfo {author} {\bibfnamefont {A.~A.}\ \bibnamefont
  {Ogloblin}},\ and\ \bibinfo {author} {\bibfnamefont {I.~R.}\ \bibnamefont
  {Zhitnitsky}},\ }\bibfield  {title} {\bibinfo {title} {{Calculation of
  Exclusive Processes With Baryons}},\ }\href
  {https://doi.org/10.1007/BF01557664} {\bibfield  {journal} {\bibinfo
  {journal} {Z. Phys.}\ }\textbf {\bibinfo {volume} {C42}},\ \bibinfo {pages}
  {583} (\bibinfo {year} {1989})},\ \bibinfo {note} {[Yad. Fiz.48, 1398 (1988);
  Sov. J. Nucl. Phys.48, 889 (1988)]}\BibitemShut {NoStop}%
\bibitem [{\citenamefont {Lansberg}\ \emph {et~al.}(2007)\citenamefont
  {Lansberg}, \citenamefont {Pire},\ and\ \citenamefont
  {Szymanowski}}]{Lansberg:2007ec}%
  \BibitemOpen
  \bibfield  {author} {\bibinfo {author} {\bibfnamefont {J.~P.}\ \bibnamefont
  {Lansberg}}, \bibinfo {author} {\bibfnamefont {B.}~\bibnamefont {Pire}},\
  and\ \bibinfo {author} {\bibfnamefont {L.}~\bibnamefont {Szymanowski}},\
  }\bibfield  {title} {\bibinfo {title} {{Hard exclusive electroproduction of a
  pion in the backward region}},\ }\href
  {https://doi.org/10.1103/PhysRevD.75.074004} {\bibfield  {journal} {\bibinfo
  {journal} {Phys. Rev. D}\ }\textbf {\bibinfo {volume} {75}},\ \bibinfo
  {pages} {074004} (\bibinfo {year} {2007})},\ \bibinfo {note} {[Erratum:
  Phys.Rev.D 77, 019902 (2008)]},\ \Eprint
  {https://arxiv.org/abs/hep-ph/0701125} {arXiv:hep-ph/0701125} \BibitemShut
  {NoStop}%
\bibitem [{\citenamefont {Chernyak}\ and\ \citenamefont
  {Zhitnitsky}(1984)}]{Chernyak:1983ej}%
  \BibitemOpen
  \bibfield  {author} {\bibinfo {author} {\bibfnamefont {V.~L.}\ \bibnamefont
  {Chernyak}}\ and\ \bibinfo {author} {\bibfnamefont {A.~R.}\ \bibnamefont
  {Zhitnitsky}},\ }\bibfield  {title} {\bibinfo {title} {{Asymptotic Behavior
  of Exclusive Processes in QCD}},\ }\href
  {https://doi.org/10.1016/0370-1573(84)90126-1} {\bibfield  {journal}
  {\bibinfo  {journal} {Phys. Rept.}\ }\textbf {\bibinfo {volume} {112}},\
  \bibinfo {pages} {173} (\bibinfo {year} {1984})}\BibitemShut {NoStop}%
\bibitem [{\citenamefont {Workman}\ and\ \citenamefont
  {Others}(2022)}]{Workman:2022ynf}%
  \BibitemOpen
  \bibfield  {author} {\bibinfo {author} {\bibfnamefont {R.~L.}\ \bibnamefont
  {Workman}}\ and\ \bibinfo {author} {\bibnamefont {Others}} (\bibinfo
  {collaboration} {Particle Data Group}),\ }\bibfield  {title} {\bibinfo
  {title} {{Review of Particle Physics}},\ }\href
  {https://doi.org/10.1093/ptep/ptac097} {\bibfield  {journal} {\bibinfo
  {journal} {PTEP}\ }\textbf {\bibinfo {volume} {2022}},\ \bibinfo {pages}
  {083C01} (\bibinfo {year} {2022})}\BibitemShut {NoStop}%
\bibitem [{\citenamefont {Pire}\ \emph {et~al.}(2015)\citenamefont {Pire},
  \citenamefont {Semenov-Tian-Shansky},\ and\ \citenamefont
  {Szymanowski}}]{Pire:2015kxa}%
  \BibitemOpen
  \bibfield  {author} {\bibinfo {author} {\bibfnamefont {B.}~\bibnamefont
  {Pire}}, \bibinfo {author} {\bibfnamefont {K.}~\bibnamefont
  {Semenov-Tian-Shansky}},\ and\ \bibinfo {author} {\bibfnamefont
  {L.}~\bibnamefont {Szymanowski}},\ }\bibfield  {title} {\bibinfo {title}
  {{QCD description of backward vector meson hard electroproduction}},\ }\href
  {https://doi.org/10.1103/PhysRevD.91.094006} {\bibfield  {journal} {\bibinfo
  {journal} {Phys. Rev. D}\ }\textbf {\bibinfo {volume} {91}},\ \bibinfo
  {pages} {094006} (\bibinfo {year} {2015})},\ \bibinfo {note} {[Erratum:
  Phys.Rev.D 106, 099901 (2022)]},\ \Eprint {https://arxiv.org/abs/1503.02012}
  {arXiv:1503.02012 [hep-ph]} \BibitemShut {NoStop}%
\bibitem [{\citenamefont {Bolz}\ and\ \citenamefont
  {Kroll}(1996)}]{Bolz:1996sw}%
  \BibitemOpen
  \bibfield  {author} {\bibinfo {author} {\bibfnamefont {J.}~\bibnamefont
  {Bolz}}\ and\ \bibinfo {author} {\bibfnamefont {P.}~\bibnamefont {Kroll}},\
  }\bibfield  {title} {\bibinfo {title} {{Modeling the nucleon wave function
  from soft and hard processes}},\ }\href
  {https://doi.org/10.1007/s002180050186} {\bibfield  {journal} {\bibinfo
  {journal} {Z. Phys. A}\ }\textbf {\bibinfo {volume} {356}},\ \bibinfo {pages}
  {327} (\bibinfo {year} {1996})},\ \Eprint
  {https://arxiv.org/abs/hep-ph/9603289} {arXiv:hep-ph/9603289} \BibitemShut
  {NoStop}%
\bibitem [{\citenamefont {Braun}\ \emph {et~al.}(2006)\citenamefont {Braun},
  \citenamefont {Lenz},\ and\ \citenamefont {Wittmann}}]{Braun:2006hz}%
  \BibitemOpen
  \bibfield  {author} {\bibinfo {author} {\bibfnamefont {V.~M.}\ \bibnamefont
  {Braun}}, \bibinfo {author} {\bibfnamefont {A.}~\bibnamefont {Lenz}},\ and\
  \bibinfo {author} {\bibfnamefont {M.}~\bibnamefont {Wittmann}},\ }\bibfield
  {title} {\bibinfo {title} {{Nucleon Form Factors in QCD}},\ }\href
  {https://doi.org/10.1103/PhysRevD.73.094019} {\bibfield  {journal} {\bibinfo
  {journal} {Phys. Rev. D}\ }\textbf {\bibinfo {volume} {73}},\ \bibinfo
  {pages} {094019} (\bibinfo {year} {2006})},\ \Eprint
  {https://arxiv.org/abs/hep-ph/0604050} {arXiv:hep-ph/0604050} \BibitemShut
  {NoStop}%
\bibitem [{\citenamefont {Kim}\ \emph {et~al.}(2021)\citenamefont {Kim},
  \citenamefont {Kim},\ and\ \citenamefont {Polyakov}}]{Kim:2021zbz}%
  \BibitemOpen
  \bibfield  {author} {\bibinfo {author} {\bibfnamefont {J.-Y.}\ \bibnamefont
  {Kim}}, \bibinfo {author} {\bibfnamefont {H.-C.}\ \bibnamefont {Kim}},\ and\
  \bibinfo {author} {\bibfnamefont {M.~V.}\ \bibnamefont {Polyakov}},\
  }\bibfield  {title} {\bibinfo {title} {{Light-cone distribution amplitudes of
  the nucleon and \ensuremath{\Delta} baryon}},\ }\href
  {https://doi.org/10.1007/JHEP11(2021)039} {\bibfield  {journal} {\bibinfo
  {journal} {JHEP}\ }\textbf {\bibinfo {volume} {11}},\ \bibinfo {pages}
  {039}},\ \Eprint {https://arxiv.org/abs/2110.05889} {arXiv:2110.05889
  [hep-ph]} \BibitemShut {NoStop}%
\bibitem [{\citenamefont {Musatov}\ and\ \citenamefont
  {Radyushkin}(1997)}]{Musatov:1997pu}%
  \BibitemOpen
  \bibfield  {author} {\bibinfo {author} {\bibfnamefont {I.~V.}\ \bibnamefont
  {Musatov}}\ and\ \bibinfo {author} {\bibfnamefont {A.~V.}\ \bibnamefont
  {Radyushkin}},\ }\bibfield  {title} {\bibinfo {title} {{Transverse momentum
  and Sudakov effects in exclusive QCD processes: Gamma* gamma pi0
  form-factor}},\ }\href {https://doi.org/10.1103/PhysRevD.56.2713} {\bibfield
  {journal} {\bibinfo  {journal} {Phys. Rev. D}\ }\textbf {\bibinfo {volume}
  {56}},\ \bibinfo {pages} {2713} (\bibinfo {year} {1997})},\ \Eprint
  {https://arxiv.org/abs/hep-ph/9702443} {arXiv:hep-ph/9702443} \BibitemShut
  {NoStop}%
\bibitem [{\citenamefont {Pire}\ \emph {et~al.}(2013)\citenamefont {Pire},
  \citenamefont {Semenov-Tian-Shansky},\ and\ \citenamefont
  {Szymanowski}}]{Pire:2013jva}%
  \BibitemOpen
  \bibfield  {author} {\bibinfo {author} {\bibfnamefont {B.}~\bibnamefont
  {Pire}}, \bibinfo {author} {\bibfnamefont {K.}~\bibnamefont
  {Semenov-Tian-Shansky}},\ and\ \bibinfo {author} {\bibfnamefont
  {L.}~\bibnamefont {Szymanowski}},\ }\bibfield  {title} {\bibinfo {title}
  {{QCD description of charmonium plus light meson production in $\bar{p}-N$
  annihilation}},\ }\href {https://doi.org/10.1016/j.physletb.2013.06.015}
  {\bibfield  {journal} {\bibinfo  {journal} {Phys. Lett. B}\ }\textbf
  {\bibinfo {volume} {724}},\ \bibinfo {pages} {99} (\bibinfo {year} {2013})},\
  \bibinfo {note} {[Erratum: Phys.Lett.B 764, 335--335 (2017)]},\ \Eprint
  {https://arxiv.org/abs/1304.6298} {arXiv:1304.6298 [hep-ph]} \BibitemShut
  {NoStop}%
\bibitem [{\citenamefont {Sawada}\ \emph
  {et~al.}(2016{\natexlab{b}})\citenamefont {Sawada}, \citenamefont {Chang},
  \citenamefont {Kumano}, \citenamefont {Peng}, \citenamefont {Sawada},\ and\
  \citenamefont {Tanaka}}]{PhysRevD.93.114034}%
  \BibitemOpen
  \bibfield  {author} {\bibinfo {author} {\bibfnamefont {T.}~\bibnamefont
  {Sawada}}, \bibinfo {author} {\bibfnamefont {W.-C.}\ \bibnamefont {Chang}},
  \bibinfo {author} {\bibfnamefont {S.}~\bibnamefont {Kumano}}, \bibinfo
  {author} {\bibfnamefont {J.-C.}\ \bibnamefont {Peng}}, \bibinfo {author}
  {\bibfnamefont {S.}~\bibnamefont {Sawada}},\ and\ \bibinfo {author}
  {\bibfnamefont {K.}~\bibnamefont {Tanaka}},\ }\bibfield  {title} {\bibinfo
  {title} {{Accessing proton generalized parton distributions and pion
  distribution amplitudes with the exclusive pion-induced Drell-Yan process at
  J-PARC}},\ }\href {https://doi.org/10.1103/PhysRevD.93.114034} {\bibfield
  {journal} {\bibinfo  {journal} {Phys. Rev. D}\ }\textbf {\bibinfo {volume}
  {93}},\ \bibinfo {pages} {114034} (\bibinfo {year}
  {2016}{\natexlab{b}})}\BibitemShut {NoStop}%
\bibitem [{\citenamefont {Singh}\ \emph {et~al.}(2015)\citenamefont {Singh}
  \emph {et~al.}}]{PANDA:2014qiz}%
  \BibitemOpen
  \bibfield  {author} {\bibinfo {author} {\bibfnamefont {B.~P.}\ \bibnamefont
  {Singh}} \emph {et~al.} (\bibinfo {collaboration} {\={P}ANDA}),\ }\bibfield
  {title} {\bibinfo {title} {{Experimental access to Transition Distribution
  Amplitudes with the \={P}ANDA experiment at FAIR}},\ }\href
  {https://doi.org/10.1140/epja/i2015-15107-y} {\bibfield  {journal} {\bibinfo
  {journal} {Eur. Phys. J. A}\ }\textbf {\bibinfo {volume} {51}},\ \bibinfo
  {pages} {107} (\bibinfo {year} {2015})},\ \Eprint
  {https://arxiv.org/abs/1409.0865} {arXiv:1409.0865 [hep-ex]} \BibitemShut
  {NoStop}%
\bibitem [{\citenamefont {Singh}\ \emph {et~al.}(2017)\citenamefont {Singh}
  \emph {et~al.}}]{PANDA:2016scz}%
  \BibitemOpen
  \bibfield  {author} {\bibinfo {author} {\bibfnamefont {B.}~\bibnamefont
  {Singh}} \emph {et~al.} (\bibinfo {collaboration} {\={P}ANDA}),\ }\bibfield
  {title} {\bibinfo {title} {{Feasibility study for the measurement of $\pi N$
  transition distribution amplitudes at \={P}ANDA in $\bar{p}p\to
  J/\psi\pi^0$}},\ }\href {https://doi.org/10.1103/PhysRevD.95.032003}
  {\bibfield  {journal} {\bibinfo  {journal} {Phys. Rev. D}\ }\textbf {\bibinfo
  {volume} {95}},\ \bibinfo {pages} {032003} (\bibinfo {year} {2017})},\
  \Eprint {https://arxiv.org/abs/1610.02149} {arXiv:1610.02149 [nucl-ex]}
  \BibitemShut {NoStop}%
\bibitem [{\citenamefont {Lansberg}\ \emph
  {et~al.}(2012{\natexlab{a}})\citenamefont {Lansberg}, \citenamefont {Pire},
  \citenamefont {Semenov-Tian-Shansky},\ and\ \citenamefont
  {Szymanowski}}]{Lansberg:2011aa}%
  \BibitemOpen
  \bibfield  {author} {\bibinfo {author} {\bibfnamefont {J.~P.}\ \bibnamefont
  {Lansberg}}, \bibinfo {author} {\bibfnamefont {B.}~\bibnamefont {Pire}},
  \bibinfo {author} {\bibfnamefont {K.}~\bibnamefont {Semenov-Tian-Shansky}},\
  and\ \bibinfo {author} {\bibfnamefont {L.}~\bibnamefont {Szymanowski}},\
  }\bibfield  {title} {\bibinfo {title} {{A consistent model for $\pi N$
  transition distribution amplitudes and backward pion electroproduction}},\
  }\href {https://doi.org/10.1103/PhysRevD.85.054021} {\bibfield  {journal}
  {\bibinfo  {journal} {Phys. Rev. D}\ }\textbf {\bibinfo {volume} {85}},\
  \bibinfo {pages} {054021} (\bibinfo {year} {2012}{\natexlab{a}})},\ \Eprint
  {https://arxiv.org/abs/1112.3570} {arXiv:1112.3570 [hep-ph]} \BibitemShut
  {NoStop}%
\bibitem [{\citenamefont {Hakioglu}\ and\ \citenamefont
  {Scadron}(1991)}]{Hakioglu:1991pn}%
  \BibitemOpen
  \bibfield  {author} {\bibinfo {author} {\bibfnamefont {T.}~\bibnamefont
  {Hakioglu}}\ and\ \bibinfo {author} {\bibfnamefont {M.~D.}\ \bibnamefont
  {Scadron}},\ }\bibfield  {title} {\bibinfo {title} {{Vector meson dominance,
  one loop order quark graphs, and the chiral limit}},\ }\href
  {https://doi.org/10.1103/PhysRevD.43.2439} {\bibfield  {journal} {\bibinfo
  {journal} {Phys. Rev. D}\ }\textbf {\bibinfo {volume} {43}},\ \bibinfo
  {pages} {2439} (\bibinfo {year} {1991})}\BibitemShut {NoStop}%
\bibitem [{\citenamefont {Schildknecht}(2006)}]{Schildknecht:2005xr}%
  \BibitemOpen
  \bibfield  {author} {\bibinfo {author} {\bibfnamefont {D.}~\bibnamefont
  {Schildknecht}},\ }\bibfield  {title} {\bibinfo {title} {{Vector meson
  dominance}},\ }\href@noop {} {\bibfield  {journal} {\bibinfo  {journal} {Acta
  Phys. Polon. B}\ }\textbf {\bibinfo {volume} {37}},\ \bibinfo {pages} {595}
  (\bibinfo {year} {2006})},\ \Eprint {https://arxiv.org/abs/hep-ph/0511090}
  {arXiv:hep-ph/0511090} \BibitemShut {NoStop}%
\bibitem [{\citenamefont {Li}\ \emph {et~al.}(2019)\citenamefont {Li} \emph
  {et~al.}}]{JeffersonLabFp:2019gpp}%
  \BibitemOpen
  \bibfield  {author} {\bibinfo {author} {\bibfnamefont {W.~B.}\ \bibnamefont
  {Li}} \emph {et~al.} (\bibinfo {collaboration} {Jefferson Lab
  F\ensuremath{\pi}}),\ }\bibfield  {title} {\bibinfo {title} {{Unique Access
  to $u$-Channel Physics: Exclusive Backward-Angle Omega Meson
  Electroproduction}},\ }\href {https://doi.org/10.1103/PhysRevLett.123.182501}
  {\bibfield  {journal} {\bibinfo  {journal} {Phys. Rev. Lett.}\ }\textbf
  {\bibinfo {volume} {123}},\ \bibinfo {pages} {182501} (\bibinfo {year}
  {2019})},\ \Eprint {https://arxiv.org/abs/1910.00464} {arXiv:1910.00464
  [nucl-ex]} \BibitemShut {NoStop}%
\bibitem [{\citenamefont {Haisch}\ and\ \citenamefont
  {Hala}(2021)}]{Haisch:2021nos}%
  \BibitemOpen
  \bibfield  {author} {\bibinfo {author} {\bibfnamefont {U.}~\bibnamefont
  {Haisch}}\ and\ \bibinfo {author} {\bibfnamefont {A.}~\bibnamefont {Hala}},\
  }\bibfield  {title} {\bibinfo {title} {{Semi-leptonic three-body proton decay
  modes from light-cone sum rules}},\ }\href
  {https://doi.org/10.1007/JHEP11(2021)144} {\bibfield  {journal} {\bibinfo
  {journal} {JHEP}\ }\textbf {\bibinfo {volume} {11}},\ \bibinfo {pages}
  {144}},\ \Eprint {https://arxiv.org/abs/2108.06111} {arXiv:2108.06111
  [hep-ph]} \BibitemShut {NoStop}%
\bibitem [{\citenamefont {Qiu}\ and\ \citenamefont {Yu}(2023)}]{Qiu:2022pla}%
  \BibitemOpen
  \bibfield  {author} {\bibinfo {author} {\bibfnamefont {J.-W.}\ \bibnamefont
  {Qiu}}\ and\ \bibinfo {author} {\bibfnamefont {Z.}~\bibnamefont {Yu}},\
  }\bibfield  {title} {\bibinfo {title} {{Single diffractive hard exclusive
  processes for the study of generalized parton distributions}},\ }\href
  {https://doi.org/10.1103/PhysRevD.107.014007} {\bibfield  {journal} {\bibinfo
   {journal} {Phys. Rev. D}\ }\textbf {\bibinfo {volume} {107}},\ \bibinfo
  {pages} {014007} (\bibinfo {year} {2023})},\ \Eprint
  {https://arxiv.org/abs/2210.07995} {arXiv:2210.07995 [hep-ph]} \BibitemShut
  {NoStop}%
\bibitem [{\citenamefont {Deja}\ \emph {et~al.}(2023)\citenamefont {Deja},
  \citenamefont {Martinez-Fernandez}, \citenamefont {Pire}, \citenamefont
  {Sznajder},\ and\ \citenamefont {Wagner}}]{Deja:2023ahc}%
  \BibitemOpen
  \bibfield  {author} {\bibinfo {author} {\bibfnamefont {K.}~\bibnamefont
  {Deja}}, \bibinfo {author} {\bibfnamefont {V.}~\bibnamefont
  {Martinez-Fernandez}}, \bibinfo {author} {\bibfnamefont {B.}~\bibnamefont
  {Pire}}, \bibinfo {author} {\bibfnamefont {P.}~\bibnamefont {Sznajder}},\
  and\ \bibinfo {author} {\bibfnamefont {J.}~\bibnamefont {Wagner}},\
  }\bibfield  {title} {\bibinfo {title} {{Phenomenology of double deeply
  virtual Compton scattering in the era of new experiments}},\ }\Eprint
  {https://arxiv.org/abs/2303.13668} {arXiv:2303.13668 [hep-ph]}  (\bibinfo
  {year} {2023})\BibitemShut {NoStop}%
\bibitem [{\citenamefont {Stefanis}(1999)}]{Stefanis:1997zyh}%
  \BibitemOpen
  \bibfield  {author} {\bibinfo {author} {\bibfnamefont {N.~G.}\ \bibnamefont
  {Stefanis}},\ }\bibfield  {title} {\bibinfo {title} {{The Physics of
  exclusive reactions in QCD: Theory and phenomenology}},\ }\href
  {https://doi.org/10.1007/s1010599c0007} {\bibfield  {journal} {\bibinfo
  {journal} {Eur. Phys. J. direct}\ }\textbf {\bibinfo {volume} {1}},\ \bibinfo
  {pages} {7} (\bibinfo {year} {1999})},\ \Eprint
  {https://arxiv.org/abs/hep-ph/9911375} {arXiv:hep-ph/9911375} \BibitemShut
  {NoStop}%
\bibitem [{\citenamefont {{L. Pentchev}}()}]{LubomirTrento}%
  \BibitemOpen
  \bibfield  {author} {\bibinfo {author} {\bibnamefont {{L. Pentchev}}},\
  }\bibfield  {title} {\bibinfo {title} {{New opportunities for $J/\psi$ (and
  beyond) photoproduction studies in Hall D with the CEBAF upgrade}},\
  }\bibinfo {note} {a talk presented at ``Opportunities with JLab Energy and
  Luminocity Upgrade'', ECT$^*$ Trento, Italy, 26-30 September 2022;
  \url{https://indico.ectstar.eu/event/152/contributions/3133/attachments/2001/2612/LPentchev\_Jpsi\_Trento.pdf}}\BibitemShut
  {NoStop}%
\bibitem [{\citenamefont {Machleidt}(2001)}]{Machleidt:2000ge}%
  \BibitemOpen
  \bibfield  {author} {\bibinfo {author} {\bibfnamefont {R.}~\bibnamefont
  {Machleidt}},\ }\bibfield  {title} {\bibinfo {title} {{The High precision,
  charge dependent Bonn nucleon-nucleon potential (CD-Bonn)}},\ }\href
  {https://doi.org/10.1103/PhysRevC.63.024001} {\bibfield  {journal} {\bibinfo
  {journal} {Phys. Rev. C}\ }\textbf {\bibinfo {volume} {63}},\ \bibinfo
  {pages} {024001} (\bibinfo {year} {2001})},\ \Eprint
  {https://arxiv.org/abs/nucl-th/0006014} {arXiv:nucl-th/0006014} \BibitemShut
  {NoStop}%
\bibitem [{\citenamefont {Pasquini}\ \emph {et~al.}(2009)\citenamefont
  {Pasquini}, \citenamefont {Pincetti},\ and\ \citenamefont
  {Boffi}}]{Pasquini:2009ki}%
  \BibitemOpen
  \bibfield  {author} {\bibinfo {author} {\bibfnamefont {B.}~\bibnamefont
  {Pasquini}}, \bibinfo {author} {\bibfnamefont {M.}~\bibnamefont {Pincetti}},\
  and\ \bibinfo {author} {\bibfnamefont {S.}~\bibnamefont {Boffi}},\ }\bibfield
   {title} {\bibinfo {title} {{Parton content of the nucleon from distribution
  amplitudes and transition distribution amplitudes}},\ }\href
  {https://doi.org/10.1103/PhysRevD.80.014017} {\bibfield  {journal} {\bibinfo
  {journal} {Phys. Rev. D}\ }\textbf {\bibinfo {volume} {80}},\ \bibinfo
  {pages} {014017} (\bibinfo {year} {2009})},\ \Eprint
  {https://arxiv.org/abs/0905.4018} {arXiv:0905.4018 [hep-ph]} \BibitemShut
  {NoStop}%
\bibitem [{\citenamefont {Yu}\ and\ \citenamefont {Kong}(2019)}]{Yu:2018ydp}%
  \BibitemOpen
  \bibfield  {author} {\bibinfo {author} {\bibfnamefont {B.-G.}\ \bibnamefont
  {Yu}}\ and\ \bibinfo {author} {\bibfnamefont {K.-J.}\ \bibnamefont {Kong}},\
  }\bibfield  {title} {\bibinfo {title} {{Features of $\omega$ photoproduction
  off proton target at backward angles : Role of nucleon Reggeon in $u$-channel
  with parton contributions}},\ }\href
  {https://doi.org/10.1103/PhysRevD.99.014031} {\bibfield  {journal} {\bibinfo
  {journal} {Phys. Rev. D}\ }\textbf {\bibinfo {volume} {99}},\ \bibinfo
  {pages} {014031} (\bibinfo {year} {2019})},\ \Eprint
  {https://arxiv.org/abs/1810.11645} {arXiv:1810.11645 [hep-ph]} \BibitemShut
  {NoStop}%
\bibitem [{\citenamefont {Strakovsky}\ \emph
  {et~al.}(2023{\natexlab{a}})\citenamefont {Strakovsky}, \citenamefont
  {Briscoe}, \citenamefont {Becerra}, \citenamefont {Dugger}, \citenamefont
  {Goldstein}, \citenamefont {Kashevarov}, \citenamefont {Schmidt},
  \citenamefont {Solazzo},\ and\ \citenamefont {Yu}}]{Strakovsky:2022jnx}%
  \BibitemOpen
  \bibfield  {author} {\bibinfo {author} {\bibfnamefont {I.~I.}\ \bibnamefont
  {Strakovsky}}, \bibinfo {author} {\bibfnamefont {W.~J.}\ \bibnamefont
  {Briscoe}}, \bibinfo {author} {\bibfnamefont {O.~C.}\ \bibnamefont
  {Becerra}}, \bibinfo {author} {\bibfnamefont {M.}~\bibnamefont {Dugger}},
  \bibinfo {author} {\bibfnamefont {G.}~\bibnamefont {Goldstein}}, \bibinfo
  {author} {\bibfnamefont {V.~L.}\ \bibnamefont {Kashevarov}}, \bibinfo
  {author} {\bibfnamefont {A.}~\bibnamefont {Schmidt}}, \bibinfo {author}
  {\bibfnamefont {P.}~\bibnamefont {Solazzo}},\ and\ \bibinfo {author}
  {\bibfnamefont {B.-G.}\ \bibnamefont {Yu}},\ }\bibfield  {title} {\bibinfo
  {title} {{Pseudoscalar and scalar meson photoproduction interpreted by Regge
  phenomenology}},\ }\href {https://doi.org/10.1103/PhysRevC.107.015203}
  {\bibfield  {journal} {\bibinfo  {journal} {Phys. Rev. C}\ }\textbf {\bibinfo
  {volume} {107}},\ \bibinfo {pages} {015203} (\bibinfo {year}
  {2023}{\natexlab{a}})},\ \Eprint {https://arxiv.org/abs/2211.12029}
  {arXiv:2211.12029 [hep-ph]} \BibitemShut {NoStop}%
\bibitem [{\citenamefont {Strakovsky}\ \emph
  {et~al.}(2023{\natexlab{b}})\citenamefont {Strakovsky}, \citenamefont
  {Briscoe}, \citenamefont {Chudakov}, \citenamefont {Larin}, \citenamefont
  {Pentchev}, \citenamefont {Schmidt},\ and\ \citenamefont
  {Workman}}]{Strakovsky:2023kqu}%
  \BibitemOpen
  \bibfield  {author} {\bibinfo {author} {\bibfnamefont {I.}~\bibnamefont
  {Strakovsky}}, \bibinfo {author} {\bibfnamefont {W.~J.}\ \bibnamefont
  {Briscoe}}, \bibinfo {author} {\bibfnamefont {E.}~\bibnamefont {Chudakov}},
  \bibinfo {author} {\bibfnamefont {I.}~\bibnamefont {Larin}}, \bibinfo
  {author} {\bibfnamefont {L.}~\bibnamefont {Pentchev}}, \bibinfo {author}
  {\bibfnamefont {A.}~\bibnamefont {Schmidt}},\ and\ \bibinfo {author}
  {\bibfnamefont {R.~L.}\ \bibnamefont {Workman}},\ }\bibfield  {title}
  {\bibinfo {title} {{Is the LHCb $P_c(4312)^+$ plausible in the GlueX $\gamma
  p\to J/\psi p$ total cross sections ?}},\ }\Eprint
  {https://arxiv.org/abs/2304.04924} {arXiv:2304.04924 [hep-ph]}  (\bibinfo
  {year} {2023}{\natexlab{b}})\BibitemShut {NoStop}%
\bibitem [{\citenamefont {Laget}(2021)}]{Laget:2021qwq}%
  \BibitemOpen
  \bibfield  {author} {\bibinfo {author} {\bibfnamefont {J.~M.}\ \bibnamefont
  {Laget}},\ }\bibfield  {title} {\bibinfo {title} {{Unitarity constraints on
  meson electroproduction at backward angles}},\ }\href
  {https://doi.org/10.1103/PhysRevC.104.025202} {\bibfield  {journal} {\bibinfo
   {journal} {Phys. Rev. C}\ }\textbf {\bibinfo {volume} {104}},\ \bibinfo
  {pages} {025202} (\bibinfo {year} {2021})},\ \Eprint
  {https://arxiv.org/abs/2104.13078} {arXiv:2104.13078 [hep-ph]} \BibitemShut
  {NoStop}%
\bibitem [{\citenamefont {Lansberg}\ \emph
  {et~al.}(2012{\natexlab{b}})\citenamefont {Lansberg}, \citenamefont {Pire},
  \citenamefont {Semenov-Tian-Shansky},\ and\ \citenamefont
  {Szymanowski}}]{Lansberg:2012ha}%
  \BibitemOpen
  \bibfield  {author} {\bibinfo {author} {\bibfnamefont {J.~P.}\ \bibnamefont
  {Lansberg}}, \bibinfo {author} {\bibfnamefont {B.}~\bibnamefont {Pire}},
  \bibinfo {author} {\bibfnamefont {K.}~\bibnamefont {Semenov-Tian-Shansky}},\
  and\ \bibinfo {author} {\bibfnamefont {L.}~\bibnamefont {Szymanowski}},\
  }\bibfield  {title} {\bibinfo {title} {{Accessing baryon to meson transition
  distribution amplitudes in meson production in association with a high
  invariant mass lepton pair at GSI-FAIR with \={P}ANDA}},\ }\href
  {https://doi.org/10.1103/PhysRevD.86.114033} {\bibfield  {journal} {\bibinfo
  {journal} {Phys. Rev. D}\ }\textbf {\bibinfo {volume} {86}},\ \bibinfo
  {pages} {114033} (\bibinfo {year} {2012}{\natexlab{b}})},\ \bibinfo {note}
  {[Erratum: Phys.Rev.D 87, 059902 (2013)]},\ \Eprint
  {https://arxiv.org/abs/1210.0126} {arXiv:1210.0126 [hep-ph]} \BibitemShut
  {NoStop}%
\bibitem [{\citenamefont {Lutz}\ \emph {et~al.}(2009)\citenamefont {Lutz} \emph
  {et~al.}}]{PANDA:2009yku}%
  \BibitemOpen
  \bibfield  {author} {\bibinfo {author} {\bibfnamefont {M.~F.~M.}\
  \bibnamefont {Lutz}} \emph {et~al.} (\bibinfo {collaboration} {\={P}ANDA}),\
  }\bibfield  {title} {\bibinfo {title} {{Physics Performance Report for
  \={P}ANDA: Strong Interaction Studies with Antiprotons}},\ }\Eprint
  {https://arxiv.org/abs/0903.3905} {arXiv:0903.3905 [hep-ex]}  (\bibinfo
  {year} {2009})\BibitemShut {NoStop}%
\end{thebibliography}%
\end{document}